\documentclass[twocolumn,twocolappendix]{aastex631}
\usepackage{newtxtext,newtxmath}
\DeclareRobustCommand{\VAN}[3]{#2}
\let\VANthebibliography\thebibliography
\def\thebibliography{\DeclareRobustCommand{\VAN}[3]{##3}\VANthebibliography}

\usepackage{graphicx}	
\usepackage{amsmath}	
\usepackage{gensymb}
\usepackage{makecell}
\usepackage{comment}

\begin{document}

\title[Planes of Satellites in NewHorizon]{Zippers and Twisters: Planes of satellite galaxies emerge from whirling and shocking gas streams in the cosmic web}

\correspondingauthor{Janvi P. Madhani, Charlotte Welker}
\email{jmadhan1@jhu.edu, cwelker@citytech.cuny.edu}

\author[0000-0002-0913-991X]{Janvi P. Madhani$^*$}
\affiliation{Johns Hopkins University, 
3400 N. Charles Street
Baltimore, MD 21218}
\affiliation{NYC College of Technology,
City University of New York,
NY, USA}

\author[0000-0001-5576-0144]{Charlotte Welker$^*$}
\affiliation{NYC College of Technology,
City University of New York,
NY, USA}
\affiliation{Graduate Center,
City University of New York,
NY USA}

\author{Sneha Nair}
\affiliation{NYC College of Technology,
City University of New York,
NY, USA}
\affiliation{Graduate Center,
City University of New York,
NY USA}

\author{Daniel Gallego}
\affiliation{NYC College of Technology,
City University of New York,
NY, USA}

\author{Lianys Feliciano}
\affiliation{NYC College of Technology,
City University of New York,
NY, USA}

\author{Christophe Pichon}
\affiliation{Institut d'Astrophsyique de Paris,
UMR 7095, CNRS, UPMC Univ.,
Paris VI 98 bis boulevard Arago,
Paris, France}
\affiliation{Kyung Hee University, Dept. of Astronomy \& Space Science, Yongin-shi, Gyeonggi-do 17104, Republic of Korea}

\author{Charlotte Olsen}
\affiliation{NYC College of Technology,
City University of New York,
NY, USA}
\affiliation{LSST-DA Catalyst Fellowship,
LSST Discovery Alliance}

\author{Yohan Dubois}
\affiliation{Institut d'Astrophsyique de Paris,
UMR 7095, CNRS, UPMC Univ.,
Paris VI 98 bis boulevard Arago,
Paris, France}

\author{Sugata Kaviraj}
\affiliation{Centre for Astrophysics Research, University of Hertfordshire, Hatfield, AL10 9AB, UK}

\author{Katarina Kraljic}
\affiliation{Aix Marseille Univ, CNRS, CNES, LAM, Marseille, France}


\begin{abstract}
We investigate dwarf satellite systems around massive centrals analogs (Milky Way to Centaurus A) in the NewHorizon simulation. Using simple estimators limiting over-detection, we identify planes of satellites comparable to observations in 30$\%$ to 70$\%$  of cases. The full sample is strongly biased towards arrangements more elongated and co-rotating than their dark-matter host, as early as $z=1$. We identify cosmic filaments and relics of local gas streams outside each system at $z\approx0$ with \texttt{DisPerSE}. We find that the thinner the local stream plane, the thinner the system is. The two align significantly for planar systems. Streams around isotropic systems are not planar. Our analysis reveals two plane types. Ultrathin planes lie orthogonally to their single nearest cosmic filament and align to coherent vortical flows within 3 Mpc, reminiscent of $z>2$ whirls. A second group of  planar systems align to their cosmic filaments. All planes are found in single cosmic filaments skirted by coherent vortical whirls while isotropic systems are found in turbulent flows at the intersection of filaments. We conclude that planes are frequent in $\Lambda$CDM simulations providing the cosmic environment is resolved. Tracking filaments back in time, we show a tight connection between a single, stable filament down to $z\approx0$ and the existence of a plane. “In-filament” planes typically get enhanced by a single, edge-on filament merger at $z<2$ ({\it zipper}) while “vortical” planes’ filaments undergo single twisters (high-orbital momentum zippers) preventing the formation of a core along the filament. In contrast, isotropic systems’ filaments undergo multiple misaligned mergers.
\end{abstract}



\section{Introduction}

In the last two decades, observations have confirmed the existence of thin streams of satellites in planar arrangements around many massive local galaxies \citep[see][for a review]{santossantos2019analysis}. The apparent ubiquity of such well-defined planar structures is at odds with how rare an occurrence they are in cold dark matter cosmological simulations. Thus, the question arises of whether or not observations are compatible with the standard, $\Lambda$ Cold Dark Matter ($\Lambda$CDM), model of cosmology \citep{2005A&A...431..517K}; particularly, whether observations are compatible with what $\Lambda$CDM predicts for the evolution of matter and the formation of galaxies. 

The discovery of planes of dwarf satellites in the Local Group began with our own Milky Way (MW). \cite{1976MNRAS.174..695L} identified a vast polar structure by finding a number of satellite galaxies within the orbital plane of the Magellanic Clouds. More observations in recent years (\cite{2005A&A...431..517K}, \cite{2014ApJ...790...74P}) have confirmed the orbit of the 11 most brightest satellites of the MW to be aligned, on a plane with thickness 20 kpc. 

Soon after, the Pan-Andromeda Archaeological Survey (PAndAS) provided accurate distance measurements to the known
population of M31 satellites using a sensitive homogeneous method
\citep{2012ApJ...758...11C}, revealing their three-dimensional distribution. This allowed for robust confirmation of the previously suspected Great Plane of Andromeda, later followed by a secondary plane in M31 \citep{santossantos2019analysis}. 
Similar planar configurations have been found around other local galaxies like Centaurus A \citep{2015ApJ...802L..25T} and the list is growing. A recent summary of the satellite alignments, or at least stark anisotropies found around close-by galaxies can be found in \citet{2021Galax...9...66P}. Another addition is discussed in \cite{Pawlowski24b}.

Simulations have attempted to recover these thin streams of satellites with limited success. Past simulations have found that MW-type planes exist in no more than a few percent of their systems \citep{2021Galax...9...66P}. This is at odds with their apparent ubiquity in our local universe, as we observe that most massive galaxies with a resolved distribution of dwarf satellites in our neighborhood hosts a plane of satellites. Yet, thin planes that are also as kinematically coherent as the one we observe around our Milky Way remain elusive in simulations. This long-standing tension, a persistent small-scale problem for $\Lambda$CDM, forms the premise of the ``planes of satellites problem" (\cite{2005A&A...431..517K}, \cite{2012PASA...29..395K}, \cite{2021NatAs...5.1185P}). 

Such occurrence rates, of course, may depend on the definition of a Milky Way/Andromeda analog and the combination of parameters with which one can define a plane (spatial extent, orbital pole clustering, kinematic coherence, timescale of a coherent structure, etc.). In simulations, the underlying baryonic physics and/or resolution might erase the signal, but this would grant the identification of specific mechanisms acting on satellites across the virial radius. Past numerical studies can roughly be grouped into two categories: small volumes (box length $<$ 2 Mpc) with high mass and spatial resolution ($<50$ pc, $m_* < 10^4 \, \rm \mathbf{M_\odot}$), and large, cosmological volumes (box length $\approx$ 100 Mpc) with lower resolution of particles ($m_* > 10^6 \, \rm \mathbf{M_\odot}$). 

At this larger, ``cosmological" end of simulation size, \cite{2020MNRAS.491.3042P} finds less than $0.1 \%$ of MW analogs in IllustrisTNG to have orbital pole clustering similar to the satellites co-orbiting on a plane around the Milky Way on the Vast Polar Structure. \cite{2014ApJ...784L...6I} finds an occurrence rate of $0.04\%$ of kinematically coherent and spatially thin alignments around M31 analogs in Millenium-II. This is coherent with many previous dark matter and hydrodynamic studies \citep{Cautun_2015, Forero_Romero_2018, 2018MNRAS.476.1796S}, which find similarly low occurrence rates.

More recently, several small-zoom studies have also investigated this tension in spacial environments, typically using a couple systems similar to the configuration of the Local Group as small-scale probes of cosmology (\citet{Pawlowski_2014}, \citet{2015ApJ...809...49B}, \citet{10.1093/mnras/stv1302}, and \citet{2019ApJ...875..105P}). More recent works \citep{2017MNRAS.466.3119A, 2020MNRAS.491.1471S, 2023NatAs...7..481S, 2024ApJ...975..100G} have shown encouraging results involving specific Local Group dynamics to explain apparent or real planar alignments around the Milky Way. However, these very specific scenarios fail to extend their validity past the narrow focus of Milky Way plane(s) and fall short of explaining the ubiquity of stark anisotropy among the satellite distributions of local massive galaxies.

In a nutshell, suggested solutions to produce planes of satellites include galaxy mergers (\cite{2016ApJ...818...11S},\cite{Kanehisa_2023}), tidal dwarf galaxies (\cite{Wang_2020},\cite{Banik_2022}), the anisotropic accretion of satellites along cosmic filaments (\cite{Aubert_2004}, \cite{Libeskind_2010}, \cite{2011MNRAS.413.3013L}), and the infall of groups of satellites (\cite{2008ApJ...686L..61D}, \cite{Li_2008}, \cite{2013MNRAS.429.1502W}, \cite{2019MNRAS.488.1166S}). These can be broadly classified into ``large-scale, environmental" and ``local dynamical context" solutions. However, the apparent failure of modern simulations (addressing either one of these two mechanistic categories) at reproducing the combination of spatial thinness, kinematic coherence, and ubiquity of satellite planes has maintained their status of cosmological discrepancy. 

 In the present study, we revisit the ``anisotropic infall from cosmic filaments" hypothesis with NewHorizon, a modern ``cosmological zoom" simulation. In particular, we argue that previous simulations lacked either a sufficient resolution or a sufficient volume to properly describe this interplay between filaments that typically feed galaxies more massive than the Milky Way and dwarf satellites, and that a ``local dynamical context" cannot accurately be established without ``large-scale, environmental" processes also resolved at the same time.

Indeed, to properly describe the web-satellite interaction, a simulation must first properly produce dwarf satellites ($M_* < 10^8 \mathbf{M_\odot}$). Previously used cosmological simulations formed stellar particles through a stochastic process with typical minimum masses $>10^6 \mathbf{M_\odot}$, and thus, dwarf galaxies with $M_* < 10^8 \mathbf{M_\odot}$, were not resolved. As a result, simulators used over-produced single stellar particles as a proxy for dwarf satellites, which does not accurately trace luminous satellites. Typically, one requires a 100-fold increase in stellar mass resolution and a maximal spatial resolution of at least 50 pc to properly model dwarf satellites \citep{2021A&A...651A.109D}.

Along with high galaxy-scale resolution, sufficiently large cosmic volumes ($>$ 20 Mpc), hosting full cosmic filaments resolved at least at the scale of plane-of-satellites heights might also be needed. Indeed, small zoom volumes may harbor heavily truncated filaments that could easily be disrupted by the inflow of low-resolution material and by tidal interactions from the lower-resolution regions they were embedded in. 

For our study, we, therefore, look for planes around massive local galaxies analogs (Milky Way to Centaurus A mass) in the NewHorizon simulation \citep{2021A&A...651A.109D}, which lies at the soft spot that ideally combines appropriate volume and resolution.\footnote{A recent work by \cite{Uzeirbegovic24} also uses NewHorizon simulation data with a novel indirect, integrated "planarity" quantification measure. This definition of planarity is broader as it is not restricted to structures centered on the central galaxy. Direct comparison with our work is not straightforward, as this approach also captures a wider range of anisotropies, such as structures associated with infalling groups.}

 We further implement new techniques to analyze and compare different scales of the cosmic web and their possible role in the origin of planes. To better allow for comparisons with future observations, we concentrate on signatures of this interplay existing at $z<0.2$.  

The paper is organized as follows. In Section \ref{sec:Numerical Models}, we summarize the details of the NewHorizon simulation, its relevance to the present study, and discuss the identification of halos and galaxies in the simulation. The methods used to characterize the planar alignments and orbital properties of satellites are then discussed in Section \ref{sec:statistical methods}. The last methodological section, Section \ref{sec:hydro}, details hydrodynamic analysis tools we use in the study, in particular the identification and characterization of cosmic gas filaments and smaller-scale gas streams. Section \ref{sec:census} then presents a census of co-rotating planes of satellites around MW-type hosts in NewHorizon and discuss its compatibility with existing halo population models and observations. In Section  \ref{sec:CosmicWeb}, we show how planar alignments can be understood as an interplay between different scales of gaseous filaments and streams across time and traceable at $z\approx0$.  Finally, in Section \ref{sec:discussion}, we present a general model of formation for satellite planes, and we explore key differences with previous simulations that explain former contradictory results. Our conclusions are summarized in Section \ref{sec:Conclusions}. 
\section{Numerical Methods}
\label{sec:Numerical Models}

In this section, we first describe the NewHorizon simulation and review the numerical methods used to investigate the properties and significance of planes of simulated dwarf satellites.
\subsection{The NewHorizon Simulation}
\label{sec:NewHorizon}
\subsubsection{Technical Summary}
NewHorizon  is a high-resolution hydrodynamic, cosmological zoom run with RAMSES \citep{2002A&A...385..337T}, centered on a sphere of average cosmic density with a co-moving diameter of 20 Mpc. It is re-simulated from the kpc-resolution hydrodynamic run Horizon-AGN \citep{Dubois_2014}. We only briefly summarize its main characteristics here. A complete description can be found in \citep{2021A&A...651A.109D}.

NewHorizon follows $\Lambda$CDM cosmology, with $\Omega_m$ = 0.272, $\Omega_{\Lambda}$ = 0.728,  $\Omega_b$ = 0.045, and $\sigma_8$ = 0.81,  $H_0$ = 70.4 km s\textsuperscript{-1} Mpc \textsuperscript{-1} and $n_s$ = 0.967 following \cite{2011ApJS..192...18K}. 
  DM particle mass is $m_{DM}$ = 1.2 $\times 10^{6} \mathbf{M_\odot}$, while stellar particle mass is $m_{*}= 1.3 \, \times 10^{4} \mathbf{M_\odot}$. A quasi-Lagrangian AMR scheme provides a maximal spatial resolution of $\Delta x \approx$ 35 pc across redshifts. 

It includes elaborate prescriptions for cooling below 10\textsuperscript{4} K, heating from a uniform UV background after z\textsubscript{reion} = 10, turbulent star formation, thermal and mechanical feedback from supernovae Type Ia feedback and stellar winds following \cite{2015MNRAS.451.2900K}, and bimodal (quasar and jet modes), spin-dependent AGN feedback. 
\subsubsection{Relevance of NewHorizon to study satellite planes}

\begin{figure*}
    \centering
    \includegraphics[width=\linewidth]{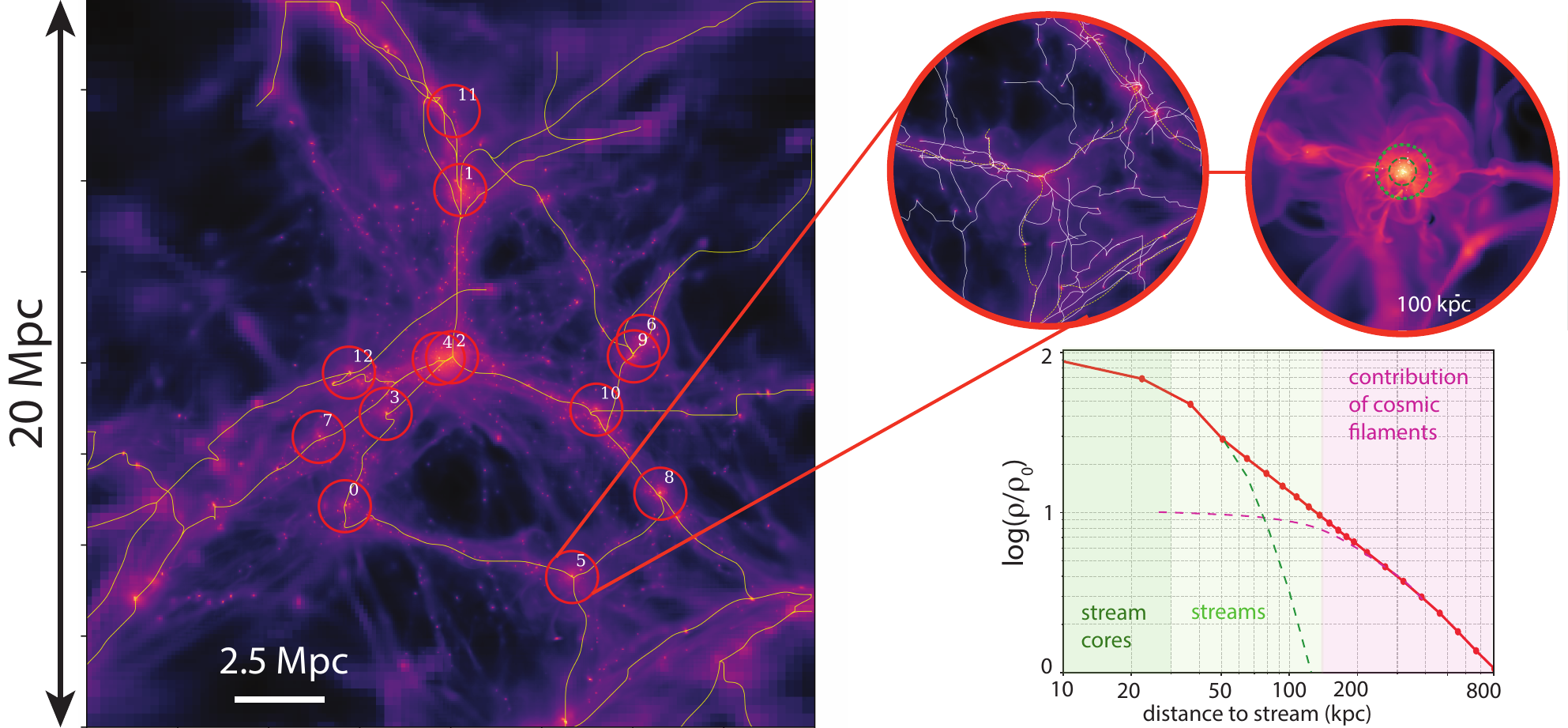} \caption{The 13 MW mass systems at $z=0.17$ overlaid as red circles on the full projected gas density map of NewHorizon. Yellow lines indicate large-scale cosmic filaments. Zoom inserts focus on $\approx (5\, \rm Mpc)^3$ around one example system. White lines correspond to local gas streams, yellow dashed lines to cosmic filaments. First insert shows a projection of the gas density across 5 Mpc. Second insert shows a 14-kpc thick slice of gas density, centered on the system, with satellites as white circles and $R_{\rm vir}$ and 2 $R_{\rm vir}$ as green dashed circles. The stacked gas density profile of all streams between $R_{\rm vir}$ and 3 Mpc of the system is also displayed, revealing both the sharp, $<30$ kpc core of streams and the wider Mpc-scape thickness of cosmic filaments around them.}
    \label{fig:mainmap}
\end{figure*}

This volume contains dozens of systems above Milky-Way mass at $z=0.17$ (the lowest available redshift). Its large volume allows for the full reconstruction of both the cosmic and fine-grained cosmic web around these galaxies while its maximal spatial resolution of 35 pc allows for precise modeling of the formation of dwarf satellite streams along cosmic filaments and of the individual structures of galaxies, as can be seen on Figure \ref{fig:mainmap}. 

Indeed, NewHorizon resolves dwarfs down to stellar masses of $M_* \approx 10^{5.8} \mathbf{M_\odot}$ with at least 50 stellar particles, amounting to more than 5200 dwarfs at $z<0.2$. Galaxies with ${M_* > 10^{6.5} \mathbf{M_\odot}}$ are resolved with at least 240 particles, allowing for accurate reconstruction of the dwarfs angular momentum and structure. It is also notable that the resolution in filaments several Mpc away for Milky Ways and more massive galaxies remains within 35 kpc, ensuring that thinner gas streams comparable to the height of satellite planes can be resolved, as is visible on the right inserts of Fig.~\ref{fig:mainmap}.

\subsection{Halo and Galaxy Identification}
\label{sec:sample selection}

We identify halos and galaxies using the density-based structure finder \texttt{HALOMAKER} \citep{2009A&A...506..647T}, a modified 3D friend-of-friend program using the \texttt{AdaptaHOP} algorithm \cite{Aubert_2004} to identify both structures and substructures. Galaxies are identified from stellar particles down to $6.5\,\times10^{5}\, \mathbf{M_\odot}$ (50 particles), while halos are identified using DM particles down to $7\,\times10^{7}\, \mathbf{M_\odot}$ (50 particles).

We identify local massive galaxy analogs conservatively, as any system containing a central galaxy with $M_* > 10^{10.5} \mathbf{M_\odot}$ and at least 4 identified luminous satellites within the virial radius $R_{\rm vir}$ of a halo with $ 10^{11.5} \mathbf{M_\odot} \leq M_{DM} \leq 10^{13} \mathbf{M_\odot}$. The higher mass end allows to include analogs of systems like Centaurus A. For each of these systems, we identify all luminous satellite galaxies that are within 1 $R_{\rm vir}$ and within 2 $R_{\rm vir}$ of the host DM halo. We excluded all substructures, including substructures of another satellite/satellite halo from the analysis to not artificially boost the alignments/co-rotation. Most of these satellites are dwarf galaxies with $10^{6.5} \mathbf{M_{\odot}}< M_{*} <10^9 \mathbf{M_{\odot}}$. We control that the level of contamination from low-resolution particles within 5 $R_{\rm vir}$ is $<5\%$. Using this set of requirements, we identify 13 MW type systems in NewHorizon at $z<0.2$, as shown in Figure \ref{fig:mainmap}. 12 out of 13 systems have more than 7 satellites within 2 $R_{\rm vir}$ and will constitute our main sample. System 11 had only 4 satellites. Because of this low number of satellites, its specific effect on the results is analyzed separately whenever included in the study. 

Virial radii for our systems are calculated as $R_{\rm 200}$ and typically range from 200 to 330 kpc, with only the three most massive systems 1, 2 and 4 ($M_{*}>10^{11.3} {\rm M_{\odot}}, M_{\rm DM}>10^{12.7} \, \rm M_{\odot}$ ) between 350 and 450 kpc.
In the following study, results are presented for satellite populations identified within 2 $R_{\rm vir}$, in accordance with typical volumes probed to identify satellite planes in observations \citep[see][for a review]{Pawlowski21}. For instance, the Vast Polar structure of the Milky Way has members out to 250 kpc ($\approx 1.3 \, R_{\rm vir}$), the Great Plane of Andromeda includes satellites out to $\approx 1.5 \, R_{\rm vir}$ ($>$300 kpc) and planes around Centaurus A extend to 800 kpc, approximately $2\, R_{\rm vir}$. However, as is the case for these observed systems, planar configurations in NewHorizon tend to be dominated by satellites within 1 $R_{\rm vir}$, which are significantly more numerous, as is visible in Fig.~\ref{fig:corotplane} and in Appendix.~\ref{appendix:embedded}. Our main results restricted to satellites within 1 $R_{\rm vir}$ are fully compatible with the rest of our analysis and are presented in Appendix.~\ref{appendix:1rvir}. Note that all systems with more than 7 satellites within 2 $R_{\rm vir}$ also have more than 7 satellites within 1 $R_{\rm vir}$, with the exception of System 7 which has 6. In addition, 9 systems have more than 10 satellites within $R_{\rm vir}$.

\begin{figure*}
    \centering
    \includegraphics[width=\linewidth]{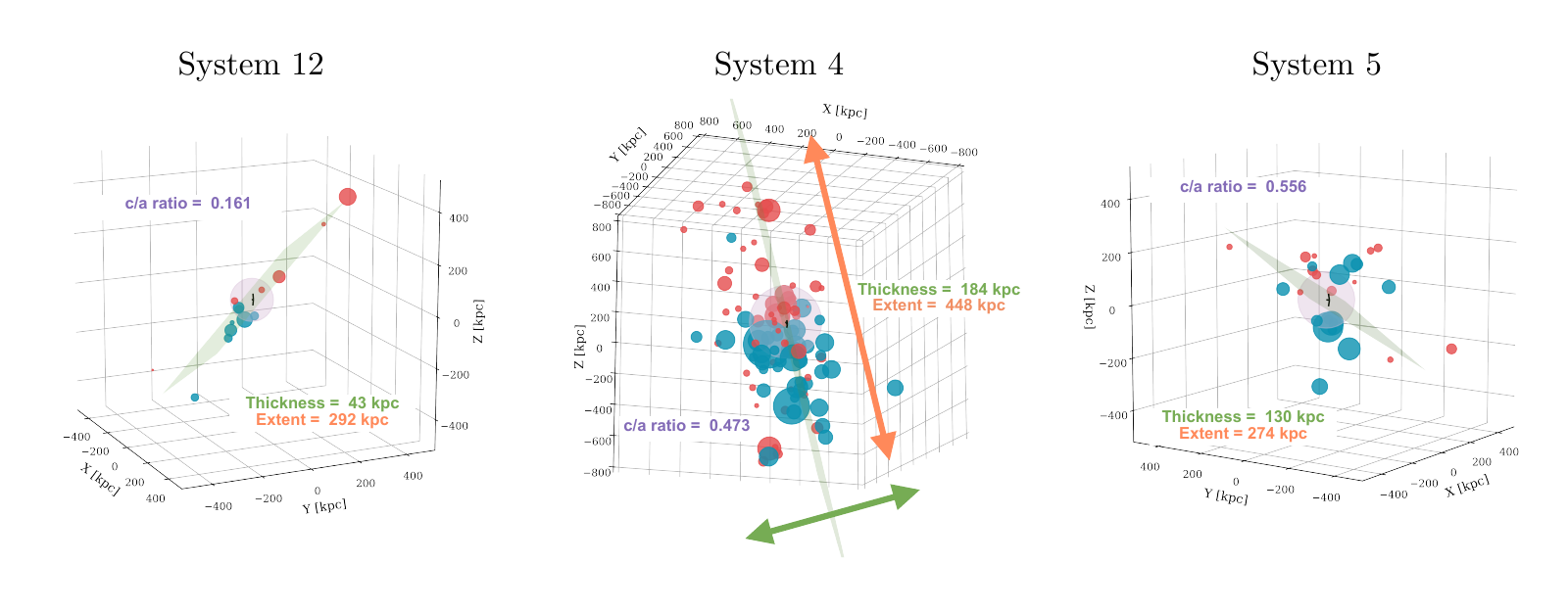}
    \caption{Example of best-fit planes identified for three Local massive galaxy analogs in NewHorizon covering the full range of possible satellite distributions: System 12 has an ultra-thin co-rotating disc, system 4 is a borderline system with strong anisotropy and signs of corotation, system 5 is mostly isotropic. Satellites and their sub-halos appear as circles color-coded by their velocity along a line of sight (blue:approaching,  red:receding). The central galaxy appears as light purple. The best-fit satellite plane is shown in light green.  Its thickness and extent are annotated in orange and green.}
    \label{fig:corotplane}
\end{figure*}

Note that populations of dwarf galaxies in New Horizon have been thoroughly studied, generally showing a very good match with observations in terms of mass function, Sersic profiles and ellipticities, and a reasonable match in terms of sizes and surface brightness, although discrepancies of a factor $\approx$ 1.5 can still exist for the latter due between $10^{7.5}\,\rm M_{\odot}$ and $10^{9}\,\rm M_{\odot}$ \citep{Jackson21, Watkins25, Martin25} due to high sensitivity to specific feedback implementations.

\section{Spatial Analysis of Satellites: Methods}
\label{sec:statistical methods}

To characterize the anisotropy and co-rotation of the distribution of satellites around MW analogs, we developed a plane-finding algorithm that mimics the methods used in observations. This algorithm is described hereafter. 
We further characterize the properties, co-rotation features and statistical significance of planes with estimators detailed in this section. 

\subsection{Best-fit Plane Finder}
We identify the best-fit plane candidate as the plane that minimizes the root mean square distances of satellites from the plane, $d_{\rm RMS}$.

Conceptually, the best-fit algorithm draws and tests random planes (characterized by their normal vector) until the minimum distance from the satellites to a plane is found. This is achieved by drawing random unit vectors as candidate normal vector $\vec{n} $ for a plane from a spherically uniform distribution. However, this brute-force approach is computationally expensive and convergence may be hindered by the presence of local minima.
To boost efficiency and avoid this bias, we sample planes following an evolutionary approach, described in Appendix ~\ref{appendix:EA}.
With this approach, satisfactory convergence is obtained after $\sim 500$ iterations as opposed to up to $10,000$ with a brute-force algorithm.

Note that these planes are all centered on the central host galaxy. In practice, relaxing this constraint did not lead to significant differences in the best-fit planes.

Note also that this method could, in theory, be used iteratively to identify several, potentially thinner planes for each system. However, we refrain from this as such a method can lead to increase false positive rates and overestimate the significance and thinness of planes characterized by too few satellites. In this study, our aim is to focus on the primary anisotropy of satellite systems.

\subsection{Measuring the anisotropy of satellite distributions}

\subsubsection{Planarity}

The {\it thickness}, $\Delta h$, of a plane is defined as twice the RMS of the vertical heights of satellites in the frame of the plane. While the {\it extent},  $e$, of a plane is defined as twice the RMS of the horizontal (planar) components of their separation vector, the vector pointing from the center of the satellite to the center of the host galaxy.

For each system, we then define its {\it thinness}, $\pi_{\rm sat}$,  as the ratio between the thickness of the plane versus its extent:
 \[ \pi_{\rm sat}=\frac{\Delta h}{e}. \]
 The smaller the thinness, the thinner the plane. This definition is in line with height and extent often presented in observations of local galaxies such as M31 \citep{2013Natur.493...62I}.

\subsubsection{Inertia Tensor Analysis}
\label{sec:inertiatensor}

Another widely used shape metric for both satellite systems and host halos consists in deriving the ellipsoidal axis ratios from the inertia tensor of the structure. The true inertia tensor of a satellite system is defined as:
$ I_{ij} = \frac{1}{M}\sum_{n} m_{(n)}x_{i}^{(n)} x_{j}^{(n)}$ with $i,j =  \{ 1,2,3\}$ ranges the three cartesian axes, $m_{(n)}$ is the mass of the individual satellite and $M$ is the total mass of all the satellites.

However, it tends to give too much importance to rare massive satellites, therefore overestimating the anisotropy of a system. Moreover, it is not always derivable for observed systems. To remain consistent with observations, we therefore use the geometric (non mass-weighted) inertia tensor: 
\[  I^{\rm G} _{ij} = \sum_{n} x_{i}^{(n)} x_{j}^{(n)} \]
The axis lengths are calculated from the eigenvalues of $I$ as:
\[ \! a\! =\! \frac{5}{2}\sqrt{\lambda_a + \lambda_b - \lambda_c}, 
 b\! =\! \frac{5}{2}\sqrt{\lambda_a+\lambda_c - \lambda_b}, 
 c\! =\! \frac{5}{2}\sqrt{\lambda_b + \lambda_c - \lambda_a } \] 
where $\lambda_a > \lambda_b > \lambda_c$, so $a>b>c$.

The minor-to-major axis ratio, $c/a$ is taken as a proxy of how spatially thin the distribution of satellites is. $c/a \approx 0 $ implies a perfectly planar distribution of satellites, while a $c/a\approx 1 $ corresponds to a fully isotropic distribution. 
Results are presented with the geometric inertial axis ratio, as opposed to the mass-weighted inertial axis ratio, hereafter referred to as {\it inertial c/a} although we checked that both definitions give similar results.

Further, we also calculate the ``on-plane" inertial c/a by restricting this definition to satellites lying within 1-RMS of the best-fit plane identified with our finder (the ``on-plane" satellites) if the $c/a$ ratio of the full distribution is $>$ 0.15, consistent with observational studies that differentiate in-plane and off-plane systems in complex distributions. 

If the full-system $c/a$ ratio is $<$ 0.15, the distribution of total satellites is already more planar than observed planes. In such a case, we consider all satellites to be ``on-plane" to avoid artificially removing satellites that are not significantly off the structure. We summarize these definitions in Table \ref{table:c2a_definition}. 

Finally, while we also tested a conservative definition of ``on-plane" satellites by calculating the inertial $c/a$ ratio of strictly the satellites that lie within 1-RMS of the best-fit plane, we did not include it in our final analysis. The more conservative definition showed the same general trend except for System 7, for which it decreased the axis ratio to artificially small values (ex: System 7 went from having $c/a$ ``on-plane" of 0.13 to  $c/a$ ``within 1-RMS" of 0.07). 

\begin{table*}
\begin{tabular}{ccccccccc}
\hline \hline
Systems & $N_{\rm sat}$ & $N_{\rm sat}$ [On-Plane]  & $e$ ($kpc$) & $\Delta h$ ($kpc$) &  $\Delta h$ [On-Plane] ($kpc$) & $\pi_{\rm sat}$ & $c/a$ & $c/a$ [On-Plane] \\
\hline
0 & 16 & 12 & 255.180 & 86.189 & 34.043 & 0.338 & 0.389 & 0.132 \\
1 & 60 & 44 & 375.335 & 194.871 & 81.136 & 0.519 & 0.690 & 0.340 \\
2 & 84 & 66 & 458.738 & 252.723 & 116.896 & 0.551 & 0.692 & 0.325 \\
3 & 17 & 13 & 242.766 & 94.405 & 54.675 & 0.389 & 0.456 & 0.258 \\
4 & 104 & 69 & 448.319 & 183.831 & 82.546 & 0.410 & 0.473 & 0.221 \\
5 & 26 & 20 & 273.771 & 129.777 & 72.147 & 0.474 & 0.556 & 0.251 \\
6 & 21 & 14 & 279.062 & 92.830 & 48.388 & 0.333 & 0.397 & 0.204 \\
7 & 7 & 7 & 171.849 & 20.078 & 20.078 & 0.117 & 0.130 & 0.130 \\
8 & 32 & 24 & 343.932 & 114.489 & 56.973 & 0.333 & 0.388 & 0.187 \\
9 & 17 & 15 & 286.505 & 71.416 & 40.312 & 0.249 & 0.288 & 0.129 \\
10 & 7 & 4 & 105.157 & 29.275 & 20.378 & 0.278 & 0.349 & 0.027 \\
11 & 4 & 3 & 358.495 & 88.742 & 59.862 & 0.248 & 0.270 & 0.089 \\
12 & 14 & 11 & 291.672 & 42.835 & 18.249 & 0.147 & 0.161 & 0.066 \\
\hline
\end{tabular}
\caption{Main properties of each system of satellites. This includes: total number of satellites, number of satellites ``on-plane,"  extent $e$ and height $\Delta h$ of the best-fit plane, for all satellites and only on-plane satellites, respectively, thinness, $\pi_{\rm sat}$, inertial $c/a$ ratio for all satellites and restricted to``on-plane" satellites, respectively. } 
\label{table:c2a_definition}
\end{table*}

Note that, while we explore all these anisotropy parameters for the sake of completeness, we find that thinness, inertial $c/a$, $c/a$ within 1-RMS, and  $c/a$ ``on-plane" are all highly correlated to one another in our sample, see Appendix \ref{appendix:metric_comparisons}, highlighting that systems exhibiting the thinnest planes are also the ones with the most anisotropic satellite distributions overall.
\subsection{Co-rotating Fraction}

Various methods exist to assess the kinematic coherence of satellites in a plane. For simplicity, we first choose a definition in line with most striking observational studies in the Local Universe \citep{2012ApJ...758...11C,2015ApJ...802L..25T}. It has the advantage of being easy to extend to outer Local Universe observations, unlike orbital pole-based metrics, which are confined to the spectroscopic analysis of the Milky Way and highly dependent on the measurement of the specific motion of the Milky Way in its host halo.

We define the fraction of co-rotating satellites, $R_{\rm cr}$, as follows:

We define a random line of sight ($\rm los$) that lies on the best-fit plane, $\overrightarrow{\rm los}$ and is centered on the central galaxy. We then calculate each satellite's velocity along the $\overrightarrow{\rm los}$, $v_{\rm los}$. This allows us to define:

\begin{itemize}
\item $N_L^{(-)}$: number of satellites left of the $\overrightarrow{los}$, with $v_{\rm los}<0$

\item $N_L^{(+)}$: number of satellites left of the $\overrightarrow{los}$, with $v_{\rm los}>0$

\item $N_R^{(-)}$: number of satellites right of the $\overrightarrow{los}$, with $v_{\rm los}<0$

\item $N_R^{(+)}$: number of satellites right of the $\overrightarrow{los}$, with $v_{\rm los}>0$
\end{itemize}

The clockwise and counterclockwise co-rotating fractions are, therefore, respectively:

\[  R_{\rm cr}^{\rm down} = \frac{N_R^{(+)}+ N_L^{(-)}}{N_{\rm tot}} \, {\rm and} \, \, R_{\rm cr}^{\rm up} = \frac{N_L^{(+)}+ N_R^{(-)}}{N_{\rm tot}}.\]

The {\it co-rotating fraction} $ R_{\rm cr}$ is the maximum of the two. 
$ R_{\rm cr} = 1$ corresponds to a fully co-orbiting population satellites while $ R_{\rm cr} = 0.5$ shows a complete absence of synchronicity. As illustrated in Figure \ref{fig:corotplane}, we ensure that $R_{\rm cr}$ is calculated for a line of sight, for which satellites are spatially spread out on the sky, to avoid noise due to the accumulation of satellites along one line of sight when projected distances fall mostly within $10\%$ of the center. Such cases would be dominated by noise and would not mimic any reasonable choice of observed system. 

\subsection{Orbital pole alignment}

As mentioned previously, another metric of the degree of co-rotation of satellites commonly used for the Milky Way, for which satellite orbital reconstruction is possible, is the orbital pole dispersion. To provide a thorough analysis of co-rotation in NewHorizon, we compute the orbital pole dispersion for all MW-type systems as follows:
\[  \overline{\Delta \theta}= \sqrt{\frac{1}{N_{\rm sat}}\sum_{i}^{N_{\rm sat}} \theta_i^2} \, ,\]
with $N_{\rm sat}$ the number of satellites in the system and $\theta_i$ the angle between the orbital pole of satellite $i$ and the average orbital pole of the full system. This is directly consistent with the definition of \cite{2020MNRAS.491.3042P}.

The shape and anisotropy parameters of the 13 MW systems are summarized in Table \ref{table:census}. 

\begin{table}[t]
\centering
\begin{tabular}{ccccc}
\hline \hline
System & $R_{\rm cr}$ & $\overline{\Delta \theta}$ (deg) & $R_{\rm cr}$ [On-Plane] & $\overline{\Delta \theta}$ (deg) [On-Plane] \\
\hline
0  & 0.750 & 35.50 & 0.750 & 30.44 \\
1  & 0.650 & 35.18 & 0.636 & 45.45 \\
2  & 0.571 & 39.00 & 0.591 & 36.38 \\
3  & 0.760 & 38.15 & 0.692 & 38.02 \\
4  & 0.769 & 40.66 & 0.797 & 39.53 \\
5  & 0.654 & 32.69 & 0.700 & 30.31 \\
6  & 0.714 & 34.33 & 0.714 & 35.46 \\
7  & 1.000 & 35.60 & 1.000 & 35.60 \\
8  & 0.688 & 31.04 & 0.708 & 28.74 \\
9  & 0.765 & 31.56 & 0.867 & 31.04 \\
10 & 0.714 & 22.26 & 1.000 & 6.16 \\
11 & 0.750 & 40.67 & 1.000 & 15.16 \\
12 & 0.929 & 17.41 & 1.000 & 17.41 \\
\hline
\end{tabular}
\caption{Census of co-rotation parameters, namely the co-rotating fraction and orbital pole dispersion for the full sample. Parameters are calculated by considering total satellites as well as just satellites defined as ``on-plane."}
\label{table:census}
\end{table}

\subsection{Statistical Methods: Significance of a Plane}
\label{sec:significance of plane}

For each system of satellites, we assess throughout the study whether the anisotropy/co-rotation of the best-fit candidate plane represents a significant departure from naive (non-baryonic) $\rm \Lambda CDM$ assembly models, typically used in observations to assess the significance of a plane. We tested best-fit planes against the two most common models for the distribution of satellites. The first one is the isotropic assumption. This emulates observational studies, which typically test their ``planar" satellite distributions against the null hypothesis that satellites distribute isotropically around the central galaxy.

The second reflects the assumption often presented as more realistic that the distribution of satellites typically relaxes with its host halo and should therefore follow the shape of said halo \citep[see references in][which however already finds that it still underestimates anisotropies]{2018A&A...613A...4W}. This takes into account the fact that the anisotropy of the halo might explain some of the anisotropy of observed satellite planes, therefore avoiding overestimating the physical significance of a planar distribution. 

We find results to be fully consistent for both models; therefore, in the rest of the study, we present results only with the latter, more realistic model: the halo-tracing hypothesis.

\subsubsection{Single system significance}

We first evaluate how outstanding the anisotropy of a system is compared to what is expected in the halo-tracing hypothesis for a halo of the same shape, satellite abundance, and satellite radial distribution. This means that the potential contribution of low sampling from luminous satellites is always taken into account in the null hypothesis. So, parameters are always compared to their expected distribution in the halo-tracing hypothesis, at similar sampling.

The null hypothesis distribution is obtained by re-sampling each system's satellite distribution 25,000 times, maintaining the number of satellites and their radial distribution constant, and constraining the satellites to follow the ellipsoidal shape of their host. The shape of each main host halo is approximated as an ellipsoid with its axis $c<b<a$ derived from the inertia tensor of DM particles within 1 $R_{\rm rvir}$. To avoid bias due to the presence of sub-halos  (associated with luminous satellites), DM particles within $R_{\rm rvir}^{sub}$ of each sub-halo are first excluded and then re-sampled to the average radial density of the host halo at the sub-halo's position. Satellite separation vectors (directions) are then drawn from a uniform distribution of points on the surface of an ellipsoid of axis $a$,$b$, and $c$ and of circularized radius $R_{\rm circ}=(abc)^{1/3}$=1. The distribution of $R_{\rm circ}$ is then sampled to match the actual radial distribution of satellites of each tested system.
We further tested that drawing $R_{\rm circ}$ from an NFW profile and an Einasto profile fitted to the DM host halo of each system did not significantly modify results. Carefully controlling for the radial distribution avoids over-estimating the significance of a ``plane" that may be mostly an artifact of a concentrated distribution of satellites.

For each synthetic sample distribution, we then identify the best-fit plane and compute the shape and thinness metrics defined in the previous sections. We therefore obtain a null distribution of every single parameter under the assumption that satellites follow the shape of their host halo. From this, we can evaluate the degree of significance for the anisotropy of each best-fit plane for the actual systems.

For the co-rotation fraction, results are presented for the null hypothesis of uniformly random line-of-sight velocities for satellites. The spin of host DM halos is not typically aligned with best-fit planes, hence drawing satellite velocities from a collective spin distribution centered on the DM halo spin did not lower significance.

For the orbital poles, results are presented for the null hypothesis obtained drawing orbital poles randomly distributed on the sphere.

In parts of our analysis, we focus on ``on-plane satellites", as defined in Sec.~\ref{sec:inertiatensor}. In this case, the null hypothesis is always obtained by applying exactly the same satellite selection procedure to the synthetic resampled distributions as to the original distribution. Hence, all parameters are calculated the same way.

\subsubsection{Population significance}

In observations (and follow-up cosmological numerical studies), the anomalous thinness of a plane and its incompatibility with $\rm \Lambda$CDM is often estimated irrespective of a given halo's specific morphology (which is usually poorly constrained), typically by simple comparison with large samples of simulated galaxies in the same mass range. As such a relatively massive system (like System 4) can be very anisotropic compared to its halo's morphology while still not being strikingly anisotropic at the population level, i.e. when simply compared to other Milky Way mass galaxies. On the other hand, some systems might be very planar because their halo is particularly flattened or spinning (recent mergers, position in the cosmic web, etc.). In this case, satellite orbits would not stand out as significantly abnormal on an individual basis when this information is available, but these would still be very planar compared to similar mass systems in general. To better estimate whether NewHorizon produces a population of Milky Way mass halos with satellite distributions that would appear anomalously thin compared to general Milky Way-type expectations, we also conduct a population-level analysis.

First, we construct a composite null distribution of thinness/co-rotation parameters for the main sample of 12 systems (excluding System 11, due to its too few satellites, which generate discrete artifacts in the co-rotation parameter distribution). To do so, we generate 300,000 synthetic distributions of satellites, under the assumption that satellites trace the shape of their halo. To properly account for the frequency of each system in the original sample, we assume that each system's structure (halo shape, number of satellites, and radial distribution) is equally likely to occur. 

We then follow the same procedure as in the previous section, identifying for each synthetic system its best-fit plane and computing its shape and co-rotation metrics, which then allows us to compute the null distributions of these parameters expected in NewHorizon (and in the nearby Universe assuming NewHorizon is representative for Milky Way mass galaxies) in the halo-tracing hypothesis. 

\section{Methods of hydrodynamic analysis}
\label{sec:hydro}

\subsection{Gas filament/stream identification and connectivity}
\label{sec:filament_identification}

We reconstruct cosmic filaments with the public, commonly used \texttt{DisPerSE} \citep{2011MNRAS.414..384S} software, which can be used across simulations and spectroscopic surveys in a consistent fashion \citep{2020MNRAS.491.2864W, 2020MNRAS.493..362K,2024MNRAS.527.1301R, 2022MNRAS.512..926K}. This topology-based structure finder identifies filaments as the ridge lines of an input density field (which can be a set of physical tracers, like a galaxy catalog, or a continuous field sampled on a regular grid). A ridge line connects a local maximum of the density field to a local filament-type saddle point, which form a ``critical pair." Noise is filtered out from significant filaments based on a persistence threshold. Persistence, here expressed as a signal-to-noise ratio, quantifies the relative density contrast of a critical pair in the field. The output is a fully connected network of filaments defined as sets of contiguous segments (or equivalently ``sampling points" ) connecting peaks to filament-type saddle points.

\subsubsection{Adapting to complexity on small scales}

In the present study, our aim is to study the interactions between different scales of the gaseous cosmic web. However, the inherently multiscale nature of the cosmic web is a source of extra complexity for the gas density field.
Although it is customary to extract \texttt{DisPerSE} filaments from galaxy/halo catalogs, its use to identify gas filaments is still limited to very large scales and heavily smoothed fields. Indeed, the complex hydrodynamics of filaments on small, non-linear scales (particularly $<5 \,\rm Mpc$) are responsible for strong deviations from the tight correlation the persistence of critical pairs and the underlying gas density display on large scales. For instance, we notice that some thinner cosmic filaments display relatively low persistence despite their high core density and length. Simple use of persistence cuts is all the more unreliable for thin streams on smaller scales.

In the present paper, we first extract the largest cosmic filaments ($\approx$1-2 Mpc wide) using the standard strong-smoothing technique. Since this technique misses some thinner yet similarly dense cosmic filaments, in particular some with a node outside of the zoomed region, or pairs of  filaments late in the process of merging into one, we also use a technique we develop and detail in a companion paper (Madhani et al., in prep). We use it to extract both thinner cosmic filaments and gas streams on smaller scales. In a nutshell, in this second technique, after first sampling the initial density field on a regular grid, we apply a minimal Gaussian smoothing over neighboring pixels followed by a correcting non-linear transformation to the resulting sampled field. The final grid is used as an input for \texttt{DisPerSE}. 

\subsubsection{ The two-scale decomposition approach}

To take into account the multiscale nature of the cosmic web, we therefore adopt a ``Two Scale decomposition" approach. This can be summarized as follows:
\begin{itemize}
    \item We reconstruct the larger scale {\bf Cosmic Filaments} by applying \texttt{DisPerSE} on a volume of $(20 \, \rm Mpc)^3$ centered on NewHorizon. The density field is sampled on a $175^3$ pixel box ($\Delta x_0= 114$ kpc)and smoothed using a Gaussian filter with a standard deviation of 7 pixels. At this scale, and without doing any transformation on the density, we extract the largest, highest persistence ($4\sigma$) cosmic filaments with a typical $\sim1-2\, \rm Mpc$ thickness. This is our main sample, typically connecting nodes (halos) $>10^{12.5} \, \mathbf{M_{\odot}}$. It is displayed on Fig.~\ref{fig:mainmap}. This best characterizes the early and  long-term tidal environment of our Milky-Way systems, which are found either halfway along one single such filament or at the node of a few ones.
    \item To extend our sample of {\bf Cosmic Filaments} to all similarly dense filaments extending over at least several Mpc, we also apply the second technique to the same initial grid. These filaments are presented in Appendix \ref{appendix:5scosmic_fils}. All our MW systems are typically nodes of 2-4 such filaments which best represent large-scale, later-time broad accretion channels.
    \item Finally, we reconstruct the {\bf Local Streams} around each local massive galaxy system (intermediate-to-small scale filaments connecting to the CGM, representing local accretion channels) by applying \texttt{DisPerSE} using a volume that typically encompasses $8 R_{\rm vir}$ around each system, with on average a volume of $(5 \, \rm Mpc)^3$ centered around the Milky Way mass halo. The density field is sampled on  $200^3$ pixels on average box (adapted to the exact size of each box to reach a resolution $\Delta x= 14$ kpc). The density field is smoothed over 2 pixels and non-linearly transformed in the same way as for the extended cosmic filaments sample. 
\end{itemize}

The two scales described are depicted in Figure \ref{fig:mainmap}. This two-scale approach is essentially similar to the approach suggested in recent analytical studies \citep{2015MNRAS.452.3369C} and in the MTNG simulation, on the galaxy field \citep{2024A&A...684A..63G}. The right panel of Fig.~\ref{fig:mainmap} further shows the stacked gas density profiles of streams identified between $R_{\rm vir}$ and 3 Mpc of System 5. This confirms the physicality of gas streams, which display a narrow core ($< 30$ kpc) and a steep decline in density within 150 kpc where the density profile becomes dominated by the larger cosmic filaments that host many of the streams. Similar profiles are obtained around all our systems, with core density contrasts typically between 100 and 200. Note that these profiles are obtained from gas density interpolated on a regular grid with pixel $\Delta x= 14$ kpc. Some streams are likely to exhibit even sharper density cores. 

We further define stream connectivity as the number of streams that breach into the virial radius of each system.

\subsection{Anisotropy of local gas streams}
\label{sec:stream-plane}

Local gas streams identified by \texttt{DisPerSE} are effectively defined as sets of sample points (the segment extremities) connecting a density peak to a saddle point. It is, therefore, straightforward to re-purpose the methods developed in previous subsections to assess the thinness and elongation of the population of local streams connecting to each individual Local massive galaxy analog.

To this end, we re-purpose the best-fit plane finder, but satellite positions are simply replaced by the sampling points of streams connected to a particular system and found between $1 R_{\rm vir}$ and $2 R_{\rm vir}$ (see Appendix \ref{appendix:stream-plane}). We, therefore, estimate the best-fit plane, thinness, and axis ratio of the system of streams connected to each individual Milky Way mass and above system. We refer to this plane as the {\bf stream-plane}. Once the stream-plane has been identified, we calculate the {\bf stream-plane angle}, $\theta_{\rm stream}$,  of the gas stream-plane with the plane of satellites, defined as the angle between the normal of both planes. An example of the procedure is presented in Appendix \ref{appendix:stream-plane}.

Studying streams between $1 R_{\rm vir}$ and $2 R_{\rm vir}$ allows for focus on nearby environmental gas flows and ensures that spatial correlations due to some satellites being at topological nodes of the web of streams are minimal as most satellites are found within $1 R_{\rm vir}$ and planes are dominated by $<1 R_{\rm vir}$ satellites.

Note that this method can also be applied to cosmic filaments connected to a given system, to establish their overall orientation relative to the plane of satellites. However, we only use the term {\it stream plane} to refer to small-scale streams, NOT large cosmic filaments.

\subsection{Cosmic Gas Vorticity}
\label{sec:vort}

Vorticity $\vec{\Omega}$ is defined as the curl of the gas velocity field $\vec{v}$: $\vec{\Omega}=\nabla \times \vec{v}$. We estimate it in NewHorizon by first interpolating the gas velocity field on a 3D regular grid of pixel extent $\Delta x= 14 \, \rm kpc$. We then use a 4-th-order centered scheme to estimate each component of the curl on the grid. Vorticity measures the local tendency of the flow to whirl and has been proposed as a key driver of the spin of structures such as galaxies \citep{2012MNRAS.427.3320C, 2014MNRAS.444.1453D, 2020MNRAS.491.2864W}.

While the $z\approx 0$ vicinity of filaments predictably appears more turbulent than in the earlier Universe, we qualitatively observe the persistence of patterns described at $z>2$ \citep{1999A&A...343..663P, 2015MNRAS.446.2744L,Hahn15}: vorticity is concentrated at the spine of densest filaments and at nodes of the cosmic web, exhibiting strand-like regions along local cosmic filaments, in which vorticity tends to align to the filament. However, we do not recover systematically well segmented Mpc scale vorticity quadrants as observed in simulations at $z>2$.

s\section{Census of co-rotating Planes}
\label{sec:census}

\subsection{Individual Plane analysis}
\label{sec:single}

We identify the ``best-fit plane" for each of the 13 MW systems with at least 4 satellites within $R_{\rm vir}$ in NewHorizon. We remind the reader that 12/13 systems actually host more than 7 satellites and constitute our primary sample. Best-fit plane properties of each system are detailed in Table~\ref{table:census} and the procedure and parameters are detailed in Section~\ref{sec:significance of plane}.

\subsubsection{Best-fit planes incompatible with halo-tracing hypothesis}

One can already notice the ubiquity of co-rotation, with more than $50\%$ of the systems showing more than $75\%$ of co-orbiting in-plane satellites. The total inertial axis ratio is also biased towards strong anisotropy, with 10/13 systems displaying a $c/a<0.5$ for their full distribution of satellites and 8/13 systems showing planarity $\pi_{\rm sat}<0.35$. As a guide, expectation is $\pi_{\rm sat}=1$ for an isotropic distribution and $\pi_{\rm sat}> 0.6$ for satellites tracing the shape of DM halos. Closer to observational methods, 8/13 systems show $c/a \le 0.2$ when restricted to in-plane satellites. This notably includes 5 systems (38$\%$ of the sample) with at least 11 in-plane satellites identified, typically amounting to 75$\%$ of their total satellite count. In contrast, underlying dark matter halos typically display much lower degrees of anisotropy ( $c/a>0.85$ within $R_{\rm vir}$). Therefore, the prevalence of markedly anisotropic distributions of satellites around MW analogs i NewHorizon is a striking baryon-only feature and readily suggests an incompatibility with the halo-tracing model.

\begin{figure*}
	\includegraphics[width=\textwidth]{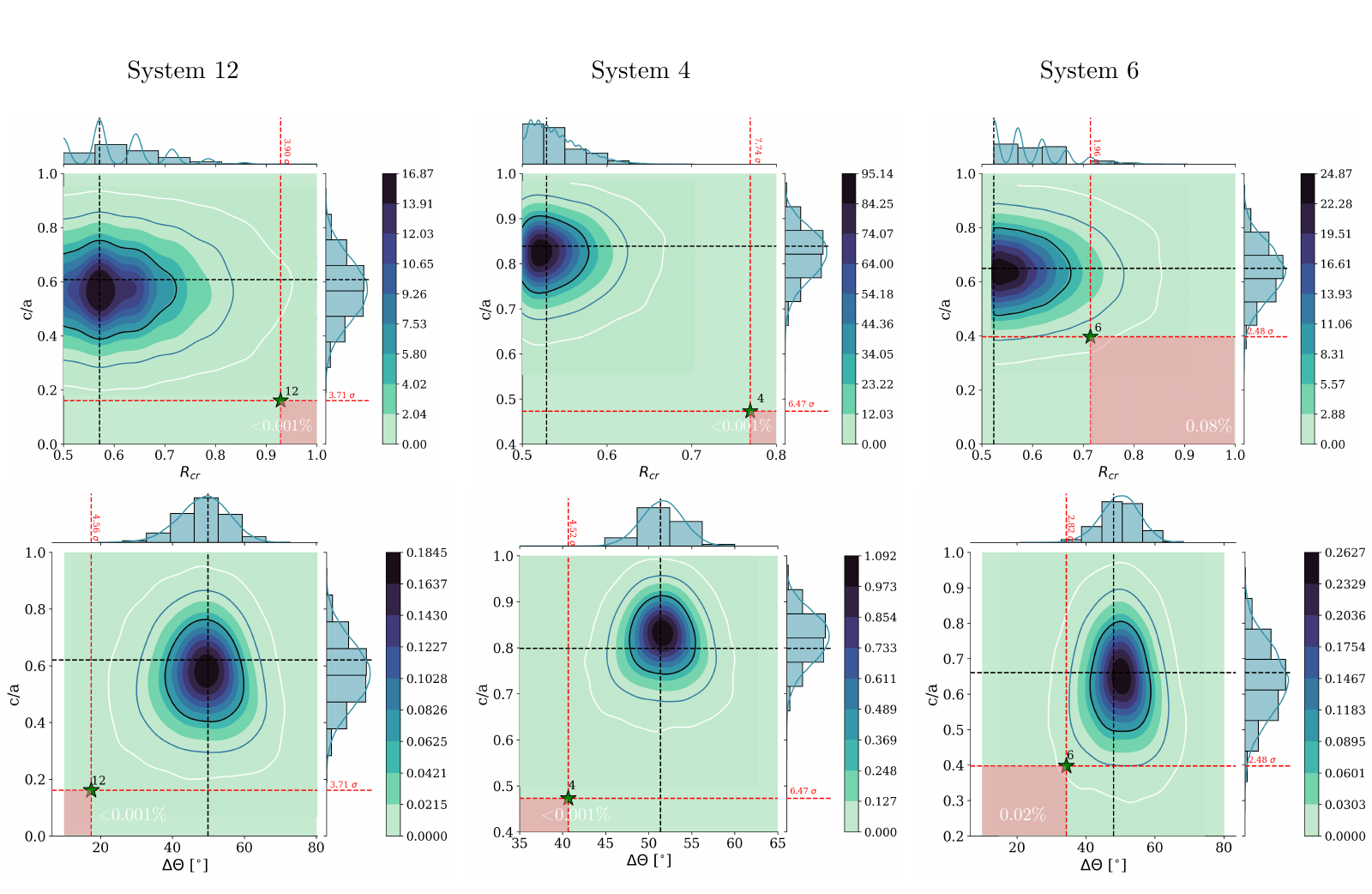}
    \caption{Expected joint distributions for $c/a$ and $R_{\rm cr}$ (upper panels) and $c/a$ and $\Delta \theta$ (lower panels) in the halo-tracing (null) hypothesis for Systems 12 (left), 4 (middle) and 6 (right). The black, navy, and white contours mark the 1$\sigma$, 2$\sigma$, and 3$\sigma$ contour, respectively. Black dashed lines mark the average expectations for each parameter. Note that due to stochasticity associated with the limited number of satellites per system, the peak fro $R_{cr}$ is $>0.5$. Green stars indicate the real parameters measured for each system, with corresponding 1D significance levels as red dashed lines and one-sided p-values (as red box excess probability) in white. These systems show marked anisotropy and co-rotation compared to the halo-tracing hypothesis.}
    \label{fig:2dhist_syst_12}
\end{figure*}

To better estimate the significance of this finding, we quantify the deviation from the peak of the null 1D distributions for $c/a$ and $R_{\rm cr}$ and for their joint 2D distribution, for each system individually. The significance of each system's planarity is, therefore, evaluated separately based on its host DM halo shape and radial distribution (see Sec.~\ref{sec:significance of plane} for details). As an example, Figure~\ref{fig:2dhist_syst_12} displays in blue to green shades the expected bi-variate distributions of $c/a$ (shape) and $R_{\rm cr}$ (co-rotation fraction, upper panels) and $c/a$ and orbital dispersion ($\Delta \theta$ , lower panels) for systems of satellites following the halo shape and radial distribution of systems 12, 4 and 6, respectively (null hypothesis). $1-\sigma$, $2-\sigma$ and $3-\sigma$ contours are highlighted as solid lines. The actual parameters of the three systems are indicated as green stars, with 1D significance levels highlighted in red. 

This shows that satellites in System 12 display a highly planar ($c/a=0.16$), highly co-rotating ($R_{\rm cr}=0.93$, $\Delta \theta =17^o$) distribution. This is inconsistent with satellites tracing the shape of the halo (at the 3.7$\sigma$ level). Overall, this system displays a $>$ 4 $\sigma$ level deviation from the joint null hypothesis. On the middle panel the same analysis for System 4 reveals that, while not a priori as thin as one of the observed planes, this massive system with 104 satellites is, in fact, strikingly anisotropic and co-rotating compared to its very isotropic dark matter host. This shows that strong baryon-only anisotropies extend to the highest mass end of the Milky Way$+$  range we consider and to richest systems. Even a relatively average system in our sample like System 6 (right panel) remains markedly anisotropic compared to halo-tracing expectations.

We repeat this analysis for every system in the sample, testing for various metrics for the shape (minor-to-major axis ratio $c/a$, planarity $\pi_{\rm sat}$) and co-rotation level (co-rotation fraction $R_{\rm cr}$, orbital dispersion $\Delta \theta$). 1D and 2D significance levels are presented in Appendix~\ref{appendix:onplane_metrics}.

Focusing on 2D significance levels for $c/a$ and $\Delta \theta$, We find that 9/13 of our systems ($70\%$) lie about the $2\sigma$ level of their null hypothesis, $46\%$ lie above the $3\sigma$ level and $23\%$ lie above the $4\sigma$ level. Note that System 11 does not appear as significantly incompatible with its halo shape due to its limited number of satellites, hence is not among these.
Consequently, one-sided p-values (often presented in numerical studies) show even more extreme levels of significance, with $50\%$ of the systems exhibiting $p<0.0001$.
Similar results are obtained with other metrics, as summarized in Figure ~\ref{fig:significance_table}, in Appendix ~\ref{appendix:onplane_metrics}, which also presents similar results obtained when restricting the analysis to on-plane satellites. Therefore, in NewHorizon individual massive systems (Milky Way mass and above) generally show strong levels of co-rotation and planarity  incompatible with a population of satellites simply relaxed with their DM halo.

To properly quantify this general incompatibility for the whole population, we estimate the incompatibility of the marginal distributions (across NewHorizon systems) of the main shape and co-rotation parameters $c/a$, $R_{\rm cr}$ and $\Delta \theta$ with KS tests comparing the distribution of these parameters in NewHorizon versus in the halo-tracing hypothesis (after re-normalizing and stacking null distributions for all systems). The methods and results are presented in Appendix.~\ref{appendix:general-inc}. The high significance of all the KS tests (p-values $<10^{-4}$) confirms that MW systems in NewHorizon overall display luminous satellite distributions strikingly incompatible with the halo-tracing model, both in terms of shape and orbit alignment, irrespective of the exact definition of a ``plane of satellites". The picture that emerges is that most MW satellite systems in NewHorizon do not relax with their DM halo but instead display a strong tendency to arrange in planar configurations and align their orbits.

\subsubsection{A behavior specific to baryons}

\begin{figure}
	\includegraphics[width=\columnwidth]{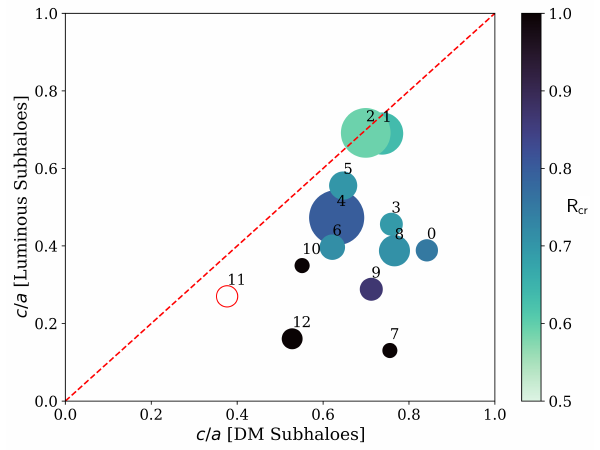}
    \caption{Minor-to-major axis ratio, $c/a$, for the distribution of all luminous satellites vs for all DM sub-halos within $R_{\rm vir}$ for all 13 MW systems in NewHorizon. Luminous satellites display significantly more flattened distributions than their DM counterparts.}
    \label{fig:c2a_dmhaloes}
\end{figure}

It should be noted that this behavior is specific of {\it luminous} satellites. Indeed, the analysis of all dark matter sub-halos with $M_{h}>7. 10^{7}\,\mathbf{M_\odot}$ within $R_{\rm vir}$ of the 12 systems reveals a completely different behavior. DM sub-halos display a distribution that is much more isotropic than their luminous counterpart, as demonstrated in Figure~\ref{fig:c2a_dmhaloes}, which shows $c/a$ for luminous satellites only vs for all DM sub-halos. All but the two most isotropic systems exhibit a luminous axis ratio, $c/a$, 2 to 4 times smaller than that of their DM sub-halos. Unsurprisingly, the most planar and co-rotating luminous satellite distributions are the ones which show the greatest deviations from the shape of their DM counterparts. This effect holds even after correcting for differences in sampling between luminous and dark matter satellites (see Appendix.~\ref{appendix:sampling} for details).

It is not unexpected that luminous structures will arrange in thinner configurations than DM ones. In fact, they grow from cosmic gaseous filaments, dissipative in nature through shocks and radiative cooling, hence typically thinner than their DM counterparts \citep{2006MNRAS.368....2D,2011MNRAS.418.2493P,2022MNRAS.512..926K,2024MNRAS.527.1301R,Lu2024}. However, the deviations observed ($>50\%$ decrease in axis ratio on average) are markedly stronger than observed for resolved satellites (i.e more massive: $M_{*}>10^9 \, \mathbf{M_{\odot}}$) in Horizon-AGN \citep{Dubois_2014}, the less resolved (kpc maximum resolution), large parent run of NewHorizon. For instance, we find that this baryon assembly bias is limited to $7\%$ for hosts with $M_{\rm h}>10^{13}\mathbf{M_{\odot}}$ in Horizon-AGN.
These findings suggest that the existence of so-called ``planes of satellites" in the Local Universe is not necessarily incompatible with $\rm \Lambda$CDM but may be the result of a multiphase, multiscale, baryon assembly bias that requires hydrodynamic simulations like NewHorizon (maintaining high resolution across a large cosmic volume) to be recovered.

\subsection{Population analysis: compatibility with observed planes in the Local Universe.}

To better estimate whether NewHorizon local massive galaxy analogs' distributions of satellites would indeed be considered anomalously planar according to observational studies, and whether the frequency of such planes would also be considered in tension with $\rm \Lambda CDM$ after comparison to larger, less resolved cosmological simulations like the ones typically used in previous studies, we perform a population analysis mimicking such studies.

For this analysis, we construct a composite null distribution for the full sample generating 300,000 synthetic systems of satellites (25,000 x 12) under the assumption that satellites trace the shape of their halo. To properly account for the frequency of each type of halo (richness = number of satellites), shape, radial distribution) in the original sample, we assume that each system's set of properties is equally likely to happen. In other words, we assume that NewHorizon ``Milky Way-mass and above" halos are representative of this mass range. 

\begin{figure*}
	\includegraphics[width=\textwidth]{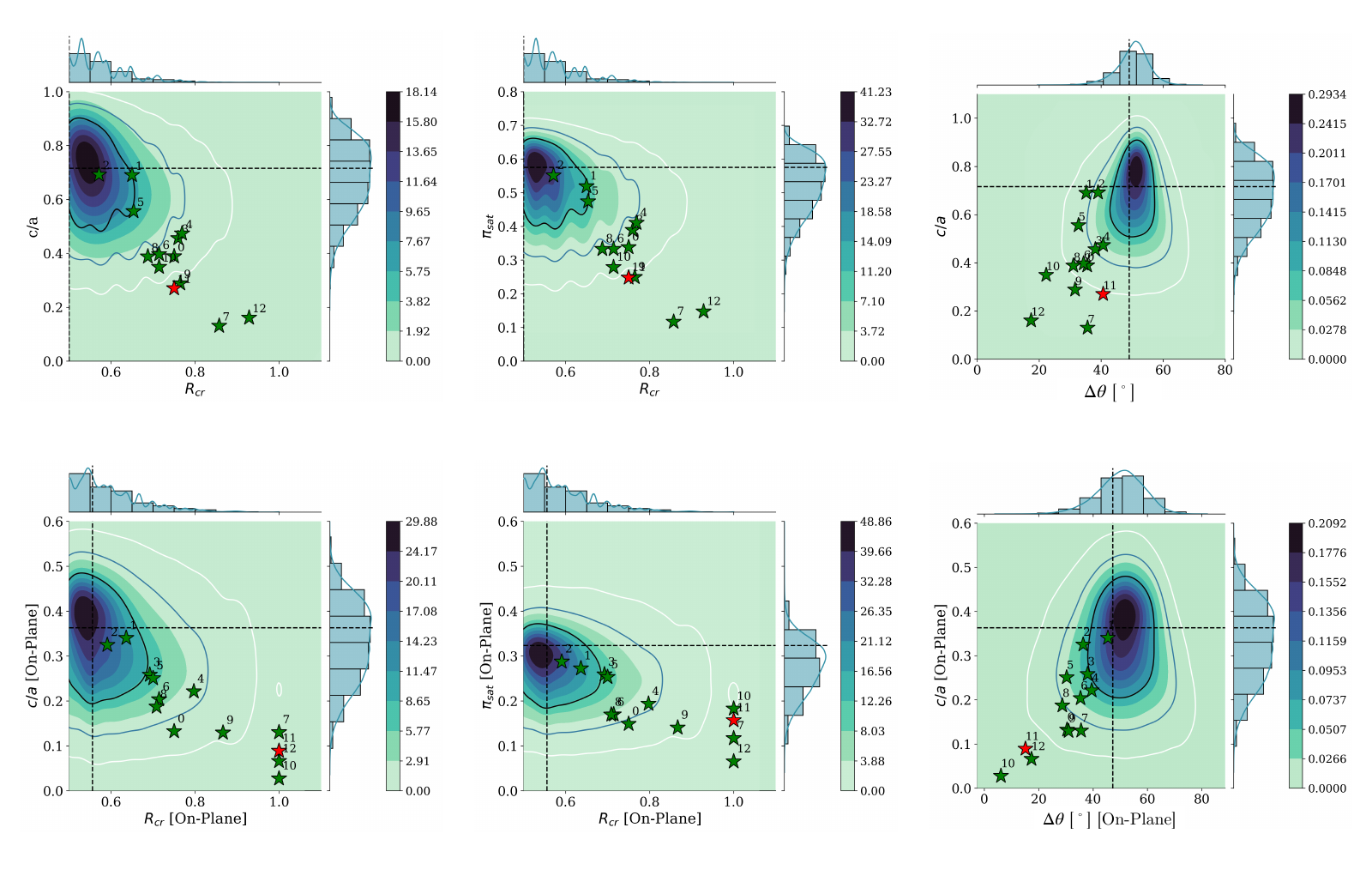}
    \caption{Expected joint distributions for $c/a$, $\pi_{\rm sat}$, $R_{\rm cr}$ and $\Delta \theta$ for full luminous satellite distributions (upper panels) and on-plane satellites only (lower panels) for a population of MW systems representative of NewHorizon, assuming satellites trace the properties of their host halo (null hypothesis). Actual values for the NH systems with at least 7 satellites are indicated as green stars. System 11 is marked in red and included for completeness. Black, navy, and white contours mark the 1$\sigma$, 2$\sigma$, and 3$\sigma$ contours, respectively. Black dashed lines mark the average expectations for each parameter.}
    \label{fig:2dhist_composite}
\end{figure*}

The resulting null 2D joint distributions for c/a, $\pi_{\rm sat}$, $R_{\rm cr}$ and $\Delta \theta$ are presented in Figure \ref{fig:2dhist_composite}. The upper panels display the full-system distributions, while the lower panels show the distributions restricted to on-plane satellites (as described in Section \ref{sec:statistical methods}). 1$\sigma$, 2$\sigma$, and 3$\sigma$ levels are indicated as solid lines. System 11 was removed from this composite distribution as it hosts too few satellites ($N_{\rm sat,11}$ = 4) to properly re-sample co-rotating fractions. The actual parameters around all local massive galaxy analogs in our main sample are displayed as green stars. System 11 is marked as a red star for reference.

Unsurprisingly, the sample of 12 systems is noticeably offset and skewed from the population-based null hypothesis, with at least $50\%$ of the systems past the 2$\sigma$ contour, irrespective of the metrics used for shape and co-rotation, and $15\%$ to $30\%$ of the sample past the 3$\sigma$ contour. This confirms that a large fraction of our systems would be considered anomalously planar and co-rotating when compared to typical subhalo distributions in their mass range as in \citep{Pawlowski_2014}.
In fact, the NewHorizon population of local massive galaxy analogs as a whole would be considered incompatible with the distributions of said halo properties, as confirmed by corresponding KS tests that do pairwise comparisons of  null distributions to NewHorizon actual distributions in Figure~\ref{fig:2dhist_composite}. Results of KS tests for all shape and co-rotation parameters are summarized in Table.~\ref{table:kstest}. 

Halo properties correspond to the parameters that are typically derived from dark-matter only runs to characterize the Milky Way and above mass range. This strongly suggests that the discrepancy between observations and $\rm \Lambda CDM$ simulations is a numerical limitation (in resolution and baryon physics) of such simulations rather than a physical model tension. It should be noted that similar conclusions can be reached for under-resolved hydrodynamic simulations as in \citep{2018MPLA...3330004P} since the identification of ``dwarf satellites" in kiloparsec-resolution runs is in practice equivalent to identification of all dwarf sub-halos due to the coarse-grained, typically overactive stochastic star formation sub-grid models used. In contrast, the increase in resolution of the CGM in NewHorizon reveals that DM sub-haloes as a whole actually follow much better the shape of their host halo than their resolved luminous counterparts, which arrange in much less isotropic configurations (see Figure \ref{fig:c2a_dmhaloes}). 

Interestingly, the set of systems that stand out as particularly planar and co-rotating in this population analysis is not limited to systems of satellites, which are the least compatible with the structure of their own halo ($>3\sigma$ individual significance as presented in the previous section). This also includes systems with a more moderate (albeit significant: $\approx 2\sigma$) tension with their host halo's structure but also smaller halos and fewer satellites to trace it ($N_{\rm sat}=7$ for System 7 and 10). These characteristics combine to produce strikingly planar structures that appear markedly anomalous compared to the overall population of local massive galaxy analogs ($>3\sigma$), while in fact reflecting in part more diverse halo shapes and more stochastic noise in their narrow mass/richness range. This effect drops sharply with the number of satellites but should prompt caution when assessing the tension with $\rm \Lambda CDM$ of planar anomalies with less than 10 satellites.

On the other hand, markedly anisotropic but more massive structures, like System 4, are washed out from significant anomalies in this population analysis. Indeed, their high number of satellites ($>100$) mechanically leads to more dispersion around the best-fit plane than low-richness systems, better represented in the population.

\begin{table}[t]
\begin{tabular}{ccc}
\hline \hline
Parameter & D-value & P-value \\
\hline
Inertial $c/a$ & 0.68 & $1.8\times 10^{-6}$  \\
Inertial $c/a$ [$1\sigma$  of Plane] & 0.53 & $0.7\times10^{-4}$  \\
Inertial $c/a$ [On-Plane] & 0.53 & $0.6\times10^{-4}$  \\
\\
$R_{\rm cr}$ & 0.74 & $6.2\times10^{-8}$ \\
$R_{\rm cr}$ [$1\sigma$  of Plane] & 0.59 & $5.4\times10^{-5}$ \\
$R_{\rm cr}$ [On-Plane] & 0.59 & $5.2\times 10^{-5}$ \\
\\
$\pi_{\rm sat}$ & 0.58 & $9.6\times10^{-5}$ \\
$\pi_{\rm sat}$ [$1\sigma$  of Plane] & 0.54 & $0.5\times10^{-3}$ \\
$\pi_{\rm sat}$ [On-Plane]& 0.55 & $0.3\times 10^{-3}$ \\
\hline
\end{tabular}
\caption{KS test results comparing the distributions of plane parameters for the NewHorizon sample and for the composite (population) null distributions displayed in Figure ~\ref{fig:2dhist_composite}. Local massive galaxy analogs in NewHorizon are as a whole incompatible with a representative population of Milky Way mass and above systems, for which luminous satellites simply trace their host halo's properties.}
\label{table:kstest}
\end{table}

To summarize, it is important to remember that -- while seemingly narrow ($\approx 1 \,\rm dex$) -- the typical $z=0$ halo/stellar mass range used in most studies to define Milky Way to Cen A analogs encompasses the turning point of several crucial transitions in galaxy evolution, including among others: gas accretion mode \citep{2006MNRAS.368....2D}, spin-filament alignment \citep{2012MNRAS.427.3320C}, stellar mass build-up \citep{2014MNRAS.444.3986R} and morphology \citep{2022MNRAS.513..439D}. In the present study, we further find that stream connectivity also displays a sharp transition at $M_{*}\approx 10^{11}\, \mathbf{M_\odot}$, with a marked impact on the shape ($c/a$) of the satellite distribution (seeAppendix.~\ref{appendix:connectivity}). More massive and isotropic halos are connected on average to 30 streams, highlighting the fact that they are often at the intersection of several large dense cosmic filaments, while this value drops to 10 below $M_{*}\approx 10^{11}\, \mathbf{M_\odot}$. These large variations in properties and environment across 1 dex of stellar/halo mass should caution against numerical estimations of significance based only on comparisons with purely mass-selected halos.

\subsection{The shape-corotation correlation}

In Fig.~\ref{fig:2dhist_composite}, it is striking that the real 12-system distribution shows a marked linear correlation between planarity and co-rotation, irrespective of the shape parameter and co-rotation parameter considered. This is also at odds with the null joint distributions.

One can clearly see a trend of increased level of co-rotation with increased planarity on every panel of Figure~\ref{fig:2dhist_composite}, particularly tight when restricting the analysis to on-plane satellites, as is often the case in observations. It is also striking that, among these distributions, even the most ``extreme" planes of satellites, and the most similar to observations (ex: Systems 9, 10,7,12), appear as the expected tip of a continuous trend in the shape/co-rotation space rather than standalone anomalies that would require a distinct formation model.

It is important to note that this linear anti-correlation between c/a and $R_{cr}$ is not a mere effect of differences in sampling (number of satellites). In fact, when correcting for sampling differences, the anti-correlation is stronger. This is explored in details in Appendix.~\ref{appendix:sampling}, where we also  discuss in detail the effect of randomizing the line-of-sight -- the main source of noise in the correlation -- in Appendix.~\ref{appendix:los}. 

\begin{figure}
    \centering
    \includegraphics[width=1\linewidth]{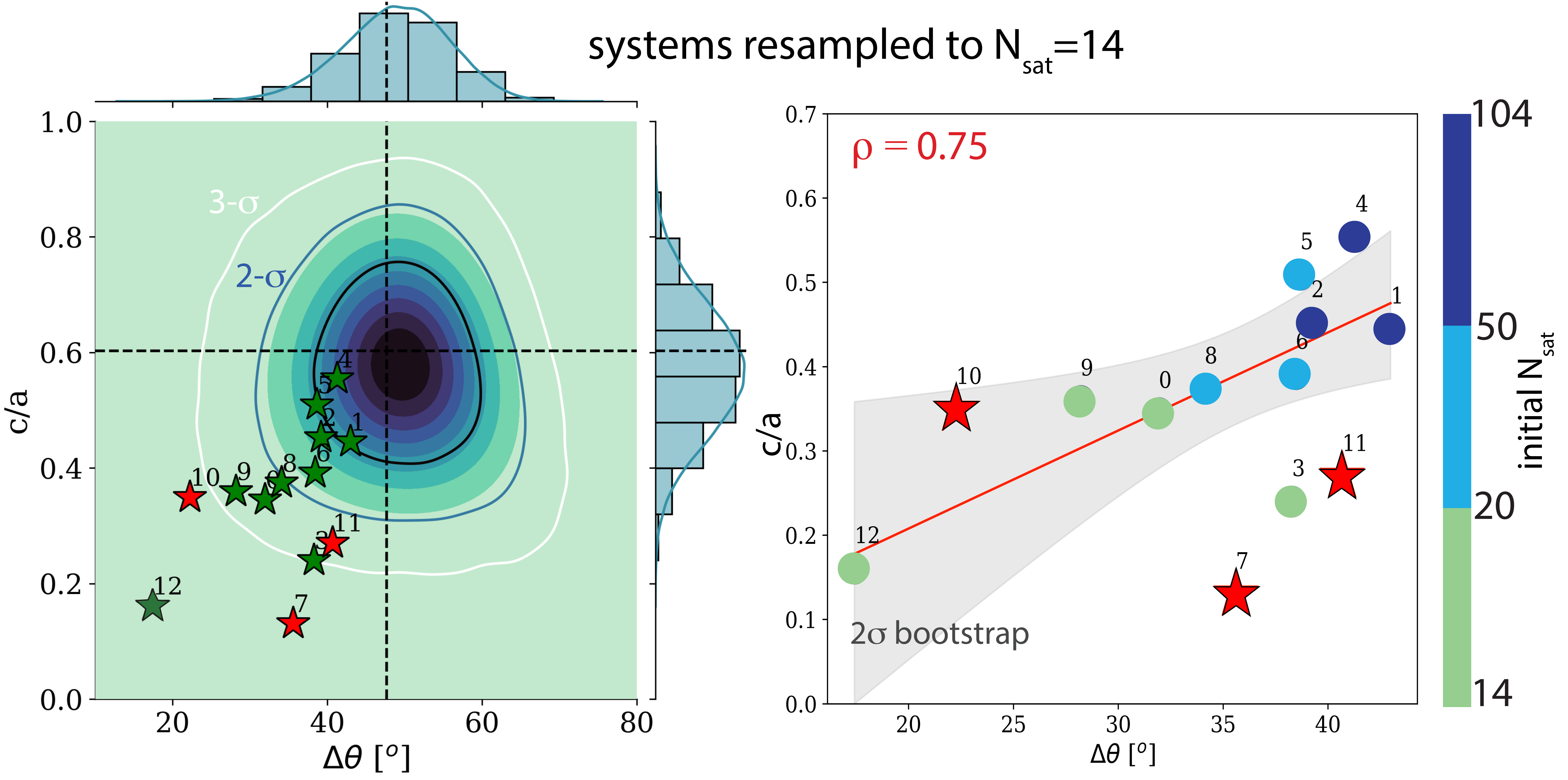}
    \caption{Same as Fig.~\ref{fig:2dhist_composite}, upper right panel, but resampling all sufficiently rich systems to $N_{sat}=14$, 1000 times each. Mean values for $c/a$ and $\Delta\theta$ appear as green stars on the left panel and blue dots on the right panel. Systems with strictly less than 14 satellites (not resampled) are indicated as red stars for reference. Resampling to a fixed $N_{sat}$ tightens the shape-corotation correlation already observed, without modifying conclusions. }
    \label{fig:orbital-resampled}
\end{figure}

To further confirm this, we now present results for the most robust, most sensitive co-rotation metric, $\Delta \theta$, which, while only available for the Milky-Way in observations, has the advantage of not being plagued by line-of-sight noise. Specifically, we reproduce the upper right panel of Fig.~\ref{fig:2dhist_composite}, but this time re-sampling every system with 14 or more satellites to $N_{sat}=14$ (This number also allows for direct comparison with M31). Each such system is re-sampled 1000 times without replacement, and for each iteration, we calculate $c/a$ and $\Delta \theta$. Median values are displayed as green stars on Fig.~\ref{fig:orbital-resampled}, left panel. This panel focuses on comparison with the corresponding null-hypothesis (obtained through similar resampling of halos). Systems with $N_{sat}<14$ satellites (not resampled) still appear as red stars for reference. The right panel focuses on linearly fitting the correlation (omitting the $N_{sat}<14$ systems) .

In both panels, it is striking that the correlation between planarity and co-rotation is enhanced when correcting for sampling variations across halos. In fact, once resampled to $N_{sat}=14$, all systems with $N_{sat}\ge 14$ fit in a very tight linear relation with a correlation Pearson coefficient of $\rho=0.75$ (p-value=0.01), while non-resampled systems are the ones that add spread to the overall correlation. It is not unexpected as, even with similar sampling, more massive halos are typically found at the intersection of several filaments (more connected), hence in a more isotropic infall region. Our previous conclusions  also remain unchanged: 5/10 ($50\%$) of the systems resampled to 14 satellites are past the $2\sigma$ contour of the corresponding null-hypothesis. This confirms once again the prevalence of flattened, co-rotating systems of luminous satellites in New Horizon, at a level way stronger than predicted by dark matter halos only. As a consequence, this also confirms that this halo-tracing hypothesis cannot be reliably used as a null hypothesis to assess the significance of systems of satellites in observations, as it systematically overestimates how unusual these systems are compared to properly resolved baryon-based predictions. Of note, since the strong correlation between planarity and co-rotation resists fixed-number resampling and seems highly driven by the mass/initial richness of the halo, the practice of sub-sampling simulated halos at fixed $N_{sat}$ to compare them to given observed systems should also be avoided. Fig.\ref{fig:orbital-resampled} confirms that a high-richness halo sub-sampled at $N_{sat}=14$ is not equivalent to a halo with inherently $N_{sat}=14$. 

\subsection{How unusual would local planes be in NewHorizon?}

 In light of our findings, one important question is whether (and how much) the ubiquity and characteristics of observed planes in the Local Universe still stand out as in tension with the $\rm \Lambda CDM$ model when taking into account our findings in the NewHorizon simulation.

To answer this without overstating the predictive power of NewHorizon's relatively small sample, we perform a Bayesian inference of the posterior joint distribution of shape $c/a$ and co-rotation fraction $R_{\rm cr}$ using the 12 NewHorizon systems. In a nutshell, we start from the maximally conservative priors that satellites should follow the shape of their host DM halo. We then factor in the likelihood of our 12 systems to update the posterior distribution, which we then sample with a Metropolis-Hastings scheme. We can then compare how planes observed in the Local Universe fit in this updated posterior distribution. The details of the methodology are specified below.

Note that, despite its intrinsically non-Gaussian marginal distribution, we use $R_{\rm cr}$ instead of the orbital pole dispersion as the latter is known only for the Milky Way, while $R_{\rm cr}$ is measured for 6 local observed ``planar" systems \citep{Pawlowski21, Pawlowski24b}. This includes the 2 planes of the Milky-Way, the Great Plane Andromeda, the main plane of Centaurus A, NGC 253 and NGC 4490. 

It should be noted that \citep{Pawlowski21} compiles systems that are listed with their “on-plane” c/a and $R_{cr}$ (M31, MW) along with some further away systems that are listed with their “full-system” c/a (i.e all satellites {\it detected}). Indeed, luminosity and line-of-sight velocity detection limits vary with distance, hence across observational systems.

Conservatively, we decide to first compare all these values to the distribution of full-system values inferred from New Horizon to avoid overstating compatibility due to varying selections in the definition of “on-plane”. This means that, in this part of the analysis, certain systems will appear more flattened, hence more incompatible than they actually are. For instance the Milky-Way is listed as having an on-plane value for Plane 1 of $c/a=0.18$, while its full-system value is actually $\approx 0.32$. Its position in our c/a - Rcr parameter space can therefore be seen as an outer bound of definition-specific values. However, if even these are considered compatible, this will provide robust evidence that these are not a challenge to $\Lambda$CDM. 

To further assess compatibility of the nearby Milky Way and Andromeda systems, for which plane parameters are derived from a robust ``on-plane" selection, we further repeat the analysis, this time using the distribution inferred from New Horizon ``on-plane" parameters.

\subsubsection{Hypothesis}

 In line with its definition, we assume a folded normal marginal distribution for $R_{\rm cr}$, folded around its mimimal value $R_{\rm cr}=0.5$. The PDF $f_{R_{\rm cr}}$ for $r\geq 0.5$ has shape parameter $c_s=\frac{\mu_1}{\sigma_1}$ with mean $\mu_1$ and dispersion $\sigma_1$:
 
\begin{equation}
\begin{aligned}
&f_{R_{\rm cr}}(r|c_s,\sigma_1) \\
&=\frac{1}{\sqrt{2\pi} \sigma_1} \cosh{(c_s(r-0.5))}\exp \left( -\frac{2(r-0.5)^2+c_s^2}{\sigma_1^2} \right)
\end{aligned}
\end{equation}

In line with what is observed in NewHorizon and Horizon-AGN across a wide range of host masses, we assume a normal distribution of PDF $f_{c/a}$ for $x=c/a$ with mean $\mu_2$ and dispersion $\sigma_2$:
\begin{equation}
\!f_{c/a}(x|\mu_2,\sigma_2) \! =\!\frac{1}{\sqrt{2\pi} \sigma_2} \exp \left( -\frac{(x - \mu_2)^2}{2\sigma_2^2} \right)\! \sim \!\mathcal{N}(\mu_2,\sigma_2)
\end{equation}
We further assume that these are correlated through a Gaussian copula $c_{\rm gauss}(u,v|\rho)$ with correlation parameter $\rho$. The joint distribution of $x=c/a$ and $r=R_{\rm cr}$, therefore, takes the form:

\begin{equation}
\begin{aligned}
&f(x,r|c_s,\sigma_1,\mu_2,\sigma_2,\rho) \\
\!&=\!f_{c/a}(\!x|\mu_2,\sigma_2\!)\! \cdot \! f_{R_{\rm cr}}\!(r|c_s,\sigma_1)\!\cdot \!c_{\rm gauss}(\!F_{c/a}\!(x)\!,\!F_{R_{\rm cr}}\!(r)|\!\rho)
\end{aligned}
\end{equation}

where $F_{c/a}$ and $F_{R_{\rm cr}}$ denote the cumulative distribution functions corresponding to PDFs $f_{c/a}$ and $f_{R_{\rm cr}}$, with conditioning parameters dropped for readibility.
Free parameters of the model are, therefore, $c_s=\mu_1/\sigma_1$, $\mu_2$, $\sigma_1$, $\sigma_2$ and $\rho$

\subsubsection{Priors}

We start with conservative priors around the expectation that luminous satellites follow the shape of their host halo. More specifically prior distributions  for $c_s$ and $\sigma_1$ (for $R_{\rm cr}$), and $\mu_2$ and $\sigma_2$ (for c/a) are chosen to be Gaussian, centered on population-based means in the halo-tracing (null) hypothesis: $\bar c=\bar \mu_1 /\bar \sigma_1$ and $\bar \sigma_1$ on the one hand, $\bar \mu_2$ and $\bar \sigma_2$ on the other hand, as displayed in Fig.~\ref{fig:2dhist_composite}. 

The standard deviations for these Gaussian priors are set at $\bar \sigma_{2}$ for $\mu_2^0$, 0.5 for $\sigma_1^0$ and $\sigma_2^0$ and 2 for $c_s^0$ (reproducing the order of magnitude of halo-to-halo variations). In practice, the effects on the posterior distribution of tightening or loosening these priors by a factor 3 are limited to modulations of the asymmetry in the upper left corner of Fig.~\ref{fig:posterior}, i.e. among the most isotropic systems. Note also that using uniform priors broadens the 2$\sigma$ contours, enhancing the compatibility of observed systems with NewHorizon. Priors for $\rho$ are chosen to be uniform between [-1,1]. In summary, our set of priors is:

\begin{equation}
\begin{aligned}
c_s^0 \sim \mathcal{N}(\bar\mu_1/\bar \sigma_1,2), \, \, \mu_2^0 \sim \mathcal{N}(\bar\mu_2,\bar \sigma_{2})\\
\sigma_{1}^0 \sim \mathcal{N}(\bar \sigma_1,0.5),\, \sigma_{2}^0 \sim \mathcal{N}(\bar \sigma_2,0.5)\\
\rho^0 \sim \mathcal{U}(-1,1)
\end{aligned}
\end{equation}

\subsubsection{Posterior distribution}

The posterior distribution $P$ of $\theta=(c_s, \mu_2, \sigma_1, \sigma_2,\rho)$ is obtained by factoring in the likelihood of the 12 NewHorizon systems $(x_i,r_i)$:
\begin{equation}
\mathcal{L}(x_1,r_1,..., x_n, r_n|\theta)=\prod_i \, f(x_i,r_i|c_s,\sigma_1,\mu_2,\sigma_2,\rho)
\end{equation}
which yields:
\begin{equation}
\begin{aligned}
&\log P(\theta| x_1,r_1,..., x_n,r_n)\\
&\, \propto \log \mathcal{L}(x_1,r_1,..., x_n, r_n|\theta)+ \log  c_s^0 (\theta)\\
&+\log \mu_2^0 (\theta)+\log  \sigma_1^0(\theta) +\log  \sigma_2^0 (\theta)
\end{aligned}
\end{equation}

We then sample the posterior distribution using a Metropolis-Hasting scheme.

\begin{figure*}
 \includegraphics[width=\linewidth]{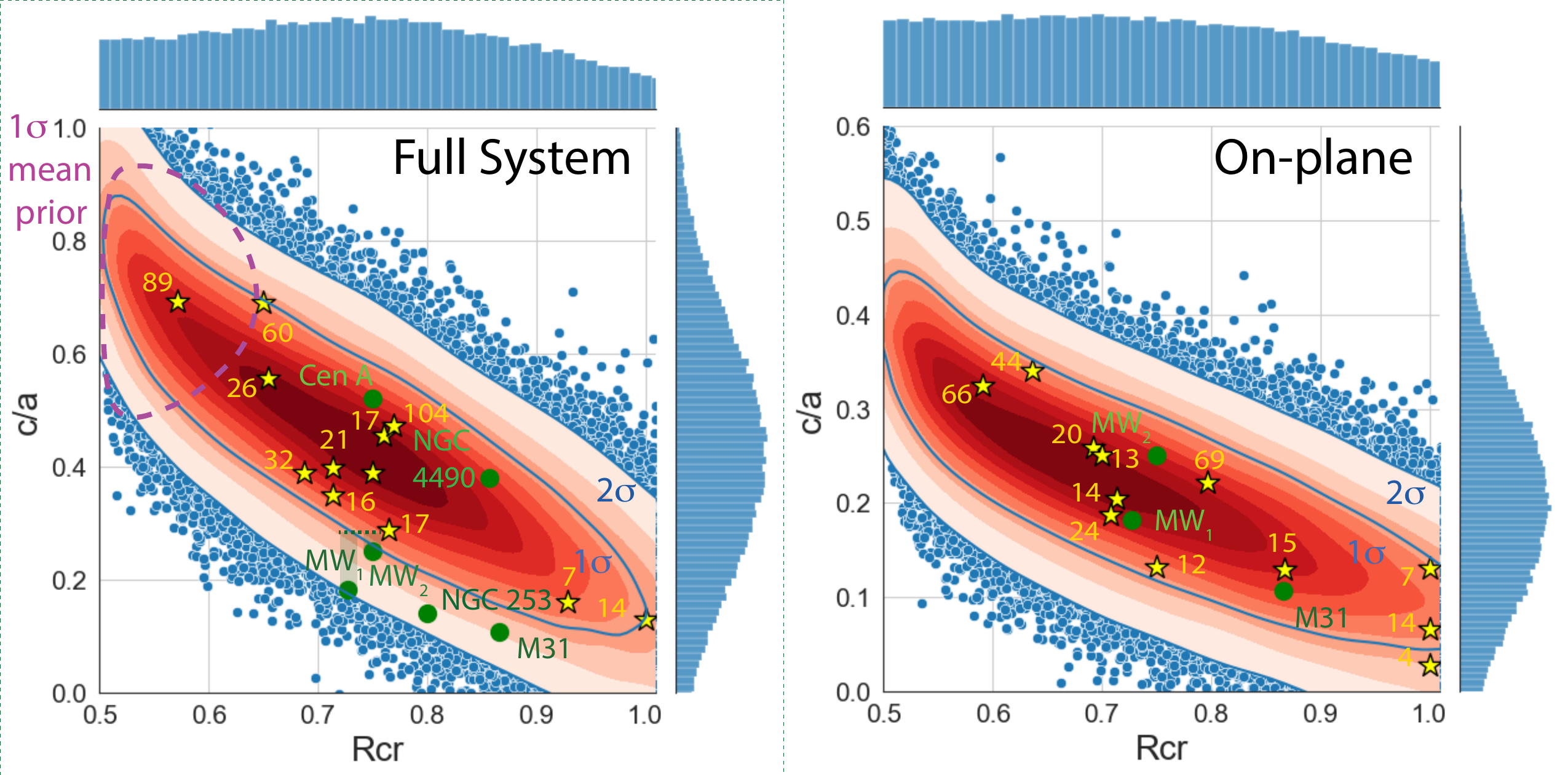}
    \caption{Mean posterior joint distribution for $c/a$ and $R_{\rm cr}$ from priors emulating the halo-tracing hypothesis in NewHorizon (mean prior contour as fuschia dashed line). The posterior is updated with the measured parameters of the 12 NewHorizon local massive galaxy analogs, either from their full-system parameters ({\it Left}) or from their on-plane parameters ({\it Right}). 1- and 2-$\sigma$ contours are shown in blue. The 12 systems appear as yellow stars. Local observed systems from \cite{Pawlowski21} and \cite{Pawlowski24b} are marked as green dots. Note that the MW and M31 are listed with their "on-plane" values, but the dashed green line shows the known full-system c/a for the MW. Local systems are fully compatible with NewHorizon. $N_{sat}$ for each NH system is indicated in yellow, showing a good agreement with typical numbers associated with observed systems.}
\label{fig:posterior}
\end{figure*}

The mean posterior distributions are presented in Figure \ref{fig:posterior}. The left panel shows the distribution inferred from New Horizon full-system values, while the right panel shows the one inferred from on-plane values. The 12 NewHorizon systems are displayed as yellow stars. We find that the correlation coefficient between $c/a$ and $R_{cr}$ is $\rho=-0.85\pm 0.1$, robust to modulations in priors. Observed systems listed in \cite{2019ApJ...875..105P}  and \cite{Pawlowski24b}, for which $R_{\rm cr}$ and $c/a$ are available are presented as green dots. The 1- and 2-$\sigma$ contours are highlighted in blue. 
On the left panel, it is easily noticed that observed ``planar" systems  do not appear as outliers in this updated distribution, all of them contained in the $2-\sigma$ contour despite tight priors. In fact, with respect to this distribution, the most planar, most co-rotating observed system (the Milky Way) only displays a $1.9\sigma$ deviation. However, even in this case, it is mostly because the Milky-Way is listed  with its ``on-plane values" in \cite{2019ApJ...875..105P}. For instance, its full system is closer to $c/a=0.32$, which alone would put the Milky Way planes back to 1$\sigma$ deviation levels (transparent green lines). When comparing the Milky Way planes with the on-plane distribution inferred from NewHorizon (right panel), the deviation is actually $<0.25\sigma$.

It should be noted that, since we are performing a population analysis, the number of satellites vary from one system used in the inference to the next. Another way to say it is that both the simulated and posterior distributions are in effect marginalized over $N_{sat}$. Since we have shown that the elongation of a system is correlated to its mass (and therefore, indirectly, to its number of satellites) a natural question is whether the number of satellites of real systems fall within the same $N_{sat}$ range as the simulated ones in their vicinity on our marginalized distributions. In particular, we need to ensure that simulated $N_{sat}$ are not significantly and systematically lower than real ones, in which case a sampling bias could not be fully ruled-out. We show $N_{sat}$ for the full systems and for the planes on Figure \ref{fig:posterior} in yellow. This shows not only a high variability of values, it also confirms that $N_{sat}$ of simulated systems are not under-sampled compared to real analogs. especially. For real systems with the best constrained $N_{sat}$ (MW1: 11, MW2: 20-30, M31: 14), it is even apparent that their sampling is strikingly close to the average sampling of the simulated NH distribution in their neighborhood on the $c/a-R_{cr}$ plane. Therefore, while a larger simulation with hundreds of massive galaxy analogs would be useful to better constrain the $N_{sat}$ distribution, we find no sign of a sampling bias at this point.

Observed elongated systems, therefore, appear to be fully compatible with the most likely distribution inferred from simulated NewHorizon systems, even when assuming tight quasi-isotropic priors. We conclude that, to date, these systems do not show any level of discrepancy with $\rm \Lambda CDM $ when including predictions from currently available cosmological runs with sufficiently high resolution.

\section{Planes as the Result of an Interplay with the Multi-Scale Cosmic Web}
\label{sec:CosmicWeb}

\subsection{The environment of Planar Systems}

The degree of anisotropy and co-rotation of satellite systems in NewHorizon is striking. In this section, we explore the potential origins of this trend, in particular, we analyze how MW systems interplay with the gas dynamics of their larger-scale environment. We focus on two different types of environmental scales: {\it ``local"} streams of gas, extracted with \texttt{DisPerSE} from resolved gas density ($\Delta x = 14$ kpc) in the first megaparsecs around the system (2-4\, Mpc, from $R_{\rm vir}$ to $6-10\, R_{\rm vir}$), and {\it ``cosmic"} filaments, extracted with \texttt{DisPerSE} from the gas density smoothed over hundreds of kpc within a 20 Mpc box corresponding to the full NewHorizon volume (see Methods \ref{sec:filament_identification} and Appendix.~\ref{appendix:stream-plane} for details).

Figure \ref{fig:mainmap} shows the main sample of cosmic filaments (in yellow) overlaid on a projected map of the gas density in NewHorizon. Zoom-in inserts further show the local streams (in white) overlaid on the resolved gas density of one example system, System 5. System 5 displays a relatively small network of streams, typical of systems with stellar mass below $10^{11}\, \mathbf{M_{\odot}}$ . 

Appendix.~\ref{appendix:connectivity} further highlights the variety of local and cosmic environments two Milky Way Mass (and above) systems can have, even in an otherwise average density simulation like NewHorizon.  Indeed, it illustrates a general trend within our sample: despite the narrow halo and stellar mass range we consider, Milky-Way to Cen A mass systems exhibit a broad range of stream connectivity (number of streams entering the virial radius of the system).  It shows that stream connectivity increases sharply with central stellar mass. Once again, this trend in variance is consistent with the fact that, albeit narrow, the Milky Way to Cen A mass range encompasses several transition masses central to galaxy evolution. Unsurprisingly, we find that slightly lower-mass, less connected systems tend to show more anisotropic distributions of satellites, in line with the fact that they are not at a node of multiple misaligned filaments.

In the next sections, we investigate the links between the geometric arrangement of local streams, cosmic filaments, and satellites.

It should be noted that the connection between accretion from the cosmic web and the anisotropy of satellite distributions was previously suggested as the origin of planar satellite alignments \citep[see for instance]{Aubert_2004}. But this model was conventionally dismissed due to the purported ``thickness" of cosmic filaments, hereby reduced to the large dark-matter cosmic filaments, typically enclosing Milky Way analogs rather than acting as connecting streams \citep{2019ApJ...875..105P}. However, this view ignored the highly multi-scale nature of the cosmic web, in particular the extreme thinness of small-scale gas filaments, i.e. local streams, and the potential geometric and tidal constraints that larger cosmic filaments impose on their configuration. This omission, or rather this confusion between the ``scale of tidal influence" and the ``scale of accretion" is largely due to the fact that most previous cosmological analysis were performed in simulations that did not resolve local streams or the CGM of Milky Way-mass and above analogs, a limitation much reduced in NewHorizon, as detailed in the next section.

\subsection{Connection to Local Gas Streams}

\begin{figure*}
\centering

  \centering
  \includegraphics[width=\linewidth]{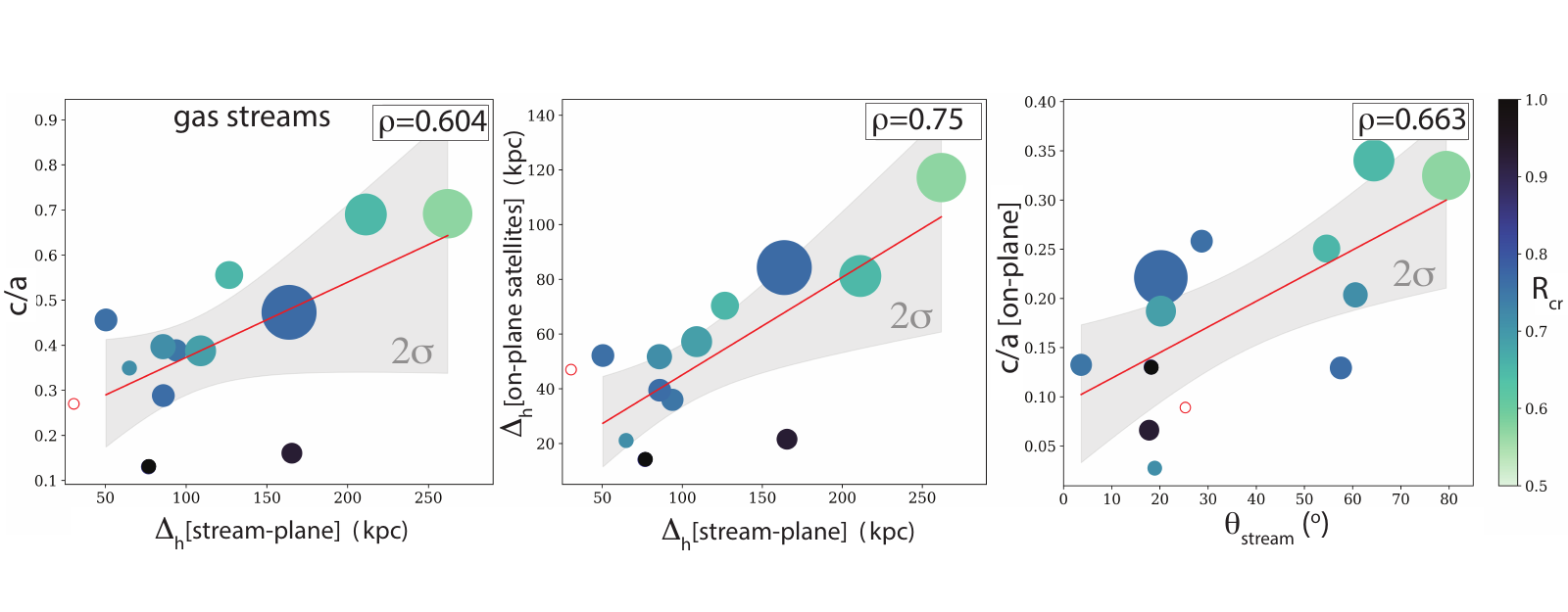}

\caption{{\it Left:} Inertial axis ratios, $c/a$ for all satellites of each system versus height of their local stream-plane, $\Delta h$[stream-plane] (gas streams within $1 \,R_{\rm vir}$ $<$ $x_{fil}$ $<$ $2 \,R_{\rm vir}$). Systems are colored by co-rotating fraction and sized by number of satellites. System 11 is indicated as a red circle for reference. The best linear fit is shown as a red solid line, with Pearson coefficient $\rho$ and $2\sigma$ bootstrap contours shaded in grey. {\it Middle:} Thickness of stream-plane $\Delta h$[stream-plane] vs. the thickness of satellite plane for on-plane satellites. {\it Right:} Angle between the stream-plane and the satellite plane as a function of $c/a$ ratio for on-plane satellites.}
\label{fig:stream_sat_align}
\end{figure*}

Figure~\ref{fig:stream_sat_align}, left panel, shows the variation of the total $c/a$ shape ratio of our satellite systems with the height $\Delta_h$ of the best-fit plane of their gas streams (stream-plane), restricted to streams between $R_{\rm vir}$ and $2 \,R_{\rm vir}$ as defined in Section.~\ref{sec:stream-plane}. The middle panel shows the height of the best-fit satellite plane versus the height of the gas stream-plane. Color encodes the co-rotation fraction and mark size the number of satellites of each system. 

Despite our limited size sample, a strong correlation emerges: more planar systems of satellites are embedded in more planar networks of gas streams (Pearson correlation coefficient of $\rho=0.60$ and $\rho=0.75$, respectively). Moreover, these also tend to correspond to more co-rotating systems.
Note that similar results are obtained for $\pi_{\rm sat}$ ($\rho \approx 0.60 $), which is expected since these metrics are very strongly correlated (see Appendix.~\ref{appendix:metric_comparisons} for a demonstration).

 The two panels confirm that the trend is robust to small changes in definition. It is also apparent, that most isotropic, less co-rotating systems tend to be the ones with very large numbers of satellites (also the most massive ones), in line with their higher connectivity. As a direct consequence, these are also clearly the ones with ``thickest" stream-plane as these systems' local environment is better described by a large number of streams reaching the halo from multiple directions than a planar configuration of streams.

To better assess the connection between the geometry of the gas streams and the existence of a satellite plane, Figure~\ref{fig:stream_sat_align}, right panel, shows the relationship between the axis ratio of on-plane satellites $c/a_{\rm [on-plane]}$ and the angle $\theta_{\rm stream}$ between the stream-plane and the best-fit satellite plane. We find that elongated distributions of satellites, also the most co-rotating, tend to be more aligned with their surrounding gas streams, In particular, 5 out of the 6 most pronounced planar and co-rotating distributions lie within 30$^o$ of their surrounding stream-plane (in particular all 3 of the ``full co-rotating" systems, for which all ``on-plane" satellites co-rotate). In comparison, all 3 of the systems with low co-rotation ($R_{\rm cr}<0.7$) show a mid-plane misaligned by more than 55$^o$ from their stream-plane. 

The Pearson $r$-coefficient of this linear trend is 0.66. However, what is perhaps more informative here is the striking bimodality in terms of orientation with local streams (aligned/orthogonal), with no system lying between 30$^o$ and 58$^o$ of their stream-plane. 
Once again, it is striking that such a trend is detected at $z \approx 0$ despite the substantial mixing and rarefaction of cosmic gas and despite the fact that satellites are the product of earlier accretion. This suggests that a relative stability of streams, hence of planar accretion of gas and satellites into the host halo, might be necessary to form a plane of satellites around a local massive galaxy analog.

\subsection{Connection to Cosmic Filaments}
\label{cos-fil}

For a network of gas streams to remain in a relatively planar conformation over extended cosmic time, a possible scenario is that its distribution be constrained by the tides of a larger cosmic filament with a relatively quiet history (limited mergers, no strong drifting). For instance previous numerical and theoretical studies \citep{1999A&A...343..663P, 2015MNRAS.446.2744L, 2015MNRAS.452.3369C, 2021MNRAS.501.4635S} predict that, at high redshift ($z>2$), large-scale flows around cosmic filaments display large vortical regions (vorticity quadrants), with vorticity typically aligned to the spine of the nearby cosmic filament (i.e the plane of the whirl orthogonal to said filament). This phenomenon arises from tidal torques from the cosmic filament and surrounding cosmic walls. These whirls could provide a mechanism for the formation of a plane of whirling gas streams around proto-Milky Way halos, then further stabilized by the growing gravitational torques of the emerging disc.

\begin{figure*}
\centering

  \centering
  \includegraphics[width=\linewidth]{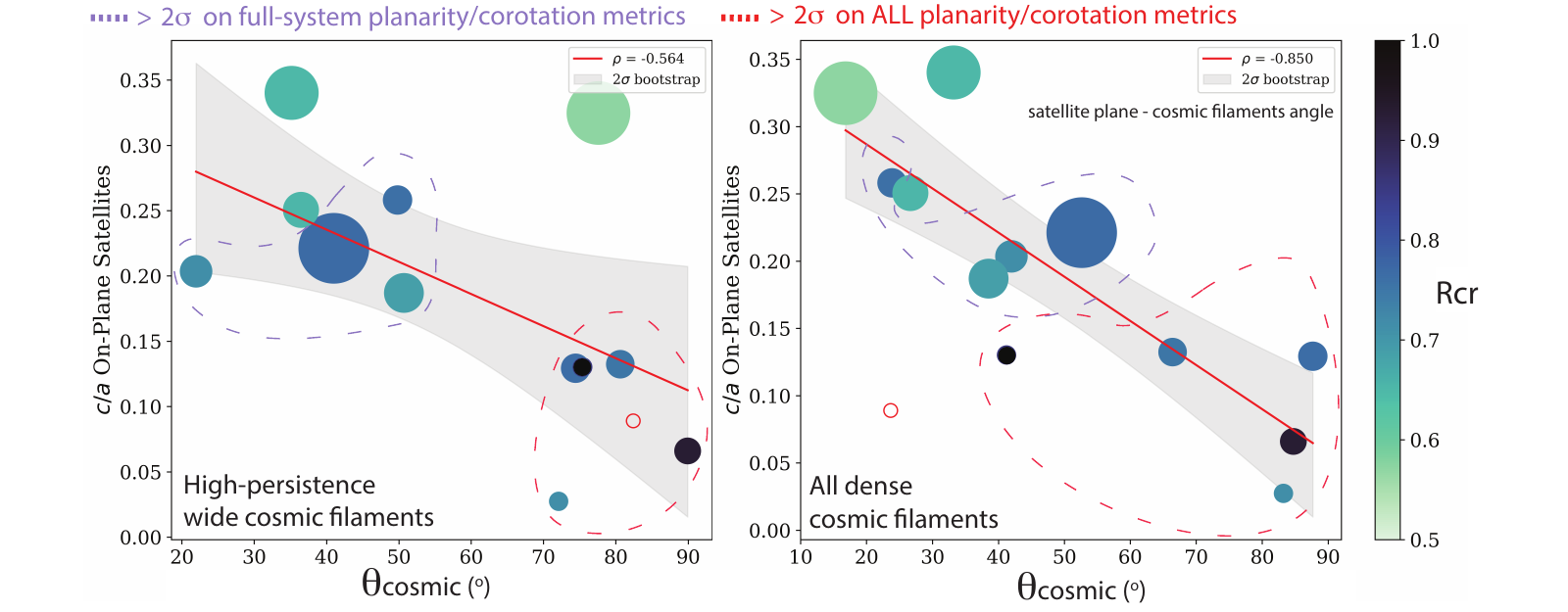}
   \caption{ Angle $\theta_{cosmic}$ between the satellite-plane and the best-fit plane of cosmic filaments as a function of $c/a$ ratio of on-plane satellites. Systems are colored by co-rotating fraction and sized by number of satellites $N_{\rm sat}$. Best-fit linear regression lines are shown in red with 2$\sigma$ error contours shown in grey. System 11 indicated as a red circle for reference. Dashed lines circle systems with high levels of planarity and co-rotation. {\it Left:} Restricted to largest, widest, most persistent cosmic filaments (see text). {\it Right:} Using all dense cosmic filaments (see text). Most significant planes of satellites lie orthogonally to their nearby cosmic filaments. However, a second population of significantly flattened systems lies parallel to their largest cosmic filaments. The divide is particularly clear with respect to largest filaments}
\label{fig:cosalign_trends}
\end{figure*}

To explore this possibility, in Figure~\ref{fig:cosalign_trends} we show the evolution of the axis ratio $c/a$ for on-plane satellites with the average angle $\theta_{\rm cosmic}$ between each satellite best-fit plane and the best-fit plane of nearest cosmic filaments. As defined in Section.~\ref{sec:CosmicWeb}, we test this for two different samples of cosmic filaments. On the left panel, we focus on the {\it main} sample, comprised of high-persistence cosmic filaments, typically 1-2 Mpc wide and connecting halos $>10^{12.5} \mathbf{M_{\odot}}$. On the right, we show results for the {\it extended} sample, which includes all cosmic filaments of similar core density, including somewhat thinner ones, or merging ones, often of lower persistence.

To identify nearest filaments for each system, we select  all filaments with a sampling point that crosses $R_{\rm vir}$. For the main sample, we then apply the plane finder to all sampling points on these half-filaments (node to saddle-point) found between $R_{\rm vir}$ and $2 \, R_{\rm vir}$. This avoids being influenced by the central disc's direction as these filaments are obtained from a strongly smoothed field. For the extended sample, we consider all sampling points along these half-filaments as these are more resolved, more sampled and sensitive to local peaks, hence usually they spontaneously end at relay saddle-points within 2 or $3\,R_{\rm vir}$.

We find a surprisingly tight anti-correlation ($\rho = -0.85$) between $c/a$ and the angle of the best-fit satellite plane with nearby cosmic filaments from the {\bf extended sample} (right panel). The most pronounced co-rotating and planar distributions lie markedly orthogonal to their nearby filaments, with 4/5 of the most planar, most co-rotating planes lying at $> 65^o$ from their nearest cosmic filament and 3 $> 82^o$. By contrast, the best-fit mid-plane of the most isotropic systems is generally more aligned with the nearby cosmic filaments, with the 3 non-corotating, most isotropic systems ($R_{\rm cr}<0.6$) lying at less than $35^o$ from their nearest cosmic filament. This indicates clear residual alignment of these systems's shape with the densest ($=$ most sampled filament) large-scale accretion channel despite being connected to a number of misaligned cosmic filaments.

These results are consistent with the idea that most strongly planar satellite distributions are formed and remain in peri-filament large-scale whirls (vortical regions tidally constrained by cosmic filaments) across cosmic time, as illustrated in the red frame of Figure \ref{fig:vorticity_alignment}. Migration at the spine of cosmic filaments, off-plane galaxy mergers or merging/disruption of such cosmic filaments at time of assembly may, on the contrary, perturb the formation of said planes.

However, it is also notable that, between these two extremes, a number of planes for more massive systems, less thin but still very significant given their number of satellites ($N_{\rm sat}=21-104$) are substantially more aligned with their cosmic filaments (typically $< 45^o$). Unlike isotropic systems, most of these intermediate systems also harbor satellite planes strikingly aligned with their local stream-plane ($< 25^o$), which indicate these were fed by contrasted streams lying mostly along cosmic filaments (see blue frame in Figure \ref{fig:vorticity_alignment}). This may call for a different mechanism to explain the formation of thin, dense accretion channels along the wide cosmic filament. It is discussed in Section.~\ref{sec:evolution}.

Now using the {\it main} sample of cosmic filaments (left panel), the anti-correlation persists albeit reduced ($\rho=-0.56$), owing mostly to the fact that one of the most isotropic systems is now at $90^o$ from its nearest, high-persistence (but not necessarily densest) filaments.
More interestingly, a clearer separation now emerges between the two groups of planar systems described above. The``ultra-thin" group already described (dashed red contour) now stands out with ALL its members displaying planes of satellites at $>70^o$ from their high-persistence cosmic filaments. The second group (dashed mauve contour) also stands out with members' planes markedly more aligned to their cosmic filaments.

This clear bimodality of planes along high-persistence filaments (i.e filaments with high contrast between the saddle point and the node) readily suggests that the strong tidal influence of one large cosmic filament is key in the formation of planes. Indeed, this behavior is particularly reminiscent of the bimodality in angular momentum orientation predicted by \cite{2015MNRAS.452.3369C} from constrained Tidal Torque Theory for collapsing regions near the saddle point (lower mass) or near the node (higher mass) of a single cosmic filament. Indeed, this model not only predicts the spin-filament alignment of regions near the saddle point of the filament (hence whirls), it also predicts that more massive regions closer to the spine and further towards the node will on the contrary spin orthogonally to the filament.

\subsection{Connection to Cosmic gas vorticity}

To further test the vorticity scenario for the thinnest planes, we measure the degree of alignment between each satellite system's best-fit plane and the cosmic vorticity field, in which it is embedded. More precisely, for each system we compute the gas vorticity field on a fixed grid centered on the system and extending to 5 $R_{\rm vir}$, corresponding to an extent of $\approx$ 2.5 Mpc for most systems, with a spatial resolution of $\Delta x= \rm 14\, kpc$. 

\begin{figure*}
	\centering\includegraphics[width=\linewidth]{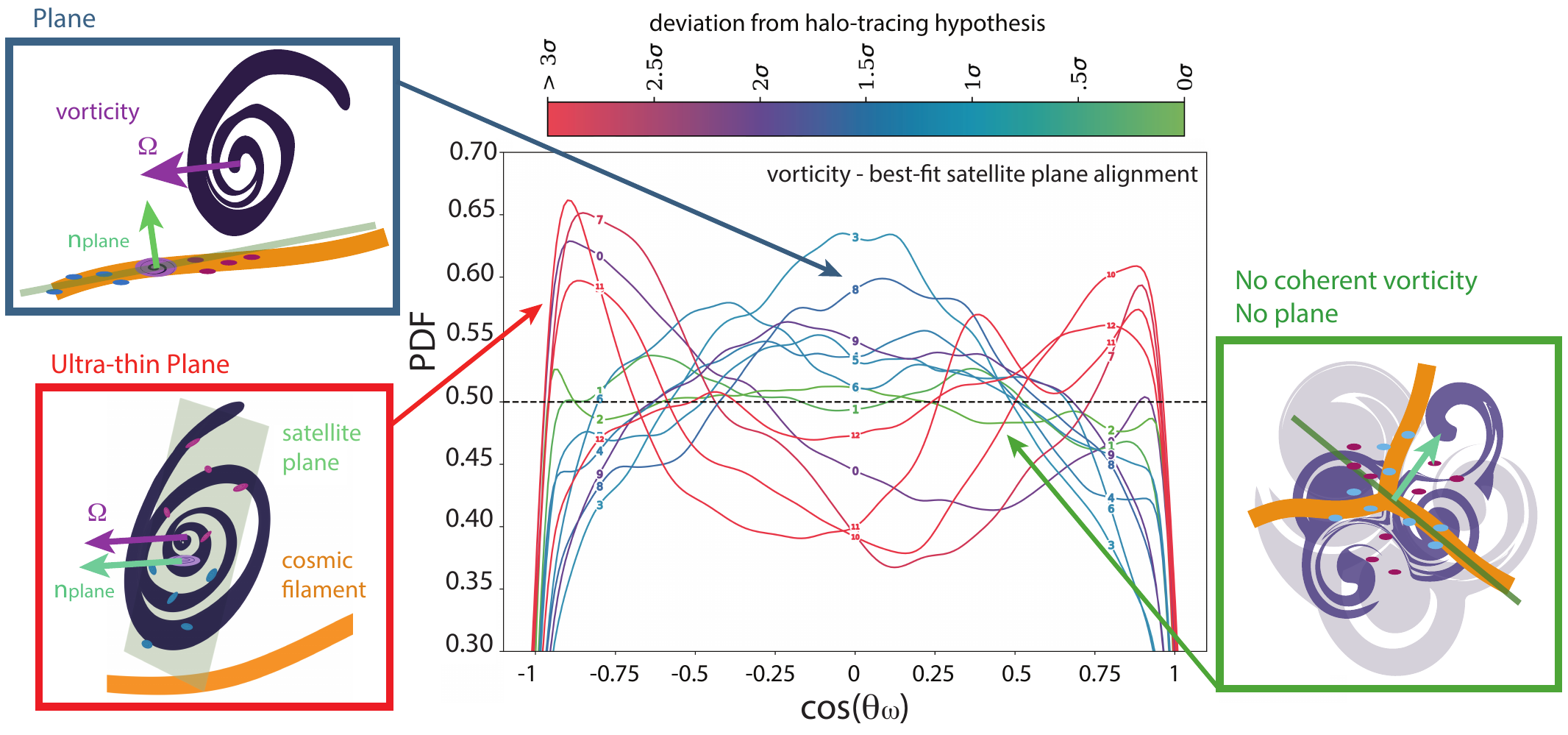}
    \caption{{\it Right:} PDF (KDE estimate) of the cosine of the angle $\theta_\omega$ between the environmental vorticity (1-5 $R_{\rm vir}$) and the normal of the best-fit satellite plane for all 13 systems. Colors rank systems from most significant plane to least significant (red to green). The uniform expectation is indicated as a black dashed line. Most planar systems lie aligned to spatially coherent vortical regions (whirls). Most isotropic systems' (green) environmental vorticity shows little to no spatial coherence. Intermediate planes (blue) lie orthogonal to vorticity, consistent with being in the filament. {\it Sketches:} illustrations for the two planar types and for an isotropic system.}
    \label{fig:vorticity_alignment}
\end{figure*}

We then compute the PDF of the cosine angle, $\theta_{\omega}$, between the vorticity of the gas cells and the normal of the best-fit plane. We excluded gas cells within 1 $R_{\rm vir}$ of each host halo to focus on environmental vorticity and exclude the halo of each galaxy. Figure \ref{fig:vorticity_alignment} displays the results for each system.
The PDFs (KDE estimates) for the most co-rotating and planar systems (ranked by standardized deviation from the mode of the null hypothesis in the 2D composite analysis, see Figure \ref{fig:2dhist_composite}) are displayed as red solid lines. Intermediate systems range from more planar in purple to less planar in blue. Finally, the most isotropic systems appear as green solid lines. Note that while individual curves are colored by their significance in the ``on-plane"  $c/a$ vs. $R_{\rm cr}$ plane, a ranking using an average of 2D $\sigma$ levels across the parameter distributions of Figure \ref{fig:2dhist_composite} gives the same rankings.  

 It is noticeable that systems displaying the most significant co-rotating planes (in red) show a specific trend towards alignment of their plane with cosmic vortical regions (marked excess probability for $\cos(\theta_\omega)=\pm 1$ and deficit probability around $\cos(\theta_{\omega})=\pm 0$). This effect is still noticeable for the most significant intermediate system. 
 Intermediate systems with more satellites and thicker yet still prominent planes show the opposite tendency. These tend to lie orthogonal to their surrounding Mpc-scale vorticity field (although the strength of the signal is reduced), consistent with their orientation within the cosmic filament hence orthogonal to the whirl. Nonetheless, the whirl visibly retains some spatially coherence.
 This bimodality in plane formation orientation is very reminiscent of the so-called ``spin-flip" phenomenon now thoroughly described in theoretical, numerical and observational work \citep{2015MNRAS.452.3369C,Dubois_2014,Welker20, Kraljic21}. This describes the tendency of low-mass halos/galaxies to spin parallel to the nearby cosmic filament (owing to their location in peri-filament whirls) whereby higher-mass halos/galaxies spin perpendicular to it (owing to their migration along the spine of cosmic filament and closer to  nodes of the cosmic web). In complete coherence with this picture, we find that the alignment of satellite planes with vorticity is also a mass-dependent behavior, as we describe in Appendix.~\ref{appendix:mass-vort}.
 
Finally, for isotropic, non-corotating systems (in green) the near uniform PDF shows no alignment trend at all. This highlights an important difference compared to the two other cases: the lack of a spatially coherent vorticity field in their vicinity. Indeed, the vorticity field around these shows no preferred orientation. This is consistent with a turbulent region, of the type observed at a node where several large cosmic filaments branch from multiple directions.

These observations strongly suggest that thinnest planes of satellites emerge from accretion in coherent, steady, mega-parsec scale cosmic flows, either whirling vortical streams in the vicinity of the filament, or filamentary streams along the spine of cosmic filament. As already mentioned, such coherent vortical flows are an expected feature in the vicinity of cosmic filaments at $z>2$ \citep{2011MNRAS.418.2493P,2015MNRAS.446.2744L,2021MNRAS.501.4635S} but can be strongly disturbed by feedback and filament evolution at later times. The fact that most strongly planar $z \approx 0$ systems of satellites display strong residual alignment with their remnants suggests a quiet history of cosmic flows -- hence of cosmic filaments -- around such systems, perhaps devoid of multi-directional merging/collapsing with other filaments. Conversely, non-planar systems may have undergone a more turbulent environmental history, with significant reshaping of the geometry of surrounding cosmic flows and filaments during the last billion years.

\begin{figure}
	\includegraphics[width=\linewidth]{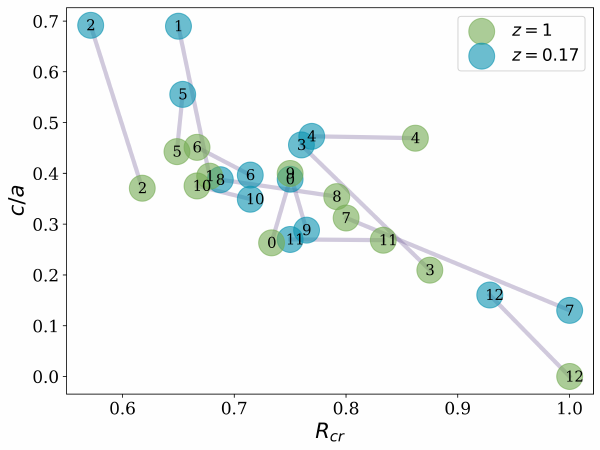}
    \caption{Relation between axis ratio c/a and co-rotating fraction $R_{\rm cr}$ for the 13 NewHorizon systems at $z=0.17$ (blue) and $z=1$ (green). Grey-shaded lines link systems across redshift. The correlation is present at $z=1$, with a trend towards thinner, more co-rotating systems at higher redshift.}
    \label{fig:zevolution}
\end{figure}

Consistently, the analysis of distributions of satellites across cosmic time reveals not only that anisotropies are long-lived but tend to be more ubiquitous at higher redshift. Figure~\ref{fig:zevolution} shows the evolution of the c/a vs. $R_{\rm cr}$ trend between $z=1$ (in green) and $z= 0.17$ (in blue) for the 13 systems in the sample. These are obtained by tracking the all satellites within a fixed physical distance of $R_{\rm vir}(z=0.17)$ around the main progenitor of each system across redshift. We checked that this did not significantly change the number of satellites around most systems. The change in $N_{sat}$ is $<20\%$ on average for most systems with the exclusion of the most populated ones (1,2,4). For most populated ones, the number can double from $z=1$ to $z=0.17$ as an effect of group merging.

It is evident that the trend for thinner distributions to also be more co-rotating is maintained across cosmic time, with most systems displaying only minor evolution in anisotropy and co-rotation across redshift.

It is also striking that $z=1$ systems are generally more co-rotating and anisotropic than their $z \approx 0$ counterparts. In fact, even the most isotropic, non-corotating $z \approx 0$ systems (1, 2 ,5) display significant anisotropy at $z=1$ and became isotropic within the last 8 billion years. This confirms that the tendency of satellites to form in planes is not a transient phenomenon but a long-lived trend, likely favored by sustained environmental dynamical features.

The coherence of the vorticity field across a large volume, maintaining an orientation in line with high-z expectations around cosmic filaments, therefore, appears as a key parameter to maintain a plane of satellites. If such is the case, the size of a vorticity quadrant (coherent whirl region) should be large enough to enclose the full system. To get an idea of the extent of coherent regions around each system, we reproduce the same analysis for a few typical systems, restraining the volume probed around each system to 1.5 Mpc in radius. Results presented in Appendix.~\ref{fig:vorticity_alignments_fixed} show that the PDFs of $\cos(\theta_{\omega})$ for the planar systems start to exhibit pronounced asymmetric peaks at either $\cos(\theta_{\omega})=1$ or $\cos(\theta_{\omega})=-1$, showing that not only vorticity's alignment with the filament but also vorticity' sign along the filament starts to self-correlate strongly across these scales. 

\subsection{Connection to filament evolution: zippers and twisters }
\label{sec:evolution}

A complete description of filament dynamics around Milky Ways and local massive galaxy analogs in general is beyond the scope of this study as full filament trackers are still in early development phase \citep{Cadiou20,2024A&A...684A..63G}. Moreover, the progressive transition from narrow gas streams to extended cosmic filaments over time is complex to characterize in this density regime. An analysis of the gaseous profiles and stability over time of cosmic filaments and streams will be the subject of a follow-up study (Madhani et al., in prep). 

However, our limited number of systems allows us to follow the main evolution of the high-persistence filaments of the gaseous cosmic web from $z=2$ to $z<0.2$ around our  systems to better understand the formation of planes of satellites. To this end, we follow the redshift evolution from $z=2$ to $z=0.17$ of the gas density of a co-moving cubic patch spanning 8 $R_{\rm vir}$ around each system at $z=0.17$. Gas density is interpolated on a 14 ckpc-resolution grid at each redshift. Note that in the rest of the article we will refer to mergers between two filaments as {\it zippers} to avoid confusion with galaxy mergers.

Figure~\ref{fig:evolution.pdf} displays 14 ckpc-deep projections of the gas density around 3 representative systems (2, 9, and 7 from top to bottom) at $z=2$, $z=1$ and $z=0.17$ (from left to right). On these gas maps, we allow a small shift of up to $\pm 2$ pixels along the line-of-sight ($\Delta x < 28$ ckpc) to recenter each system's progenitors between consecutive redshifts to optimize visibility. As a reminder:
\begin{itemize}
    \item System 2 (top row) is one of the most massive systems in our sample. At $z<0.2$, its satellite distribution is mostly isotropic with no sign of co-rotation, and the vorticity field around it shows no spatial coherence at our resolution. It lies at the node of several contrasted cosmic filaments. Nonetheless, its direction of maximal elongation is mostly aligned with the densest nearby filament.
    \item System 7 (middle row) is on the lower-mass end of our sample. It displays one of the thinnest and most co-rotating planes of satellites. It also lies at the heart of 1 single robust cosmic filament offset from major nodes. Its satellite plane is orthogonal to it. The projection shows an edge-on view of said filament.
    \item System 9 (bottom row)'s mass is mid-range in our sample. At $z<0.2$, it displays a significant plane of satellites with a demonstrated tendency to co-rotation, but its level of satellite anisotropy is not in the top quartile of our sample. It lies at the heart of 1 single robust cosmic filament -- to which the plane is well aligned --  and offset from major nodes. It is however less than 1 Mpc away from a bifurcation point where 2 dense filaments are still smoothly fusing into the major one. The projection shows the cross-section of said filament.
\end{itemize}
The differences in cosmic environments (node of the cosmic web for System 2, mid-filament position for System 7 and 9) are easily visible on right panels of Figure~\ref{fig:evolution.pdf}. Now tracking these environments back in time reveals striking differences in cosmic evolution:

\begin{figure*}
    \includegraphics[width=1\textwidth]{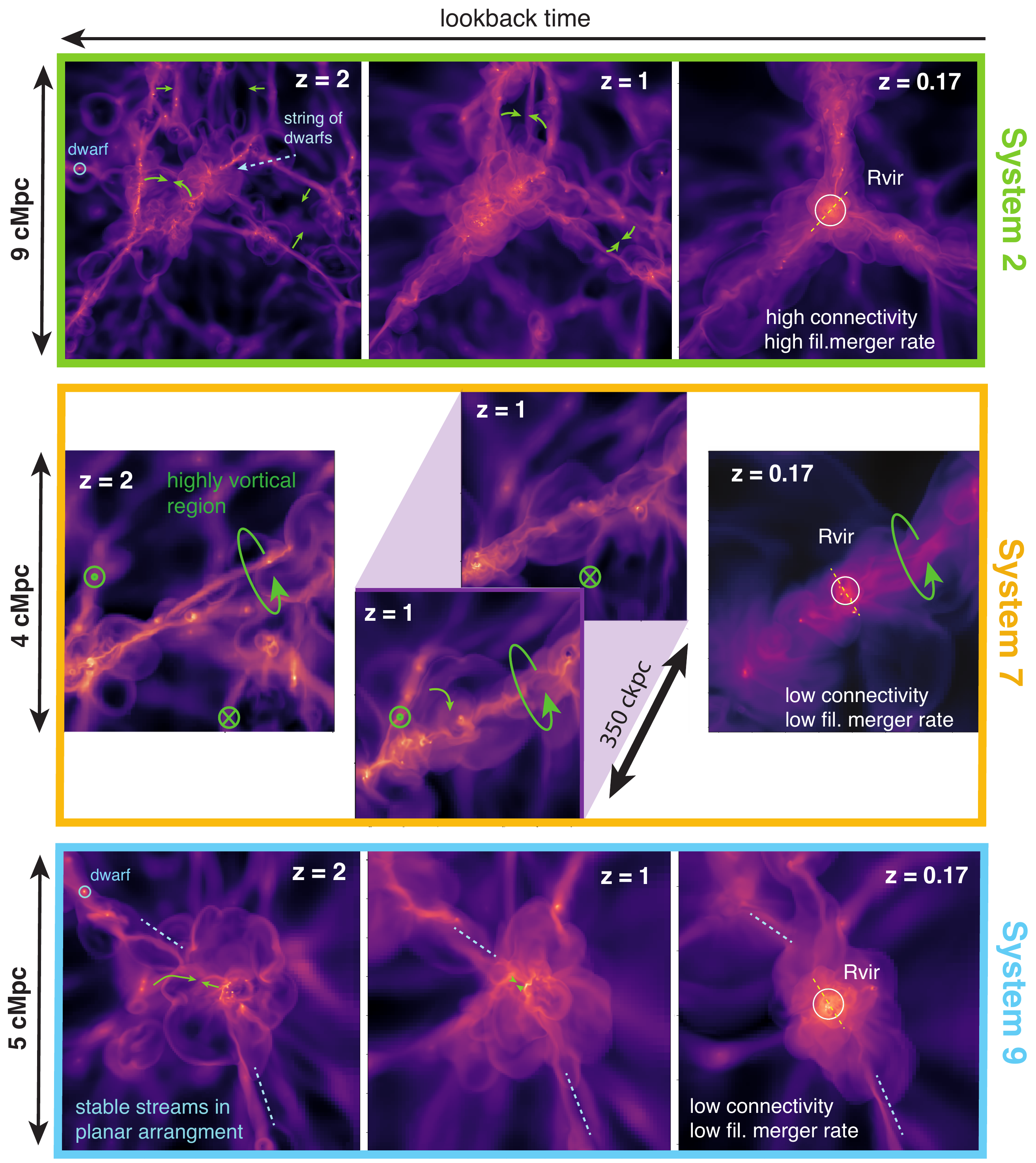}
    \caption{14-kpc wide projection of the cosmic gas density around three typical NewHorizon MW systems at $z=2$, $z=1$, and $z=0.17$. Green symbols indicate major zippers and cosmic flows. Yellow dashed lines indicate the satellites' direction of maximal anisotropy. System 2 is one of the most isotropic. It is at a dense node of the web, connected to $>3$ large feeding filaments resulting from multi-directional filament merging. System 7 has a strong plane orthogonal to its single major cosmic filament, the result of highly vortical zippers of thin streams between $z=2$ and $z=0.17$. The plane is offset from the spine and aligned with the vortical whirls. System 9 has a plane aligned with its one major, stable cosmic filament (along the LOS), undergoing one single minor filament zipper after $z=2$ and fed by stable in-wall streams. It sits at the spine of its filament.  }
    \label{fig:evolution.pdf}
\end{figure*}

\begin{itemize}
    \item {\bf System 2} originated from a web of multiple, multi-directional filaments at $z=2$. Its nodal position at $z\approx 0$ is the result of multiple filament collapses and zippers in various directions at $z<2$ and even $z<1$ (green arrows). It is striking that even for System 2 we do observe the formation of an extended planar distribution of dwarfs through the early ($z>1$) zipping of thin-cored gas-rich filaments (see $z=1$ panel). The plane is unsurprisingly found along the remnant filament. It should also be noted that the residual anisotropy of its satellites still aligns with it at $z<0.2$. This shows that, even for System 2, the late-time isotropy is actually the result of multiple competing accretion channels from multiple contrasted cosmic filaments in a fast evolving cosmic region, and is not hindered by a purported intrinsic thickness of cosmic filaments. Similar conclusions can be reached for Systems 1 and 5, which also display residual anisotropies aligned with their densest connected filament. It should be noted also that many of these residual anisotropies result indeed from the existence of one or several embedded planes within a more isotropic distribution as is described in Appendix.~\ref{appendix:embedded}.
    \item {\bf System 7} lies in a less dense region of the Universe, in the vicinity of a single dense filament, strikingly stable and away from multi-connected nodes. This region is characterized by high vorticity. Its $z=0.17$ main filament is the descendant of one single zipper between two aligned filaments at $z>1$. It is striking that this zipper is characterized by high orbital momentum and helicity. This prevented the regeneration of a dense core (hence of a dense accretion channel along the filament) and did not modify the direction of the remnant filament itself. We refer to this type of zippers as {\it twisters}). On the middle panels, the $z=1$ map is split into a ``front" and a ``back" map. Indeed, the rapid spinning of the twisting and zipping filaments did not allow to follow the evolution of the large branching side streams on the same slice. The plane of satellites, therefore, appears to emerge from sustained accretion in this coherent, stable whirling region, with vorticity typically aligned with the filament. Similar conclusions can be reached for Systems 0 and 10. It shows how such systems' striking planes emerge from the combination of a stable, highly vortical environment tidally induced by a single high-persistence cosmic filament. This vorticity-rich region both channels satellites from a few stable, thin streams and prevents the formation of thin, dense accretion channels along the filament. System 12 follows a mostly similar evolution but connects to a markedly denser side stream/intermediate scale filament that may further help with the stability of vorticity and accretion. 
    \item {\bf System 9} and all its main progenitors since $z=2$ lie at the heart of a single, strikingly stable filament with virtually no orbital momentum (here in cross-section to highlight this aspect) undergoing one single zipper between $z=2$ and $z=0.2$. It is fed by steady (non-spinning), thinner side-filaments with stream-like cores (Edge-on views reveal that these progressively re-align with the main filament at close separation). It is also visible that the plane shows marked secondary alignment with the most steady streams branching into the filament. This suggests that the plane in this case is mostly the result of direct accretion along the core of the main filament, with likely enhancement from the zipper in a fashion similar to System 2's main filament between $z=2$ and $z=1$. Indeed, by boosting shocks along the spine, the zipper boosts the formation of thin streams along the filament itself. Systems 8 and 6 follow the same environmental dynamical pattern (Note: System 4 shows similarities with this latter pattern in terms of environment dynamics but is more massive, lies at the heart of a much wider filament, and is much closer to a massive node. It has undergone more zippers and more prominent secondary accretion channels, which explains its greater isotropy, as illustrated by System 2).
\end{itemize}
This analysis, therefore, confirms a bimodal formation of satellite planes closely related to the geometry and evolution of the surrounding cosmic web and streams. It is important to note that these scenarios are not incompatible with major halo and galaxy mergers contributing to planes because these primarily occur along the same accretion channels as gas and their orbital momentum is also sensitive to cosmic flows. However, the fact that most of the systems in our sample do not have a massive close companion at $z=0.17$ or a recent major merger (in the last 2 billion years) highlights the fact that these may not be necessary contributors to satellite planes. This, therefore, better explains the ubiquity of anisotropic satellite distributions.

One can note that System 9's pattern elegantly explains Andromeda's plane(s), which are at the spine of a cosmic-flow detected cosmic filament and lie strikingly aligned to it ($4.5^o$ and $13.9^o$, while remaining away from the large Virgo node \citep{Libeskind2015}. Their secondary, near-perfect alignment with the direction of fastest collapse onto the filament ($<5^o$) is also very reminiscent of the secondary alignment with contrasted, stable side streams seen around Systems 9, 8, or 6. This is also the case for Centaurus A plane(s), which might be better explained as one thicker plane, or competing planes from competing close-by streams. The Milky Way plane displays a stronger misalignment ($38.7^o$) with the same cosmic filament but is still nearly perfectly aligned with the direction of the fastest collapse. This could, therefore, arise from any of the two patterns analyzed here.
 
\section{Discussion}
\label{sec:discussion}
\subsection{A multi-scale model for satellite plane formation}
\label{sec:Model}

\begin{figure*}
	\includegraphics[width=2.1\columnwidth]{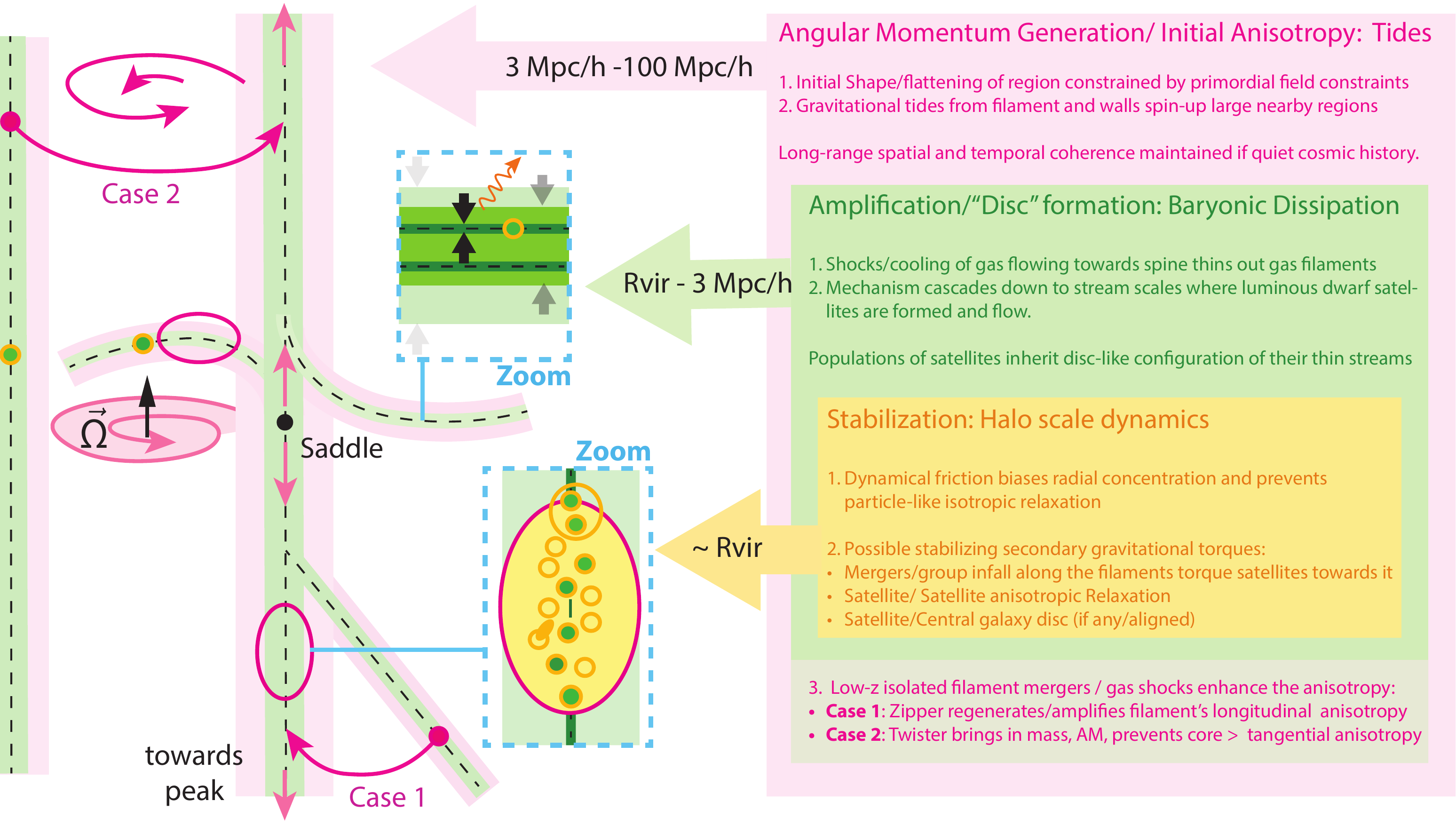}
    \caption{Illustrative sketch of our multi-scale model for satellite plane formation in a Milky Way to Cen A mass halo, centered on one major cosmic filament with possible interactions with secondary smaller filaments. DM-dominated cosmic scales are displayed in pink with fuchsia arrows for filament dynamics and lighter ones for cosmic flows within the main filament. Gas structures are displayed in green, with darker shades for denser, cooler gas and DM small-scale virialzed structures (DM sub-halos and diffuse DM within $R_{\rm vir}$) are in yellow shades. Blue zoom inserts show mechanisms that require high (NH) resolution to be properly described. The formation of a plane of satellites is akin to the formation of a galactic disc but with key mechanisms distributed across different coupled scales: initial spin build-up and flattening with long-term coherence (cosmic tidal torques near the filament-type saddle point), gas radiative dissipation (due to shocks and cooling that form streams, and enhanced by zippers), stabilizing small-scale friction and gravitational torques within $R_{\rm vir}$. Because of the coupling across scales, no fine-tuning is required, which explains their relative ubiquity.}
    \label{fig:model}
\end{figure*}

The picture that emerges from our analysis highlights that multiple scales are involved in the emergence and survival of a plane of dwarf satellites. Unsurprisingly, their formation shows similarities with the emergence of galactic and accretion discs in other contexts. It involves the triad of angular momentum build-up then conservation, gas dissipation/accretion, and possible stabilization/reinforcement by internal dynamics. However, a specific and crucial aspect is the distribution of these three classes of mechanisms across vastly different yet coupled scales, from local to cosmic. In particular, most of the necessary dissipation takes place outside of the halo. We summarize the most salient aspects of our model in Fig.~\ref{fig:model}.

 {\bf Angular momentum}: We find two types of plane, aligned or orthogonal to their nearby cosmic filament (Fig.~\ref{fig:stream_sat_align}). In all cases though, halos hosting planes are found along one single major filament, away from major nodes where several large filaments branch from misaligned directions. They are also surrounded by vorticity aligned with the filament and spatially coherent across Mega-parsecs (Fig.~\ref{fig:vorticity_alignment}), as was predicted analytically and numerically around isolated high-redshift filaments by \cite{1999A&A...343..663P}, \cite{2015MNRAS.446.2744L} and \cite{2015MNRAS.452.3369C}. This bimodal nature of satellite planarity can, therefore, be related to the bimodal nature of shape and angular momentum predicted for collapsing regions in the vicinity of a large-scale filament by \cite{2015MNRAS.452.3369C}. This analytical study finds that the initial elongation of a collapsing region in an initial (density) Gaussian random field is constrained by its location relative to the nearby proto-filament. The build-up of its spin results in turn from the correlated gravitational tides they experience either near the filament-type saddle-point (resulting in aligned spin) or near the peak (resulting in orthogonal spin). Cosmic filaments on this scale (3 - 100 Mpc/h) therefore set the initial flattening and spin of the region that will feed the MW-mass halo. In the first case, the halo will later accrete from the vortical whirl. In the second case, the dominant accretion will occur along the core of the filament itself. In \cite{2015MNRAS.452.3369C}, the spin transition is best predicted by the mass of the collapsing proto-halo as more massive halos tend to grow closer to density peaks, hence spin orthogonally to their filament. This mass dependence also holds true in our study: the flip of satellite planes from parallel to orthogonal to the filament is seen for higher-mass systems ($10^{10.7}\,M_{\odot}<M_{*}<10^{11.2}\,M_{\odot}$). The loss of planarity is seen at even higher-mass ($M_{*}>10^{11.2}\,M_{\odot}$)), for systems at dense nodes of the cosmic web. Mass, as a tracer of environment and connectivity, is therefore a good predictor of the presence of a plane. This aspect is developed in Appendix.~\ref{appendix:connectivity} and .~\ref{appendix:mass-vort}. However, as was noted by previous pure DM or lower resolution numerical studies \citep{2021Galax...9...66P}, these large-scale structures themselves are not triaxial enough on their own to explain the thinness of planes of satellites.

{\bf Dissipation}: However, planes are formed by luminous satellites, hence are more related to the dynamics of the cool gas. It was noted early in the development of hydrodynamic cosmological simulations \citep{2011MNRAS.418.2493P,Lu2024} that gaseous filaments are dissipative, hence substantially thinner than their DM counterparts. Indeed, these filaments form through shocks and radiative cooling of cosmic flows converging towards their spine. However, a key aspect missed in earlier simulations is that, with sufficient resolution, this mechanism cascades down to much smaller scales of the cosmic web, forming extremely thin streams ($< 30 $ kpc in NH) within 1-2 Mpc/h from MW-mass halos. Moreover, the collective geometry of these streams remains stable and constrained by larger scales provided that the cosmic filament is slowly evolving.

Dwarf galaxies form in these thin streams, therefore, are already accreted in a planar fashion on these scales (0.3 - 3 Mpc/h), as we showed in Fig.~\ref{fig:stream_sat_align}. As we discussed in Section.~\ref{sec:evolution}, filament mergers at $z<2$ (when filaments are contrasted and metal-rich) that do not destabilize the direction of the main filament also play a role in thinning out the accretion plane. An edge-on zipper ( case 1) can form a contrasted filament core through shocks and boost dissipation and planar accretion along the main filament. In contrast, ``twisters" (case 2, zippers with a lot of orbital momentum) prevent the formation of a cool core along the main filament (hence preventing a major accretion channel) but boost the build-up of angular momentum in the vortical whirl. Streams shocking and dissipating in this whirl therefore flatten in the plane of the whirl. 

{\bf Internal dynamics}: Once these marked planar configurations are in place, in-halo dynamics can work to sustain or further enhance them, even if not necessary to form them in the first place. In particular, while the shock-mediated dissipation typical of gas streams does not directly apply to infalling satellites, dynamical friction is expected to take over as the main dissipative term within $R_{\rm vir}$. As such, it is responsible for the accretion, hence radial concentration, of satellites and for the redistribution of their coherent angular momentum across the halo. This feature can in itself help maintain a preferential plane \citep{Sato2024}.  Its net effect on the flattening of satellite distributions in NewHorizon can be deduced from Fig.~\ref{fig:c2a_dmhaloes}, where it is noticeable that the average inertial axis ratio of all DM subhalos($\approx 0.7$) is already significantly lower than that of the diffuse DM host halos ($\approx 0.8$5). However, for most halos it is compatible with the null hypothesis of Fig.~\ref{fig:2dhist_syst_12} obtained by resampling the shape of the DM halo at the concentration of luminous satellites. This suggests that the main contributor to this flattening is the distinct radial concentration of satellites, which results from dynamical friction.

It is also likely that luminous group infall and secondary torques, for instance between satellites and a major merger or between satellites in the plane themselves, are at play to enhance or stabilize the anisotropy. Although our study does not focus on these mechanisms, their contributions have been suggested by spectroscopic observations of the Milky Way halo \citep{Julio24,Taibi24}, and analysis of small halo-scale zoom simulations have described them at length \citep{DSouza21,Samuel21,Kanehisa23,Patel24,Joshi24}. These studies showed how these could contribute to form planar configurations. In the absence of sufficiently resolved larger scale constraints, such planar configurations were however still relatively transient and infrequent in zoom simulations (up to a few $\%$ of cases) \citep{SantosSantos20,2020MNRAS.491.1471S,2023NatAs...7..481S}, albeit already $>$ 10 times more common than in less resolved large cosmological runs \citep{2019ApJ...875..105P}. Moreover, the ad-hoc addition of more massive structures, like an LMC, to simulations \citep{Samuel21} prompted worries of over-tuning in light of the ever-growing ubiquity of planes in the Local Universe \citep{Heesters21, Pawlowski21,Pawlowski24}. However, within our model, where the formation of the plane is first fully generated on a mix of environmental scales, these mechanisms can exert a steady stochastic enhancement and secular stabilization of the planarity down to $z=0$ without tension with observations. 

It is important to stress that this model does not require any fine-tuning. The key aspect that allows the cascade of mechanisms to work in concert is the sustained, predictable coupling between scales through tides and accretion. This, therefore, explains the ubiquity of satellite planes, in a mass range that corresponds to a very specific preferential large-scale environment.

\subsection{Discrepancy with previous numerical studies}
\label{sec:LowResSim}

We are now equipped to understand the seemingly unique ability of NewHorizon to recover planes in the current simulation landscape. This simply boils down to its computationally intensive hence still rare approach to force both volume and resolution down to $z\approx0$. This results in its ability to resolve dwarfs down to $6.10^{5} \, \mathbf{M_\odot}$ and 35 pc but also maintain $<35 \, \rm kpc$ resolution $> 1$ Mpc away from halos.

\subsubsection{Large cosmological simulations}

A first issue that plagues large cosmological runs like TNG100 or Eagle \citep{Pillepich15,Schaye15} is their inability to resolve dwarf galaxies. Typically, in such runs only galaxies with $M_{*}>10^9\,\mathbf{M_{\odot}}$ are resolved with more than 50 star particles, and ``galaxies" below $10^8\,\mathbf{M_{\odot}}$ drop to only a few particles. Additionally, it should be noted that the coarse stochastic nature of star formation subgrid models in such runs results in virtually ALL dark-matter halos with $M_{\rm DM} \approx 10^9 - 10^{10} \mathbf{M_{\odot}}$ hosting similarly small amounts of star particles. As a result, these simulations can only claim to trace dwarfs by selecting significantly under-sampled structures as ``galaxies" \citep[see for instance][and references therein]{2018MPLA...3330004P}, which practically amounts to selecting all DM halos within the corresponding mass range. The analyzed distribution, therefore, traces the distribution of DM subhalos and NOT that of realistic luminous satellites. This results in an underestimation of the real anisotropy of luminous satellites, which typically form in narrower cosmic gas filaments.

\begin{figure}
	\includegraphics[width=1.1\columnwidth]{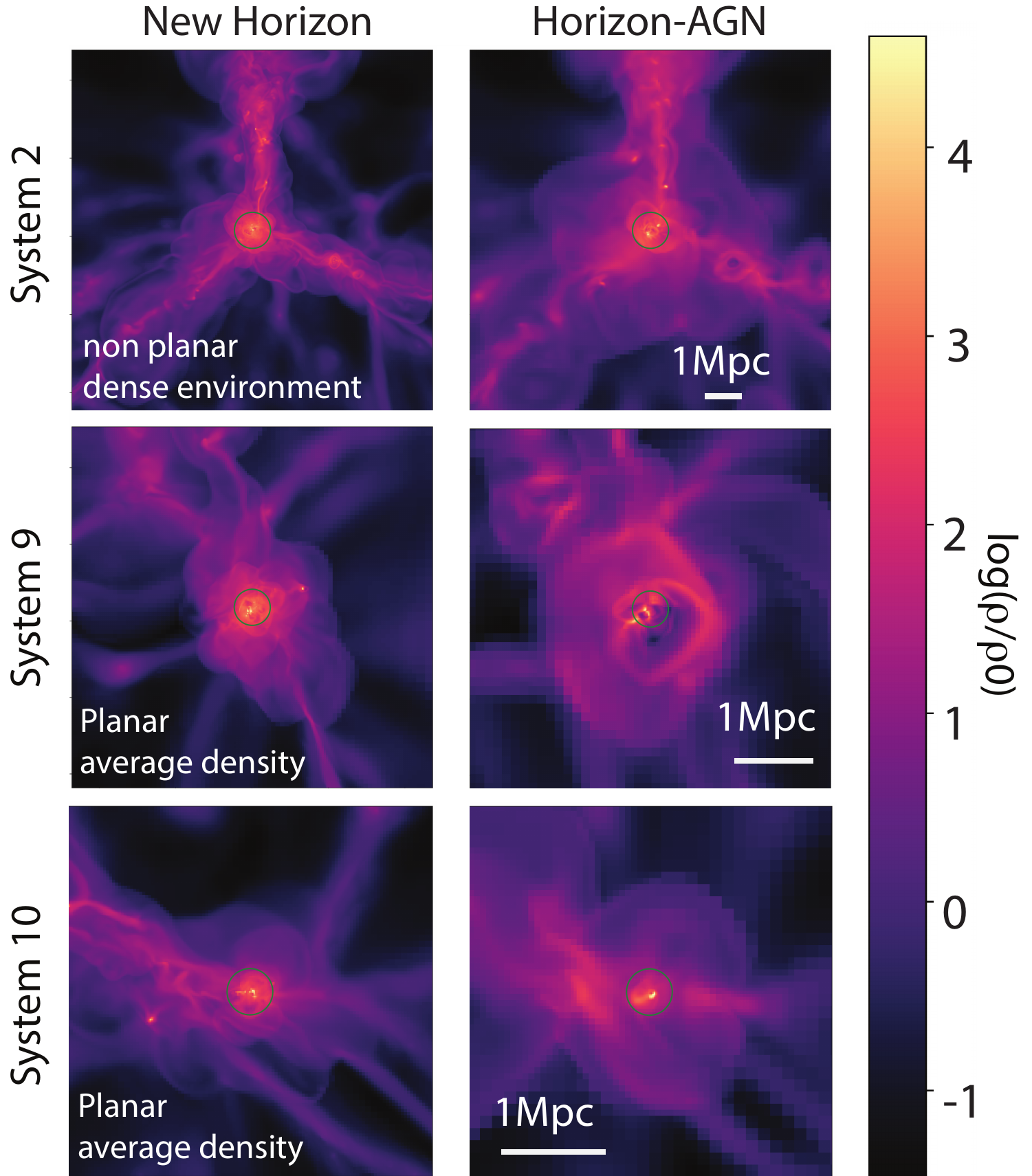}
    \caption{Projected gas density maps of three systems in NewHorizon (left panel) and Horizon-AGN (right panel) at $z=0.17$ using density interpolated on a regular grid with 14 kpc pixel size. The host halo's virial radius is marked in green. NewHorizon resolves the extent of gas streams at the core of main cosmic filaments while Horizon-AGN does not.}
    \label{fig:hagn_comparison}
\end{figure}

This effect is already detectable in Horizon-AGN \citep{Dubois_2014}, the 142 Mpc wide, kpc resolution parent run of NewHorizon, where the distribution of {\it resolved} luminous satellites ($>10^9 \, \mathbf{M_{\odot}})$) within $R_{\rm vir}$ of their host is significantly less isotropic than that of purely DM satellites within the same subhalo mass range ($>10^{10} \, \mathbf{M_{\odot}})$). For instance, stacking satellite distributions for all hosts with $M_{\rm h}>10^{11} \, \mathbf{M_{\odot}})$ in Horizon-AGN around pre-defined, physically motivated test directions (nearest cosmic filament, central's minor axis, halo's minor axis) typically reveal deviations in average angle of $3-4^o$ from isotropy for luminous satellites and deviations $<1^o$ for DM-only satellites. 
As revealed in Fig.~\ref{fig:c2a_dmhaloes}, this effect is greatly amplified for dwarf galaxies in NewHorizon, which highlights the particular importance of environmental resolution. As an illustration, Figure~\ref{fig:hagn_comparison}  shows the main differences in hydrodynamic resolution between NewHorizon and Horizon-AGN, on the example of three flagship MW systems identified in NewHorizon:
System 2 (upper panels), System 9 (middle panels) and System 10 (lower panels, evolution pattern similar to System 7)

In each case, the left panel displays a $\Delta z=14\, \rm kpc$ deep projected map of the gas density around the system of interest in NewHorizon with the virial radius indicated as a green circle. All of these maps were obtained after projecting the gas density onto a 3D grid with pixel extent $\Delta x=14\, \rm kpc$. The right panel displays the equivalent map obtained from Horizon-AGN.

These maps illustrate the extreme thinness of the density profiles of distant ($ >1 \, \rm Mpc$) streams feeding local massive galaxy analog systems in NewHorizon, typically comparable to that of satellite planes. In particular, thin dense streams are visible near the core of cosmic filaments, megaparsecs away from the central galaxy. These are absent in Horizon-AGN, where the limited resolution smooths out the vicinity of cosmic filaments and allows bubbles inherited from shocks and outflows to reach larger cosmic scales and substantially disturb the network of streams.

For instance, it is particularly striking that, in NewHorizon, the Mpc scale environment of planar systems is notoriously anisotropic and planar itself, with aligned dense and thin ($<35\, \rm kpc$) filamentary streams extending across 10 $R_{\rm vir}$ on each side of the system. Conversely, in Horizon-AGN the lack of cosmic resolution isotropizes the gas density around each system, with the thinnest features already exceeding a few 100 kpc in thickness. What is more, this effect is particularly striking around systems found to be planar in NewHorizon (Systems 9 and 10), precisely because these are found in average to low-density regions, where the adaptive resolution of kiloparsec resolution cosmological simulations like Horizon-AGN quickly degrades out of the virial radius. This highlights that the resolution of anisotropic structures on both cosmic and local scales is needed to recover planes of satellites and to properly model the assembly history of Milky Way mass and above systems. 

\subsubsection{Halo-scale zoom simulations at $<100 \,\rm pc$ resolution}

As discussed in previous sections, small-volume zoom simulations do resolve dwarf galaxies and recover stronger anisotropies than cosmological runs, sometimes reaching frequencies of $2-5\%$ for planar distributions, albeit only in heavily constrained scenarios (presence of a large neighbor and an LMC) \citep{Samuel21}. However, these efforts do not explain the ubiquity of planes of satellites in the nearby Universe \citep{SantosSantos20,2020MNRAS.491.1471S,2023NatAs...7..481S,Samuel21}. Lack of environmental resolution is the most likely culprit in these runs too, with zoom regions extending at best out to 1 or 2 Mpc from the central galaxies. Indeed, in these cases, two mechanisms can be at play in limiting the resolution and stability of the streams:

\begin{itemize}
    \item The Lagrangian patch of the gas is distinct from the lagrangian patch of dark matter that is typically extracted to produce zoom-in simulations \citep{2017ApJ...844...86L}. In particular, its exact geometry, angular momentum, and degree of anisotropy might be underestimated, and its dynamics might be modified by DM-based truncation in resolution. It should also be noted that, unlike its DM counterpart, the cooler gas and baryons flowing along the spine of filaments and radially into massive halos are not prone to backsplash. For instance, at cluster scale, filament gas has been found to outrun dark matter inwards up to Megaparsecs from their host at $z=0$ \citep{Rost21,2022MNRAS.512..926K}, in large part due to the absence of backsplash of this component. In massive halos like Milky Way hosts, this can further increase the mismatch between the DM and gas patches depending on the strategy used to define the boundaries of the resimulated patch.
    \item Tidal interactions from cosmic structures are a major driver of anisotropic dynamics. Given the observed thinness of dense streams around planar Milky Way mass and above systems in NewHorizon and their high degree of planarity, precisely resolving density anisotropies out to cosmic scales is likely necessary to form and stabilize these streams. In particular, contamination by low-resolution DM particles and/or under-resolved tides might be a major factor in disturbing planes of streams across cosmic time, therefore breaking the dynamical and spatial coherence between satellites accreted at vastly different times. As an example, even in the Latte runs \citep{2020MNRAS.491.1471S}, which boasts the most impressive compromise between volume and resolution among halo-scale MW zooms  (1-2 Mpc wide boxes, 1 pc maximum resolution), the uncontaminated region around Milky Way halos at $z=0$ only extends out to $\approx$ 600 kpc. Low-resolution particles can, for instance, emulate ``massive accretion events," which have been found to significantly torque low-mass satellites out of their initial orbits \citep{DSouza21}. In this context, sufficient resolution of tides is not guaranteed for late-entry satellites, possibly leading to spatial decoherence of planes.
    \item A least known but potentially dominant effect merging considerations on tidal disruption and volume pertains to the size of the parent simulation that encloses the zoom region. Indeed, resolving the tides (or equivalently the initial velocity gradients) that significantly spin up gas and dark matter on halo scales, coherently over cosmic time, requires to include contributions from large modulations of the density field, spread out to large cosmic distances \citep[see for example][]{Cadiou21,Pontzen21}. For a box to be large enough to properly sample the tides (i.e recover most of the variance) on MW halos in LCDM cosmology, a simulation box must span at least 100 Mpc/h (private communication with the authors). Therefore, properly resolving the large-scale generation of angular momentum in a proto-Milky Way halo requires an embedding box at least as large as 100 Mpc/h, ideally twice as large. In that regard, with a parent box spanning 100 Mpc/h at $z=0$, even NewHorizon lies at the extreme lower limit of what is acceptable to properly recover long-term spin coherence. Most zoom simulations of Milky Way galaxies display boxes smaller than this, therefore potentially severely underestimating the large-scale tides. 
\end{itemize}

This highlights the fact that interfacial disturbance mechanisms and effects of box size are still understudied in zoom simulations and will require further scrutiny. 

\section{Conclusions}
\label{sec:Conclusions}

In this study, we have analyzed 12 local massive galaxy (Milky Way to Cen A) analog systems in the NewHorizon simulation, which combines a cosmic volume of (16)$^3$ Mpc$^3$, a maximal spatial resolution of 35 pc, and a resolution in stellar mass of $10^4\, \mathbf{M_\odot}$. Using a best-fit plane finder similar to what is used in existing observations, optimized with an evolutionary algorithm, we find that co-rotating planes of dwarf satellites comparable to the ones observed in the Local Universe are recovered in at least $30\%$ to $70\%$ of our systems, depending on definitions. 

Most importantly, irrespective of specific definitions, we find a marked trend towards anisotropy and co-rotation in most systems, inconsistent with the assumption that satellites of local massive galaxy analogs relax with and follow the shape of their host halo. Our results are also incompatible with the supposed transience of such configurations. As a result, planar systems observed in the Local Universe are found to be fully consistent with Bayesian expectations inferred from the NewHorizon sample. Moreover, we find a tight correlation between the planarity of a system of satellites (traced by its minor-to-major axis ratio) and its degree of co-rotation, as measured with either co-rotation fraction or angular pole dispersion.

To understand the origin of these ubiquitous anisotropies, we investigated their environmental dependence. We reconstruct two scales of the gaseous cosmic web:
\begin{itemize}
    \item  ``cosmic filaments" across 20 Mpc while typically connecting nodes above the Milky-Way mass . These regions typically have profiles that extend above average density up to 1 Mpc from their spine. 
    \item Local gas streams within 1-2 Mpc around each structure and typically trace thin filamentary flows ($< 35 \rm \, kpc$ in core thickness) that drag dwarf galaxies into Milky Way mass and above halos.
\end{itemize}

 Reconstructing the best-fit plane for each network of streams in a fashion similar to what was done for satellites, we find that:
\begin{itemize}
\item MW-type satellite systems are more planar, where gas streams between $R_{\rm vir}$ and 2 $R_{\rm vir}$ are also planar, in which case the two components also tend to align. This is not the case for isotropic systems.
\item Thinnest planes of satellites are found within a narrow range of halo mass close to numerous transition masses in galaxy evolution,
\item A striking correlation exists between the planarity/co-rotation of best-fit planes and their angle with the nearby cosmic filaments. The thinnest and most co-rotating planes of satellites lie at $80-90^o$ from their closest cosmic filament, unlike more isotropic systems, which are more aligned with their cosmic filaments ($<35^o$). 
\item A second type of plane (less thin but still significantly anisotropic) shows, on the contrary, a tendency to align with the nearby cosmic filaments ($<45^o$). These tend to correspond to more massive systems with more satellites, more similar to Centaurus A.
\end{itemize}

To investigate the dynamic origin of these trends, we derive the vorticity of cosmic gas flows around each system. This reveals that, irrespective of orientation towards the filament, planes of satellites are found in regions of spatially coherent vorticity (strongly correlated orientation across Mpc). In contrast, vorticity around isotropic systems shows no large-scale coherence. The thinnest planes (orthogonal to cosmic filaments) show a strong tendency to align within coherent environmental vortical whirls. Unsurprisingly, flipped intermediary planes (parallel to filaments) lie orthogonal to their nearby vortical whirls. 

This suggests that vortical regions around planes of satellites have maintained the high-z vortical pattern predicted around filaments by analytical Lagrangian models and cosmological simulations: large, tidally constrained whirls distributed in the immediate vicinity of cosmic filaments and aligned to them \citep{2015MNRAS.452.3369C}.  

These results suggest a cosmic origin of planes of satellites related to the dynamics of the cosmic web. Specifically, the fact that the remains of clear vortical structures are detectable down to $z=0.17$ around planes of satellites while they are more elusive around isotropic systems hints at specific conditions of formation. In particular, the presence of a sufficiently large filament with a quiet history (in terms of drifting, mergers or mechanical feedback) might be required to stabilize the plane of satellites.

We confirm this by analyzing the evolution of filaments since $z=2$ around each system. The thinnest planes form from thin streams in highly vortical flows in the vicinity of strikingly stable filaments undergoing only one zipper (filament merger) between $z=2$ and $z \approx 0$, characterized by high orbital momentum and helicity. We refer to these as twisters. Note that the high orbital momentum of these twisters seems to prevent the formation of a competing dense stream accretion channel along the filament. On the other hand, intermediary planes form specifically from accretion along a cosmic filament, boosted by one or two edge-on (no orbital momentum) zippers after $z=2$. Future studies will investigate how zipper-induced shocks at $z<2$ may contribute to the formation of such planes. These planes are also fed by thin side streams, stable across time, that realign with the cosmic filament at close separation. Isotropic systems are at the nodes of the cosmic web, accreting from multiple misaligned channels in a predominantly turbulent environment, arising from multiple, multidirectional zippers. 

The quantitative study of the exact evolutionary pathways of the cosmic web that lead to such configurations remains to be explored and requires the development of methods to track filaments and streams across cosmic time, which are now in their infancy \citep[see ][]{Cadiou20,2024A&A...684A..63G}. Larger simulated volumes are also required to obtain sufficient statistical power to disentangle the specific effects of different correlated accretion channels (gas streams, mergers). However, our study highlights the power of simulations that maintain high enough resolution to resolve dwarf galaxies and accretion streams across cosmic volumes to solve apparent discrepancies with Local Universe observations. 

\section{Acknowledgments}
We thank Corentin Cadiou for important comments in the final stage of this paper. We also thank Mia S. Bovill, Chris Hayward and Joel Primack for valuable discussions during the course of this investigation.
The core scientific team is comprised of Janvi Madhani, Charlotte Welker, Sneha Nair, Daniel Gallego, and Lianys Feliciano. JM led the analysis, implemented the pipeline in Python, produced most analysis plots, wrote parts of the draft and provided technical support to DG, SN and LF. CW designed the study and the overall  pipeline, wrote parts of the draft, led the bayesian inference and the discussion, prepared illustrative figures and supervised the core team. SN developed and optimized a Python package to analyze gas properties around filaments and produced stream profiles, with contributions from LF. DG conducted the statistical analysis of plane correlations with cosmic filaments and streams. CP participated in the discussion. All authors commented on the final draft. 
JM and CW also acknowledge the mentorship and administrative support of Susan Kassin in the early stages of this study, as well as her financial support through NSF award AST-1815251 and her 2022 STScI Director's Discretionary Research Fund award.
CO, JM and CW acknowledge the support of the LSST-DA Catalyst Fellowship funded by the Templeton Foundation. JM acknowledges the support of NRAO ALMA SOS award SOSPADA-034. CW acknowledges the support of the PSC-CUNY award Cycle 54. SN and CW acknowledge the support of NSF LEAPS-MPS Award AST-2316862. DG and LF acknowledge the support of the AstroCom NYC fellowship funded by NSF awards AST-2219090 and AST-1831412. Part of this study was conducted at the KITP CosmicWeb23 workshop, supported in part by the NSF Grant PHY-1748958.

\facilities{ Main computational resources were provided by CP and YD through access to Infinity HPC cluster at IAP managed by Stephane Rouberol and to the NewHorizon and Horizon-AGN simulations.}
%



\appendix

\section{An Evolutionary Algorithm to Identify the Best-Fit Plane of a Distribution}
\label{appendix:EA}

An evolutionary algorithm is an optimization algorithm that mimics the principles of biological evolution. Initialization starts with a population of data points, $\rm N_{\rm members}$, initially randomly selected. Their ``fitness" is tested against some optimization function, the fitness function. At each iterative step, a number $\rm N_{\rm remove}$ of members are removed from the population, based on their lowest fitness. A number $\rm N_{\rm mutants}$ of mutant members is introduced. These are selected as some combination of the removed members. Finally, a number $\rm N_{\rm random}$ of random points are added to the new population. The process is repeated $\rm N$ times until convergence to a pre-set threshold level of fitness is obtained. The strengths of this algorithm are its fast convergence and robustness to local minima. 

To find the best-fit plane for a system of satellites, we choose $\rm N_{\rm members}$= 25 random candidate plane normal vectors, sampled from a unit sphere centered on the central host galaxy, with $\rm N_{\rm remove}$=10, $\rm N_{\rm mutant}=5$ and $\rm N_{\rm random}=5$. $\rm N_{\rm mutant}=5$ are randomly selected among pairwise averages of removed members. The fitness function is the RMS distance of satellites to the candidate plane. We obtain significantly better convergence ($\Delta \alpha \approx 1.7^o$) after 500 iterations (corresponding to $\approx$ 5000 computations of the fitness function) than after 10 000 iterations of the brute force, random sampling approach ($\Delta \alpha \approx 3.5^o$).

\section{Planarity Metrics for ``On-Plane" Satellites}
\label{appendix:onplane_metrics}

\begin{figure*}
\includegraphics[width=\textwidth]{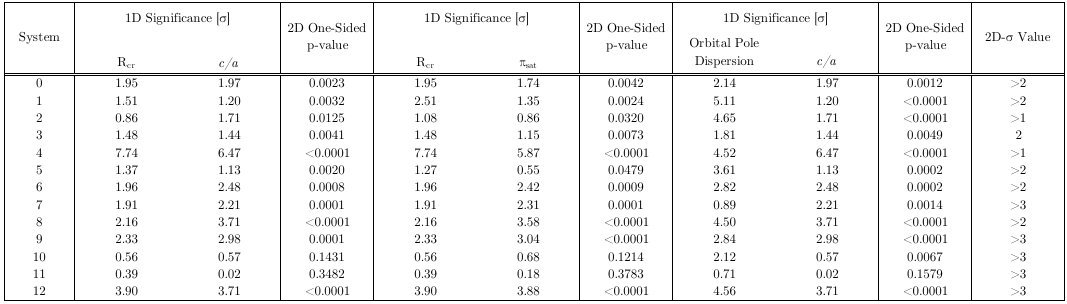}
\caption{Statistical deviations from the halo tracing hypothesis for all local massive galaxy analog systems of satellites in NewHorizon. Tests are done for various metrics: the co-rotating fraction, $R_{\rm cr}$, the spatial flattening given by the inertial $c/a$ ratio, the thinness of the plane calculated as the ratio of the extent to height of satellites from the plane, $\pi_{\rm sat}$, and finally, the orbital pole dispersion. Significance levels are presented for each 1D distribution, then for joint 2D distribution of corotation and shape parameters. The one-sided p-value measures the probability of the system occurring within the given 2D parameter space.}
\label{fig:significance_table}
\end{figure*}

In Section.~\ref{sec:significance of plane}, we tested the Milky-Way like satellite systems in NewHorizon individually against the hypothesis that they should relax in the shape of their halo and not particularly co-rotate. For this analysis, we tested various metrics for the shape (minor-to-major axis ratio $c/a$, planarity $\pi_{\rm sat}$) and co-rotation level (co-rotation fraction $R_{\rm cr}$, orbital dispersion $\Delta \theta$) of our systems. The null hypothesis was obtained was obtained by sampling 25,000 synthetic systems under the halo-tracing hypothesis for each individual system, maintaining its DM halo shape, number of satellites and radial distribution of satellites fixed. We find strong deviations that invalidate this hypothesis: Figure~\ref{fig:significance_table} summarizes  the 1D (single parameter) and 2D (joint distribution of shape and corotation) significance levels for each individual Milky-Way mass system identified in NewHorizon for all metrics measured from the full distrbution of satellites.

\begin{figure*}
\includegraphics[width=\textwidth]{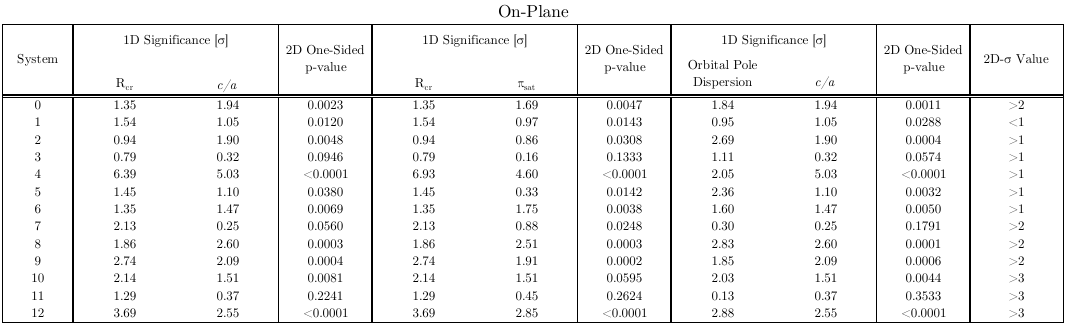}
\caption{Individual deviation levels from the halo-tracing hypothesis for all the systems in our sample, for various definitions of planarity and co-rotation parameters restricted to ``on-plane" satellites.}
\label{fig:app-sig}
\end{figure*}

In addition, Figure~\ref{fig:app-sig} summarizes the significance levels of different parameters to assess planarity and co-rotation. These are similar to these tabulated in Figure \ref{fig:significance_table} but restricted to ``on-plane" case (i.e to satellites considered on the plane). We find that the trend towards planarity and co-rotation persists regardless of our choice of definition and does not require restricting which satellites are used to observe this trend. 

Therefore, in NewHorizon individual Milky Way mass and above systems generally show strong levels of corotation and planarity incompatible with a population of satellites simply relaxed with their DM halo. This is confirmed whether using all satellites in each system or whether restricting the analysis to ``on-plane" satellites,

\section{Consistency between planarity definitions}
\label{appendix:metric_comparisons}
\begin{figure*}
	\includegraphics[width=\linewidth]{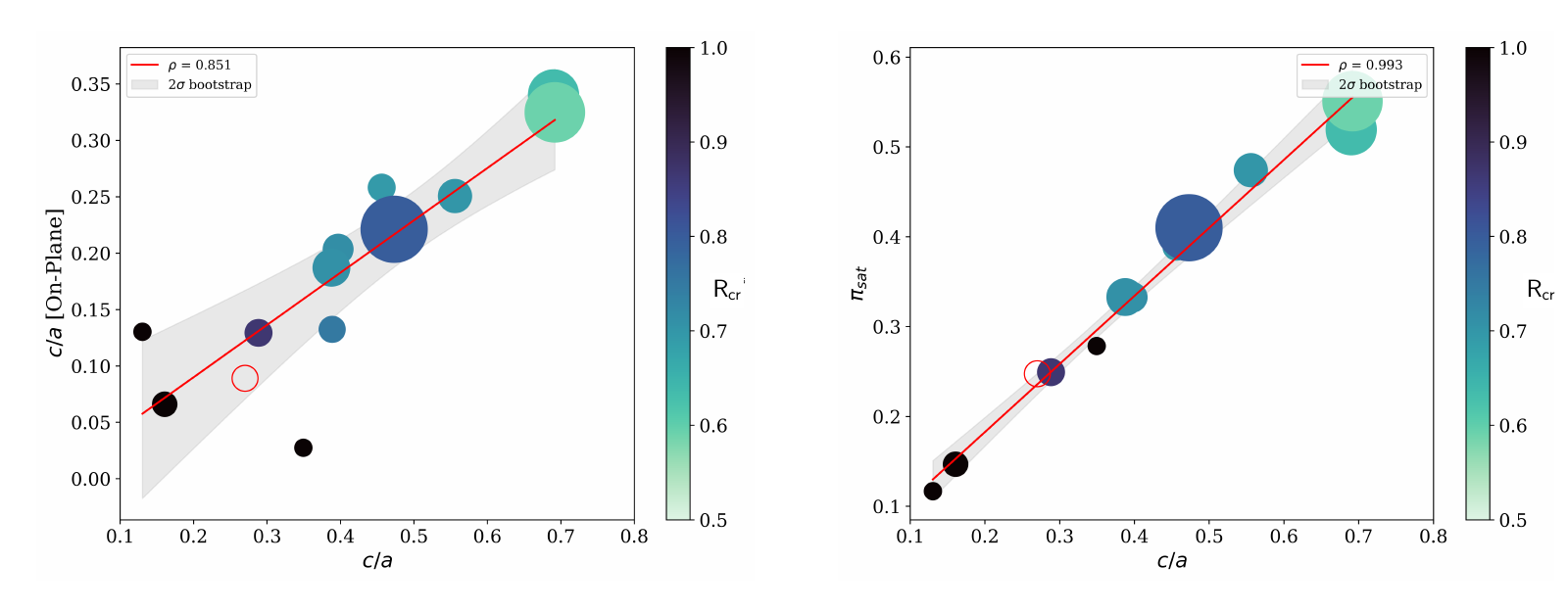}
    \caption{ ``on-plane" $c/a$  vs.  total $c/a$ (left) and $\pi_{\rm sat}$ vs. the total $c/a$ ratio (right). The best fit line is in red and $2\sigma$ bootstrapped error bars shaded in light gray. Systems are sized by $N_{\rm sat}$ and colored by the total co-rotating fraction $R_{\rm cr}$.}
    \label{fig:onplane}
\end{figure*}

In this section, we test the correlation between various definitions of planarity used in the study.
Figure~\ref{fig:onplane} displays relations between $c/a$ and $c/a_{\rm [on-plane]}$ (left) and  $c/a$ and thinness $\pi_{\rm sat}$ (right). All parameters are tightly correlated, with Pearson coefficients $\rho$ = 0.885 and $\rho$ = 0.99, respectively. This confirms that establishing the existence of planes is not dependent on specific definitions.

\section{General Incompatibility with the halo-tracing hypothesis.}
\label{appendix:general-inc}

\begin{figure}
    \includegraphics[width=\linewidth]{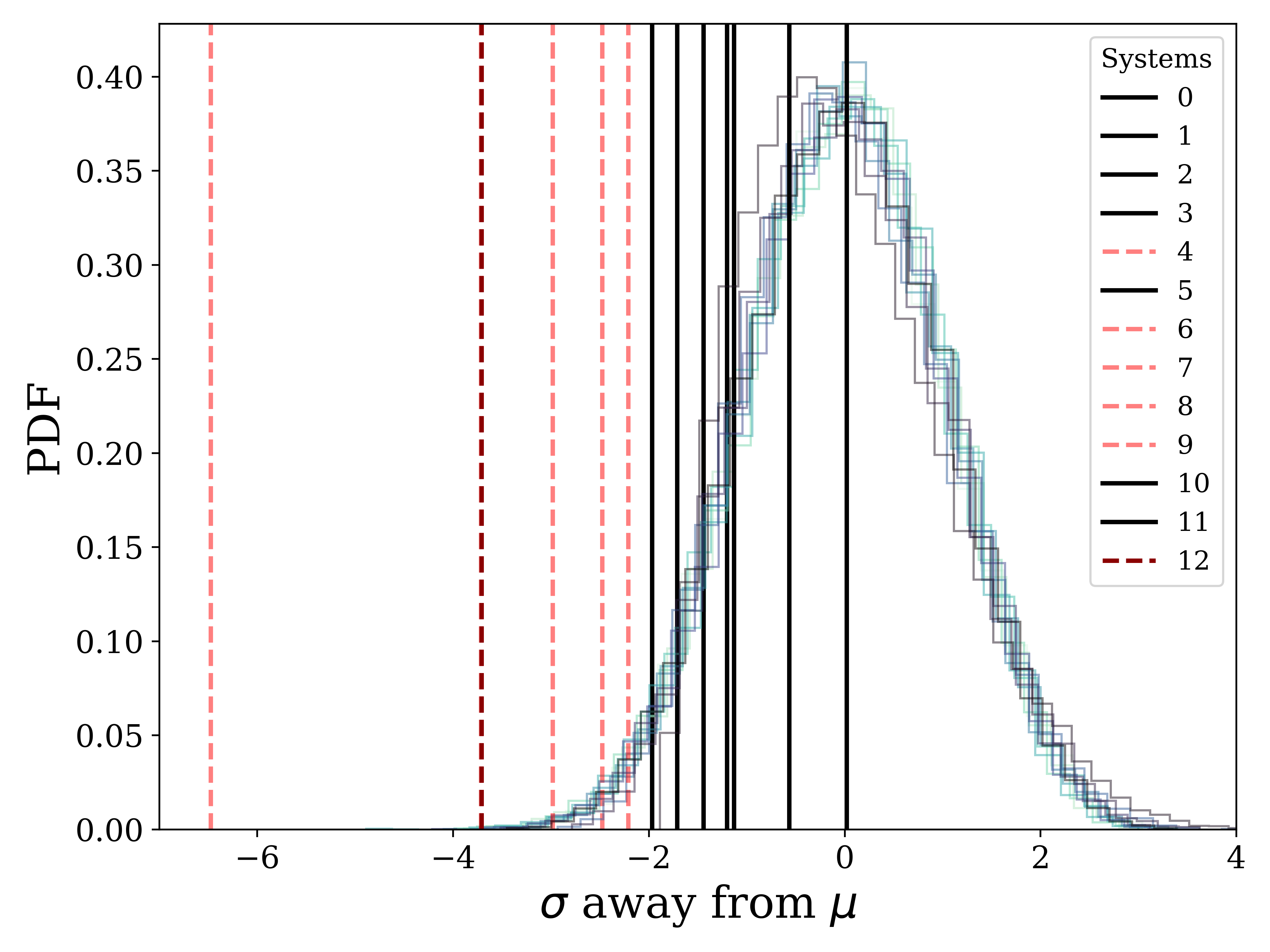}
    \caption{Stacked null distributions of normalized c/a for the 13 MW systems in NewHorizon. Vertical lines indicate the actual shape ratio of each system. Red dashed lines highlight systems with more than 2$\sigma$ deviation from the mean of the null distribution. More than $40\%$ of the sample are above the 2$\sigma$ threshold.}
     \label{fig:shape-sigma-all}
\end{figure}

To properly quantify this general incompatibility for the whole population of MW-like systems in New Horizon, we estimate the incompatibility of the marginal distributions of the main shape and co-rotation parameters $c/a$, $R_{\rm cr}$ and $\Delta \theta$ around our best-fit planes, with a KS test. We normalize and rescale the  1D null distributions as in Figure \ref{fig:shape-sigma-all} -- which shows the procedure on the example of $c/a$ -- and express the parameters of each system in terms of 1D significance. 

We then stack the null distributions (see transparent shades for c/a in Figure \ref{fig:shape-sigma-all}). Vertical lines indicate the actual shape of each system and red dashed lines highlight systems with $>2\sigma$ deviation. This allows us to perform simple two-sample KS tests for each parameter, comparing the stack null distribution to the actual NewHorizon distribution as traced by our systems.
The results are presented in Table ~\ref{table:kstest-ind}. The high significance of all the KS test confirms that MW systems in NewHorizon overall display luminous satellite distributions strikingly incompatible with the halo-tracing model, both in terms of shape and orbit alignment.

\begin{table}
\begin{tabular}{ccc}
\hline \hline
Parameter & D-value & P-value \\
\hline
Inertial $c/a$ & 0.72 & $2.8 \times 10^{-7}$  \\
$R_{\rm cr}$ & 0.57 & $1.8\times 10^{-4}$ \\
$\Delta \theta$ & 0.81 & $8.6\times 10^{-10}$ \\
\hline
\end{tabular}
\caption{KS test results comparing the distribution of main plane shape and co-rotation parameters, rescaled by the standard deviation of their individual null expectation, in the NewHorizon sample to the halo-tracing (null) expectation.}
\label{table:kstest-ind}
\end{table}

Irrespective of the exact definition of a ``plane of satellites," which often varies across studies, the picture that emerges is that most luminous satellite distributions around local massive galaxy analogs in NewHorizon are markedly anisotropic and display high degrees of co-rotation. Most systems do not relax with their DM halo but instead display a strong tendency to arrange in planar configurations and align their orbits.

\section{Behavior of Satellites within $R_{vir}$}
\label{appendix:1rvir}

\begin{figure*}
    \includegraphics[width=\linewidth]{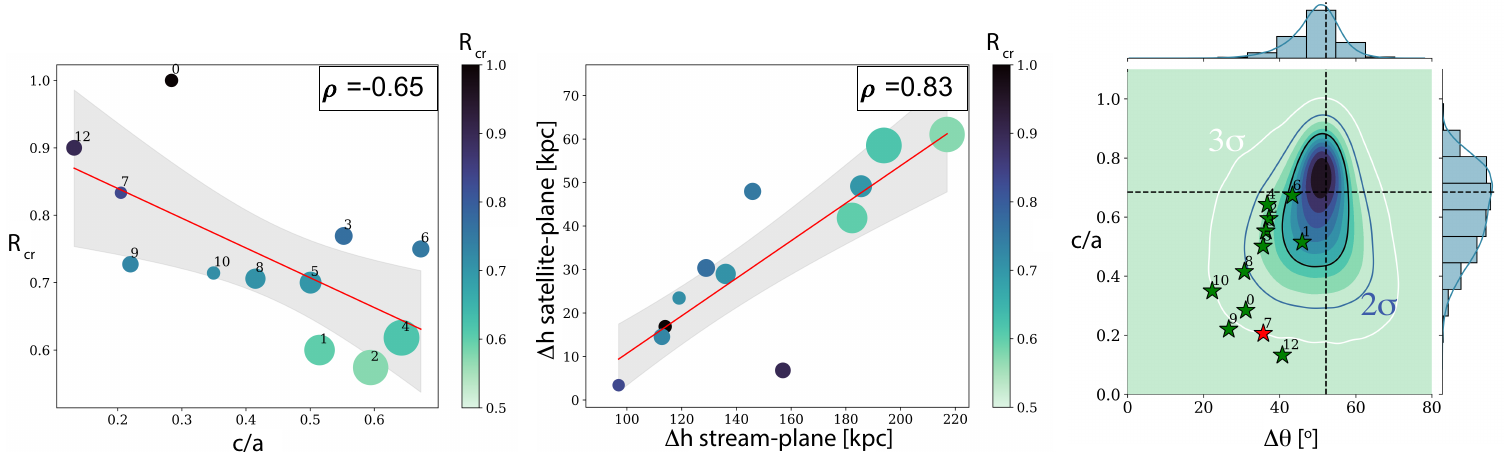}
    \caption{ {\it Left:} Co-rotating fraction versus inertial axis ratios, $c/a$ restricted to satellites within 1 $R_{vir}$, with best linear fit, Pearson coefficient $\rho$ and $2\sigma$ bootstrap contours shaded in grey. Markers size with $N_{sat}$. {\it Middle:} Thickness of satellite plane for satellites $ < 1 R_{\rm vir}$ vs. thickness of stream-plane $\Delta h$[stream-plane]  (gas streams within $1 \,R_{\rm vir}$ $<$ $x_{fil}$ $<$ $2 \,R_{\rm vir}$). Systems are colored by co-rotating fraction and sized by number of satellites. {\it Right:} Joint distribution for $c/a$ and $\Delta \theta$ with null hypothesis conoturs for the halo-tracing hypothesis in shaded of blue-green. NH systems are indicated as green stars and low-sampling System 7 is marked in red for completeness. Black, navy, and white contours mark the 1$\sigma$, 2$\sigma$, and 3$\sigma$ contours, respectively. Black dashed lines mark the average expectations for each parameter.}
     \label{fig:1rvir_results}
\end{figure*}

In our study, we investigated the behavior of satellites within a 2 $R_{\rm vir}$ volume of local massive galaxy analogs, as defined in Section \ref{sec:sample selection}. In Figure \ref{fig:1rvir_results}, we present the key results obtained with a more conservative approach: restricting our analysis to satellites that lie strictly within the $R_{\rm vir}$ volume. 
From left to right we show: the correlation between the c/a and corotation fraction (left), the relationship between satellite distribution thicknesses and the thickness of the planar network of gas streams in which they are embedded (middle, similar to Fig.\ref{fig:stream_sat_align}) and finally the equivalent of Fig.\ref{fig:2dhist_composite} for the orbital pole dispersion and c/a joint distribution (right). Note that all systems with more than 7 satellites (our pre-set threshold) within 2 $R_{\rm vir}$ also have more than 7 satellites within $R_{\rm vir}$, with the exception of system 7 which has 6. It is therefore excluded from the composite null-hypothesis but marked as a red star for reference.

We identify clearly the trend of increased co-rotation with increased planarity as defined by lower inertial axis ratio, c/a, in the leftmost panel. This anti-correlation is particularly tight with a Pearson correlation coefficient of $\rho = -0.654$. Marker size encodes number of satellites. Consistently with results in 2 $R_{\rm vir}$. the most massive systems, which host the most satellites within $R_{\rm vir}$, tend towards isotropy with lower co-rotating fractions. 

On the middle panel, the height of the stream-plane $\Delta_h$ is calculated restricting streams to those between $R_{\rm vir}$ and $2 \,R_{\rm vir}$ (defined in Section.~\ref{sec:stream-plane}). Even with a restricted sample, we confirm a strong correlation: more planar systems of satellites are embedded in more planar networks of gas streams (Pearson correlation coefficient of $\rho=0.833$). As with the original sample shown in Figure \ref{fig:stream_sat_align}, these also tend to correspond to more co-rotating systems.

Finally, the rightmost panel reproduces the 2D joint distribution of Figure \ref{fig:2dhist_composite} and present the distribution of c/a and $\Delta \theta$. As with the larger volume, the sample of systems is noticeably offset and skewed from the population-based null hypothesis of orbital pole dispersion, with at least $50\%$ of the systems past the 2$\sigma$ contour and at least 25$\%$ past $3\sigma$. This again confirms that a large fraction of our systems would be considered anomalously planar and co-rotating when compared to typical halo properties in their mass range even when considering satellites just within $R_{\rm vir}$. 

In conducting this study, we find that the tendencies of luminous satellites that lie within $R_{\rm vir}$ towards corotation, planarity, and correlation with their gaseous environment are consistent with our study of satellites within 2 $R_{\rm vir}$.

\section{Baryonic assembly biases: sampling vs. environmental anisotropy}

\subsection{Sampling of satellites.}
\label{appendix:sampling}

Within massive halos, luminous satellites are found in much lower numbers than DM sub-halos (typically $\approx$ 10$\%$ of DM sub-halos with $>10^{7.5}\,M_{\odot}$ are luminous). This difference in sampling can, in itself, bias luminous satellite distributions towards flatter shapes (smaller c/a). This is why, throughout the first part of this study, we have carefully demonstrated that this sampling-driven assembly bias is markedly insufficient to recover the elongations of the planes we observe, by systematically including sampling effects in null hypothesis (see Fig.~\ref{fig:2dhist_syst_12} and Fig.~\ref{fig:2dhist_composite} for instance). Our main conclusion is therefore that there must be an environmental source of anisotropy (beyond halo shape). However, it does not mean that luminous under-sampling brings no contribution at all to the planarity signal. This is particularly true when assessing the collective behavior of NewHorizon systems, such as the strong $c/a - R_{cr}$ anti-correlation (Fig.~\ref{fig:2dhist_composite}), as these distribute over a dex in mass and therefore in number of satellites, also. 

\begin{figure*}
\centering \includegraphics[width=0.8\linewidth]{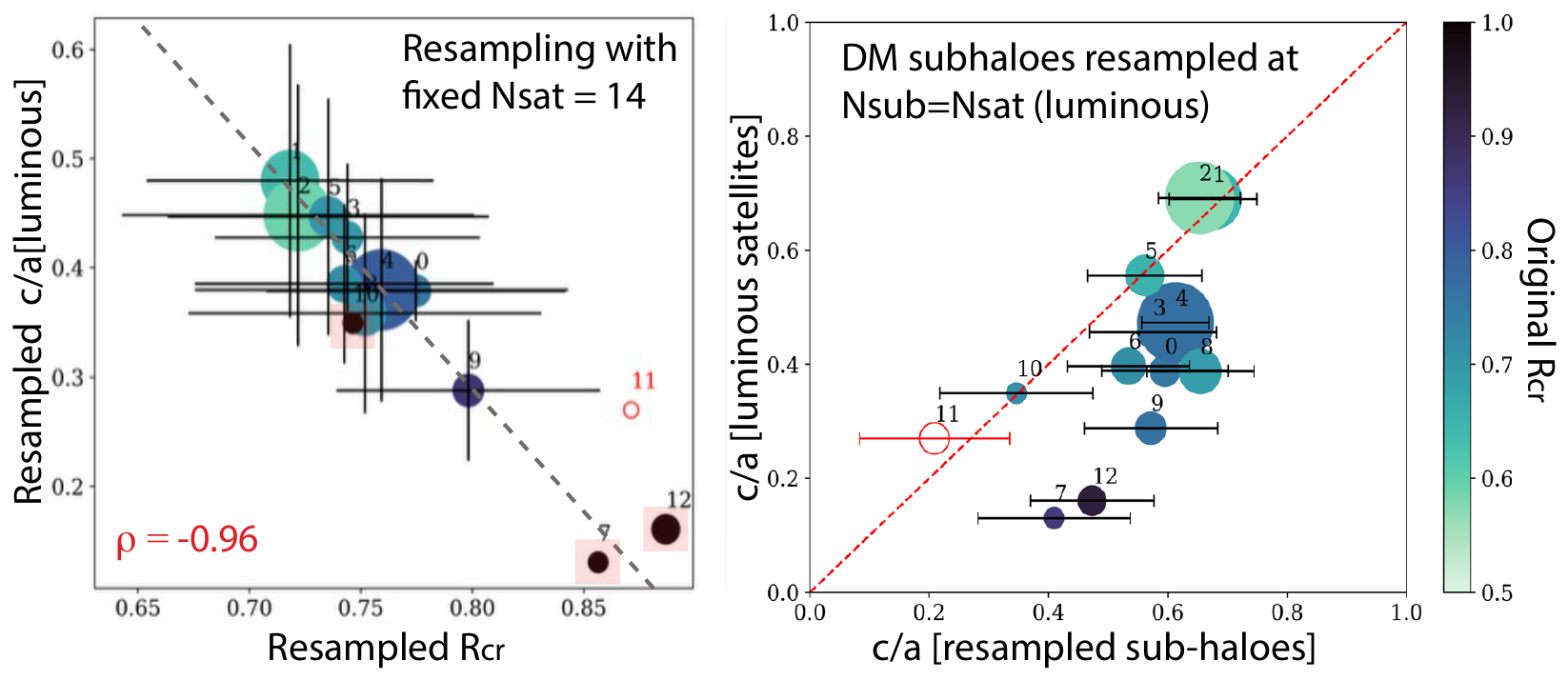}
    \caption{\it Left panel: Median c/a - $R_{cr}$ relation after resampling luminous satellites at $N_{sat}=14$ for all systems with $>14$ satellites. Non-resampled systems are highlighted by red squares. {\it Right panel:} Luminous c/a versus median ``all DM subhaloes" c/a after resampling each system's DM subhaloes at the number of luminous satellites $N_{sat}$ of each system. Colors show original $R_{cr}$ of the systems and sizes their original number of luminous satellites.}
    \label{fig:sampling}
\end{figure*}

To assess the specific contribution of sampling on this relation, we select all 9 systems with $>14$ satellites and re-sample them (without replacement) at $N_{sat}=14$ and calculate c/a and $R_{\rm cr}$ for each resampled system. Resampling is performed 1000 times for each system. The median c/a and $R_{\rm cr}$ are shown on Fig.~\ref{fig:sampling}, left panel. Standard deviations are shown as error bars. Systems with less than 14 satellites that cannot be resampled are overlaid by red squares (Note: System 12 has exactly 14 satellites). Colors indicate initial $R_{\rm cr}$ (for the full systems) and sizes indicate initial numbers of satellites.

As is visible, while resampling at fixed $N_{sat}=14$ predictably reduces c/a values by 20-40$\%$ for the few previously isotropic systems, the effect is much more limited for most systems. In any case, resampled c/a for mostly isotropic systems remain $> 0.45$, far from planar configurations. It is therefore apparent that low sampling fails to explain the thinness of systems such as 9, 12, or 7, for which is appears to contribute very little.
The linear trend between c/a and $R_{\rm cr}$ is also unchanged and is, in fact, tighter than for full-system values. This shows that if anything, variations in sampling blurred the strong correlation between c/a and $R_{\rm cr}$.

This highlights once more that an environmental source of anisotropy independent of sampling is at play in the formation of planes of satellites. To further confirm that this source of anisotropy is specific to baryons, on the right panel of Fig.\ref{fig:sampling} we display the relation between luminous and `` all-subhaloes" c/a, after resampling DM subhaloes at the number of luminous satellites for each system. Resampling is performed 1000 times per system and we display medians values, and standard deviations as error bars. Once again, even after ensuring similar sampling, it is obvious that luminous satellites display distributions much more planar than than their DM counterparts.

Note that this analysis also demonstrates that resampling higher richness systems to $N_{sat}=14$, even within a relatively narrow "Milky-Way" mass range, is NOT equivalent to studying halos with true $N_{sat}=14$. This practice should therefore be avoided when constructing null-hypothesis from simulations as it would lead to systematically overestimate the significance of any observed system in comparison.

\subsection{Additional effect of Line-of-sight variability}
\label{appendix:los}

In the main body of this study, we randomly picked one line-of-sight per system, simply optimizing it by ensuring edge-on spread of satellites, to mimic the way real systems are selected and observed. However, unlike observations, simulations allow to cast multiple lines of sight for the same system. This allows to assess the spread of results with choice of line-of-sight, including those that may cause accumulation of satellites in the centre, concentrated along the line-of-sight, thereby increasing assignment errors when parting galaxies as  "receding" or "approaching" (tangential velocity mostly orthogonal to the line of sight) when calculating $R_{\rm cr}$.

In Fig.\ref{fig:los-var}, we therefore not only resample every system to $N_{\rm sat}=14$ (25,000 times), we also pick 250 random lines of sight (LOS) for each system for the calculation of $R_{\rm cr}$. We then display $c/a$ versus the mean $R_{\rm cr}$ value (colored dots and stars, color indicates initial optimized LOS value). Red-contoured stars represent systems with $N_{sat}<14$ (not resampled, only the LOS is). Black error bars show $1-\sigma$  {\bf spread on the mean} $R_{\rm cr}$ value while grey/red error bars show the $1-\sigma$  {\bf spread on individual} $R_{\rm cr}$. 

\begin{figure*}
    \centering
    \includegraphics[width=1\textwidth]{Rcr-avg.pdf}
    \caption{\it Left panels: Median c/a - Mean $R_{cr}$ relation after resampling high-richness systems to $N_{sat}=14$. Systems with $N_{sat}<14$ appear as red stars for reference. Mean $R_{cr}$ is obtained through 250 random samplings the LOS, with 1$\sigma$ spread on $R_{cr}$ showing as grey/red error bars. 1$\sigma$ Errors on the mean  appear in black. Linear regression fit with correlation coefficient $\rho$ is found on the leftmost panel. {\it Right panels:} Same analysis restricted to on-plane satellites. The vast majority of observed systems previously deemed to have planes incompatibles with simulations (green circles) are found within 1.5$\sigma$ of similar NewHorizon systems}
    \label{fig:los-var}
\end{figure*}

The left panels display results obtained from full (resampled) system calculations. The $2\sigma$ population-based null-hypothesis contour from mean $R_{\rm cr}$ is shaded in light green. It is clear that previous results still hold for mean $R_{\rm cr}$, with 3/9 fully resampled systems out of the 2$\sigma$ contour and 6/13 when including systems with $N_{sat}$ naturally $<14$. The New Horizon distribution remains strikingly at odds with the halo tracing null-hypothesis (KS-test $<0.03$).

What's more, it is notable that, despite the unavoidable publication bias,  all systems deemed ``unusually planar" in observations display $R_{cr}$ values within 1.5$\sigma$ of the mean $R_{\rm cr}$ of NewHorizon systems with similar true number of satellites, and all but one are found within 1.5$\sigma$ of systems of similar flattening. In fact, the only outlier is NGC 4490 (14 satellites). When compared to NewHorizon systems with similar true number of satellites (e.g. systems 0, 9 and 12), it has similar shape as systems 0 and 9 but similar $R_{\rm cr}$ as system 12 (which is however much flatter). This phenomenon is likely due to the fact that NGC 4490 is actually a close-encounter gas-rich merger where, unlike in most other planes, dwarf satellites include no old dwarf but only star forming tracers of the HI flat shell surrounding the merging pair \citep{Kroupa24}. Therefore, it is perhaps better described as a filament-aligned plane in the making, where dwarfs' orbits have not had time to relax. The connection between the satellite distribution and gas streams is however undeniable in this case.

Note that many observed systems actually display values for satellites deemed ``on-and plane", not full-system values (MW1, MW2, M31, NGC253). For a fair comparison, we replicate the analysis restricted to ``on-plane" satellites on the right panel of Fig.~\ref{fig:los-var}. Now all these systems are found within $1\sigma$ of similar NewHorizon systems, and within $1.5\sigma$ of ANY NewHorizon system. 

What's more, just as in Fig.~\ref{fig:sampling} (left panel), this analysis reveals that correlation with initial $R_{cr}$ values remain strong and the linear anti-correlation between $c/a$ and $R_{cr}$ is also preserved, especially when focusing on on-plane satellites. Predictably, additional noise from reshuffling and averaging  the line-of-sight weakens the correlation (Pearson coefficient $\rho=-0.41$ for full systems). However, even in this case, the simple focus on on-plane satellites allows to recover a very significant signal (rightmost panel, $\rho=-0.72$). This correlation is of course the result of the general shape-corotation correlation we established more robustly in Fig.~\ref{fig:orbital-resampled}. This confirms that resampling higher richness systems to $N_{sat}=14$, even within a relatively narrow mass range, is NOT equivalent to studying halos with true $N_{sat}=14$. It is important to consider as this type of re-sampling on simulated datasets is commonly used to construct null-hypotheses to compare to given observed systems. Our analysis shows that this method is bound to underestimate the expected degree of planarity and co-rotation in $\Lambda$CDM, hence overestimate the significance of observed systems in comparison. We conclude that it should be avoided in future studies.

\section{Extended sample of cosmic filaments}
\label{appendix:5scosmic_fils}

\begin{figure}
	\includegraphics[width=\columnwidth]{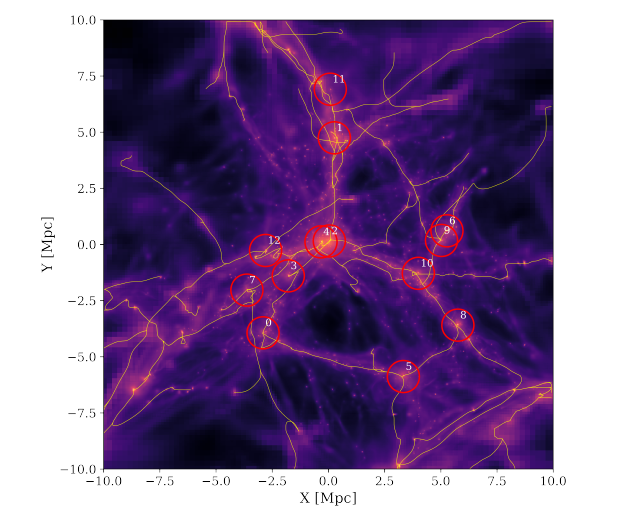}
    \caption{ The 13 MW mass systems overlaid as red circles on the Z-projected gas density map of NewHorizon. Yellow lines indicate large-scale cosmic filaments extracted by \texttt{DisPerSE}.}
    \label{fig:fil-app}
\end{figure}

Fig.~\ref{fig:fil-app} shows the {\bf extended} sample of cosmic filaments described in Section \ref{sec:filament_identification}.

 We reconstruct these filaments by applying \texttt{DisPerSE} on a large volume of $(20 \, \rm Mpc)^3$ centered on NewHorizon, the same as for the main sample of cosmic filaments. The density field is sampled on a $175^3$ pixel box and smoothed using a Gaussian filter with a standard deviation of 7 pixels. This is followed by applying a non-linear transformation to the density field to boost the persistence of thinner, dense filaments with limited contrast between their critical points. We then extract said filaments with \texttt{DisPerSE} with a 5$\sigma$ persistence threshold. 

 It is visible that, with this new set, all MW systems are nodal, connect to 2 to 4 cosmic filaments. This set still focuses on relatively long and dense filaments but includes some thinner ones or late-stage merging pairs compared to the main sample.

\section{Identifying the ``Stream-Plane"}
\label{appendix:stream-plane}

\begin{figure*}
    \includegraphics[width=\linewidth]{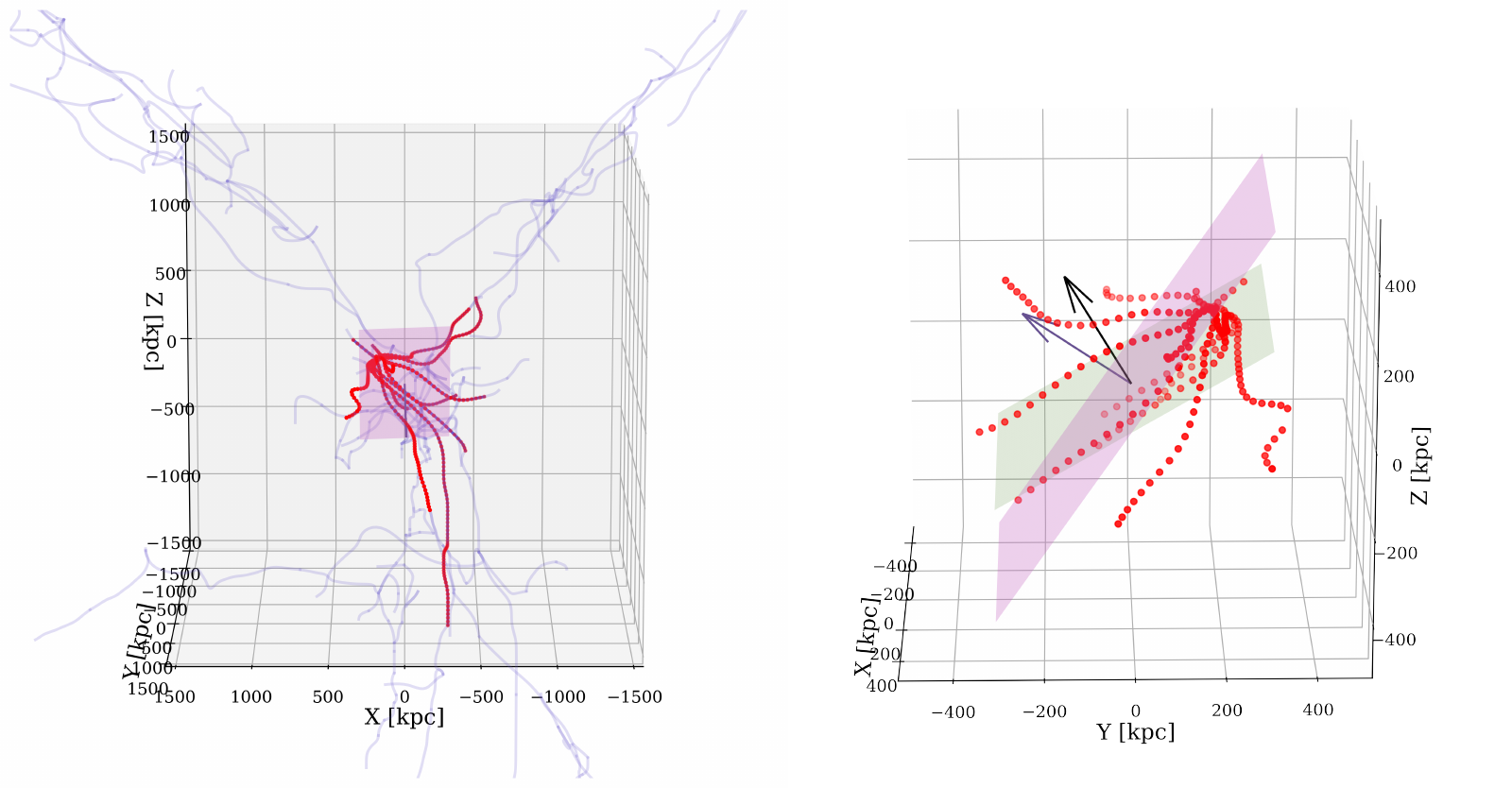}
    \caption{{\it Left:} Gas stream selection to calculate the stream-plane on the example of System 1. The complete skeleton of gas streams surrounding System 1 is shown in light purple. Sampling points of stream segments that lie within $1 R_{\rm vir}$ $<$ $x_{fil}$ $<$ $2 R_{\rm vir}$ are overlaid in red. {\it Right:} Zoom-in. The stream-plane (in pink shade, with normal vector in black) is calculated as the best-fit plane of the red sampling points. The satellite plane is shown in light green with normal vector in black. The angle between both planes naturally appears as the angle between their normal vectors.  }
    \label{fig:syst1_stream-plane_method}
\end{figure*}

We identify the ``stream-plane" using exactly the same method we used to identify the satellite plane. We simply replace the satellite positions by the positions of the DisPerse sampling points of streams connected to a particular system and found between $1 R_{\rm vir}$ and $2 R_{\rm vir}$.  Figure \ref{fig:syst1_stream-plane_method}, illustrates this method. In the left panel we show all gas streams identified by \texttt{DisPerSE} in light purple, and in a red-dotted line, restricted to the sampling points of the streams that lie within our region of interest. We use the plane finding algorithm to identify the plane of best fit for those sampling points. This allows us to measure alignment trends, for instance between the ``stream plane" (in pink shade) and the satellite plane (in green shade) for example. 

\section{Environmental drivers VS. mass}
\label{appendix:massenv}

 It is apparent in our results that the central mass (or equivalently the host halo mass, both are tightly correlated in this cosmic density regime) is a predictor of the existence of a thin plane. This correlation is not surprising since mass is a predictor of environmental density, vorticity and connectivity, all parameters central to our model. In the following sections, we provide evidence for these correlations.

\subsection{Mass and Connectivity Relationship}
\label{appendix:connectivity}

\begin{figure}
	\includegraphics[width=\columnwidth]{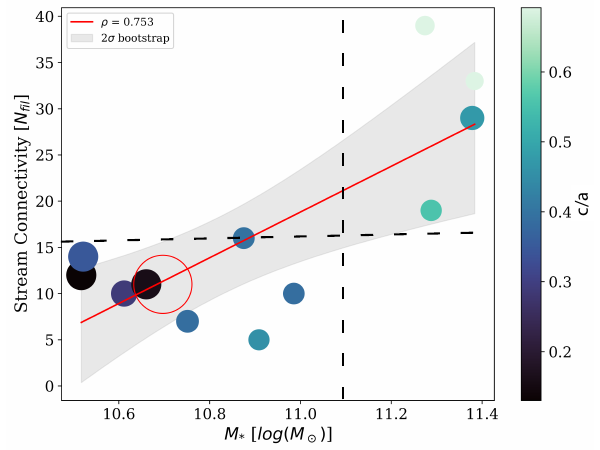}
    \caption{Stream connectivity vs. $M_{*}$, the stellar mass of the host central galaxy for each system. Systems are colored by $c/a$ ratio and sized by $R_{\rm cr}$. System 11 is added for reference as a red circle}
    \label{fig:connectivity}
\end{figure}

Figure~\ref{fig:connectivity} displays the variation of stream connectivity (number of gas streams reaching the virial radius) as a function of stellar mass across all the systems in our sample, color-coded by axis ratio.
One can see that stream connectivity increases with stellar mass, with a sharp transition at $M_*\approx 10^{11}\,\mathbf{M_\odot}$.
Below this mass, the number of streams remains below 15 (with a median at 10), and all systems of satellites show a pronounced anisotropy ($c/a <0.4$). These also correspond to the systems we found to be co-rotating and embedded in planar arrangements of streams. Above this mass, the number of streams is $>18$ for all systems, with a median at 30. Systems in this range are markedly more isotropic ($c/a >0.5$).

This corresponds to the fact that most massive systems usually lie in dense nodal regions, at the intersection of several of the most contrasted, thickest cosmic filaments in the volume, while lower-mass systems usually lie along one single well-contrasted filament, with secondary connections limited to less robust, laminar streams from nearby walls. 
This contributes to show that, as systems grow in mass, their network of cosmic filaments also becomes more contrasted and turbulent (through collapses and zippers) with several thick, high-density filaments emerging as co-dominant accretion channels. Stream connectivity increases and the distribution of streams, therefore, becomes more isotropic. As a consequence, the $z=0.17$ satellite distribution around such systems also becomes more isotropic.

This adds up to previous observations that the Milky Way mass range is one of many fast transitions, hence the difficulty to properly characterize its statistical properties.

\subsection{Mass and Vorticity Relationship}
\label{appendix:mass-vort}

\begin{figure}
	\includegraphics[width=\columnwidth]{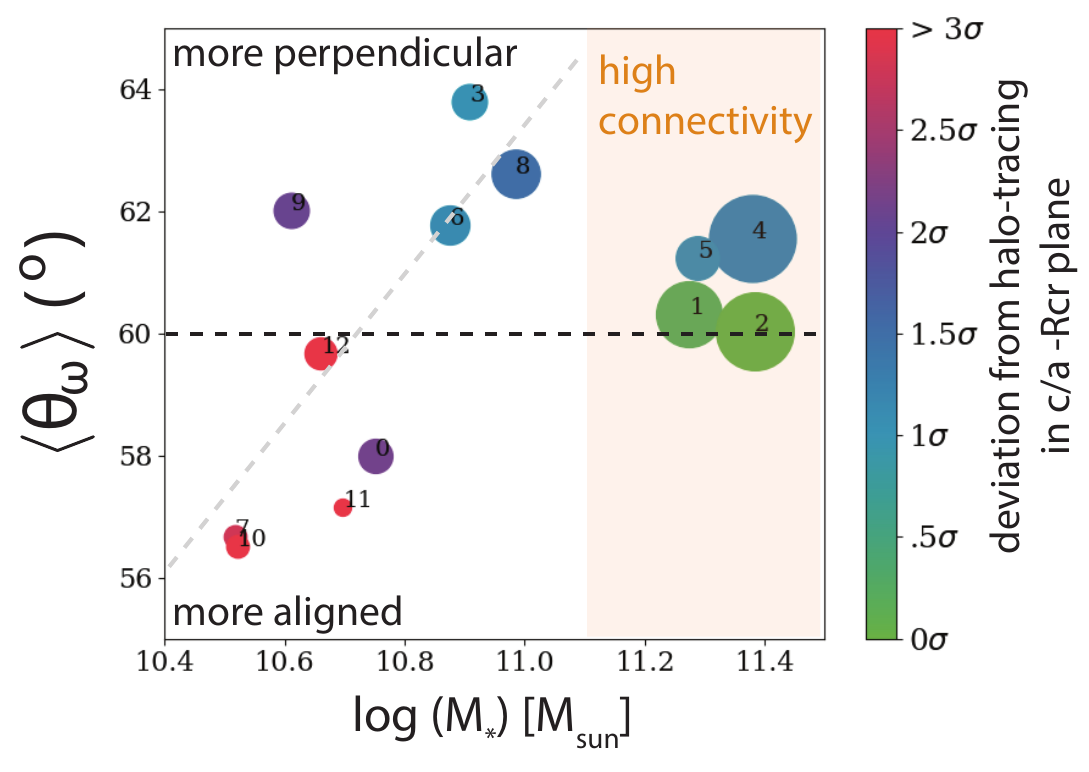}
    \caption{Mean angle between environmental vorticity and best-fit plane as a function of central stellar mass for New Horizon systems. The isotropic expectation is shown as a horizontal dashed line. The high-connectivity mass range is displayed in orange shade. Colors show significance and size show number of satellites. Least-squares Best-fit trend for systems below the connectivity transition mass is shown in dashed grey.}
    \label{fig:anglemass}
\end{figure}

Fig.~\ref{fig:anglemass} shows the evolution of the mean angle (theta) of environmental vorticity (within 3 Mpc) with the best-fit satellite plane as a function of the system’s central stellar mass. Colors show the deviations from null hypothesis of each system as defined in Fig.~\ref{fig:vorticity_alignment}. Isotropic expectation for vorticity is shown as a dashed black horizontal line. Additionally, we highlight the bimodality due to the connectivity transition mass identified in Fig.~\ref{fig:connectivity} in orange.

As can be seen, most massive systems correspond to environments with little to no spatially coherent vorticity, in accordance with their high stream connectivity because of the high-density environment they live in. This corresponds to systems at the node of 3 or more large misaligned cosmic filaments, channeling many streams across various directions.

Below $10^{[11-11.2]}\, M_{\odot}$ , stream connectivity drastically drops and mass trends with vorticity alignment, with a flip around $10^{10.7}\, M_{\odot}$. This corresponds to systems found mostly within or nearby one large filament, with sometimes a connection to a thinner side filament. Lower mass systems are further from the node and typically embedded within nearby vortical whirls. Slightly higher-mass systems are closer to the node and typically at the spine of the filament, where local whirls meet and immediate vorticity cancels out as a result. This makes these systems mostly sensitive to accretion along the filaments.

\section{Vorticity Alignments on Mpc Scale}
\label{appendix:vort1Mpc}

\begin{figure}
\includegraphics[width=0.95\linewidth]{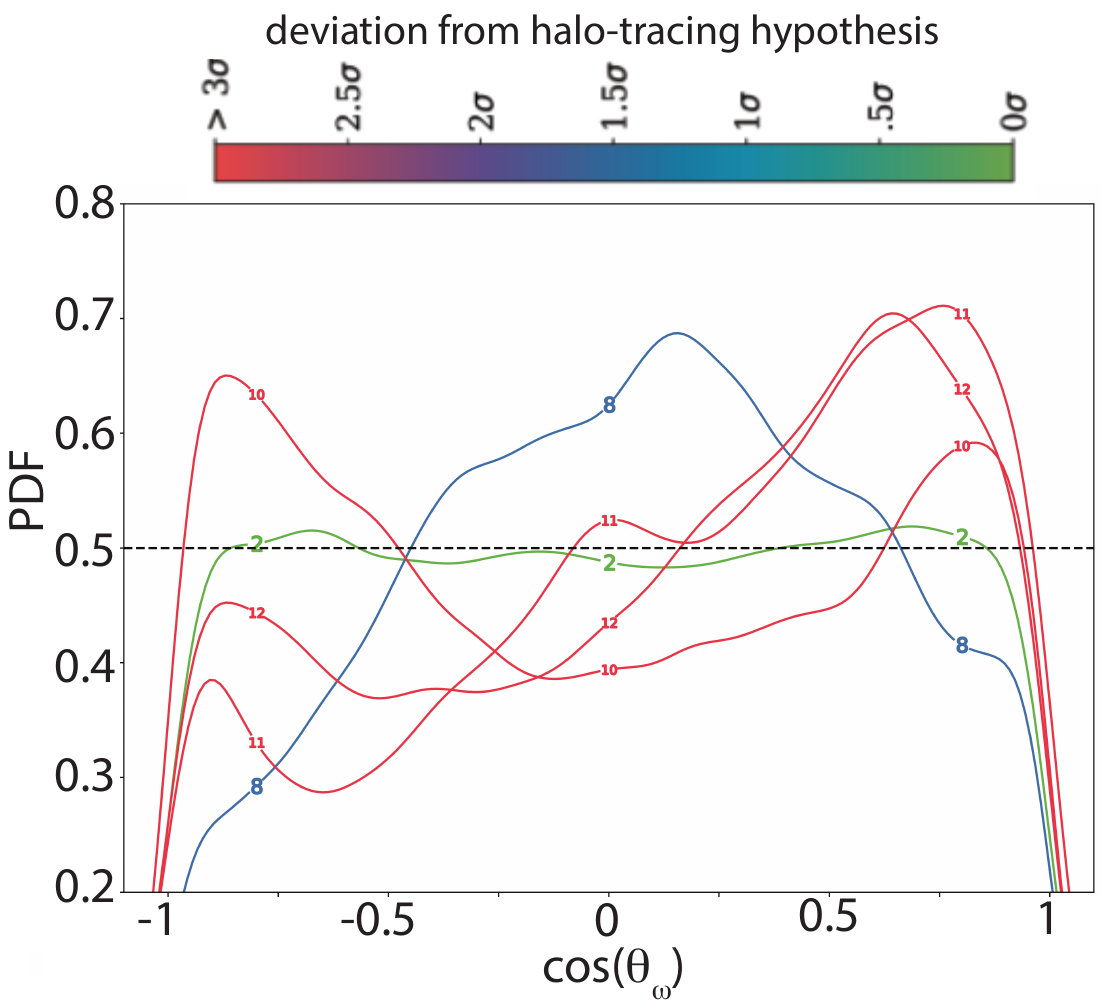}
\caption{ PDF of the cos-angle between the best-fit satellite plane normal and gas vorticity within a fixed radius of 1.5 Mpc around flagship systems. Colors show average 2D significance of the plane. The uniform expectation is marked as a black dashed line.}
\label{fig:vorticity_alignments_fixed}
\end{figure}

Fig.~\ref{fig:vorticity_alignments_fixed} repeats the analysis of Figure \ref{fig:vorticity_alignment} for a few flagship systems and restricted to vorticity between $R_{\rm vir}$ and 1.5 Mpc box around each host. Colors indicate the 2D significance the $c/a$ vs. $R_{\rm cr}$ space, as shown in Fig.~\ref{fig:2dhist_composite}. \\
While qualitatively similar to larger-scale results, it is striking that the double-horn alignment vorticity profiles around planar systems become markedly asymmetric, even being mostly limited to one-sided peaks for Systems 11 and 12. This indicates that mostly one ``whirl" (one vorticity quadrant of fixed sign) contributes to the signal. This is consistent with a typical coherence scale (``quadrant size") of $\approx$ 1-1.5 Mpc, in line with previous measurements \citep{2015MNRAS.446.2744L, 2021MNRAS.501.4635S}.

\section{Embedded planes in massive systems.}
\label{appendix:embedded}

While our study finds that most massive, nodal systems tend to be more isotropic, this conclusion only holds for their full satellite sample. Indeed, in our study, to avoid overidentification, we opted to only identify the best-fit plane from the full distribution of satellites of each system, rather than iteratively identify potential planes as flattened subsets within an otherwise isotropic distribution.

\begin{figure*}
\includegraphics[width=0.95\linewidth]{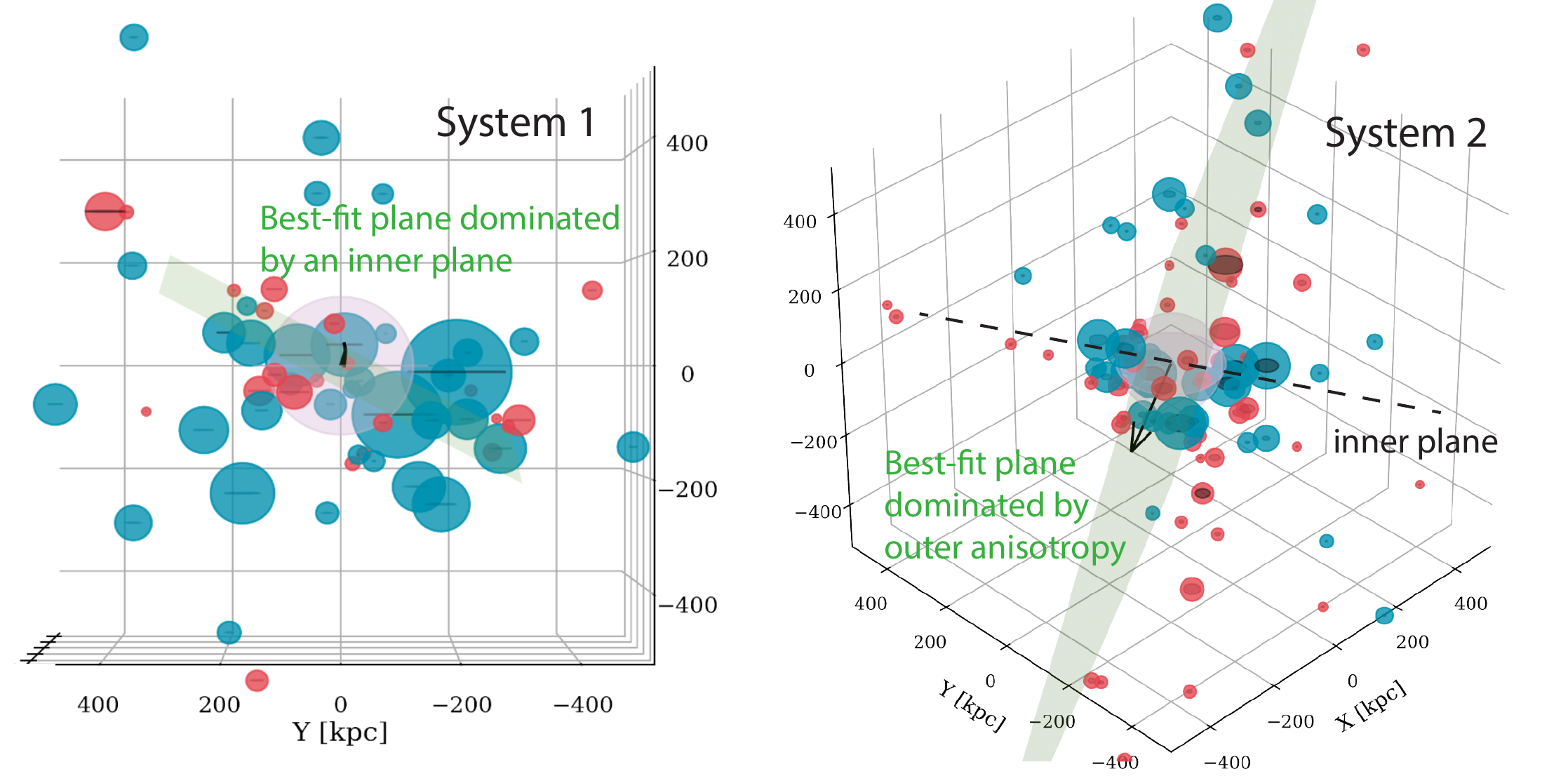}
\caption{ Distributions of approaching (blue) and receding (red) satellites around the two most massive systems in the New Horizon sample. Best-fit planes identified by our finder is shown in green shade and the random line-of-sight picked in black.}
\label{fig:large}
\end{figure*}

However, it is visually evident that even the most isotropic systems in our sample host markedly planar subsets, which could be characterized as satellite planes by certain more permissive observational procedures. For instance, Fig.~\ref{fig:large} displays the satellite distribution around Systems 1 and 2, the two most massive and most isotropic in the New Horizon sample, both at relatively dense nodes of the cosmic web. Best-fit planes are shaded in green and a random line-of-sight in the plane is indicated as a black arrow.

It is visible that system 1 hosts a number of inner satellites clustered in a planar configuration that dominates in the identification of the best-fit plane. Conversely, System 2 hosts two flattened, elongated subsets of satellites, roughly at $90^o$ from each other. The best-fit plane is, in this case, dominated by the outer subset, which also shows signs of corotation. Interestingly, each flattened structure is found to be aligned to a distinct large cosmic filament branching into the host halo.

These examples show that past channels of accretion can pervade as planar configurations even within mostly isotropic distributions of satellites. This further confirms the role of the cosmic web in shaping  planes of satellites and provides an explanation to the identification of multiple planes around systems like Centaurus A.

\bibliography{bibliography}{}
\bibliographystyle{aasjournal}



\end{document}